\documentclass[a4paper,12pt]{article}
\pdfoutput=1
\usepackage{amssymb,amsmath,bm}
\usepackage{a4wide}
\usepackage{color}
\usepackage{slashed}
\usepackage{cite}
\usepackage{stackengine}
\usepackage{hyperref}
\usepackage{units}
\usepackage{relsize}
\usepackage[latin1]{inputenc}
\usepackage{amsfonts}
\usepackage{lscape}
\usepackage{amsthm}
\usepackage{booktabs}
\usepackage{array}
\usepackage{rotating}
\usepackage{float}
\usepackage{multirow}
\usepackage{graphicx,rotating,amsmath,multirow,xcolor,colortbl,xcolor}
\usepackage{hyperref,pdflscape,slashed}
\usepackage{upgreek}

\definecolor{rosso}{cmyk}{0,1,1,0.4}
\definecolor{rossos}{cmyk}{0,1,1,0.55}
\definecolor{rossoc}{cmyk}{0,1,1,0.2}
\definecolor{blu}{cmyk}{1,1,0,0.3}
\definecolor{blus}{cmyk}{1,1,0,0.6}
\definecolor{bluc}{cmyk}{1,1,0,0.1}
\definecolor{verde}{cmyk}{0.92,0,0.59,0.25}
\definecolor{verdec}{cmyk}{0.92,0,0.59,0.15}
\definecolor{verdes}{cmyk}{0.92,0,0.59,0.4}
\definecolor{Gray}{gray}{0.95}

\font\tenrsfs=rsfs10 at 12pt
\font\sevenrsfs=rsfs7
\font\fiversfs=rsfs5
\newfam\rsfsfam
\textfont\rsfsfam=\tenrsfs
\scriptfont\rsfsfam=\sevenrsfs
\scriptscriptfont\rsfsfam=\fiversfs
\def\mathscr#1{{\fam\rsfsfam\relax#1}}

\usepackage[small,bf]{caption}
\hypersetup{colorlinks,bookmarksopen,bookmarksnumbered,
linkcolor=black,pdfstartview=FitH,urlcolor=black,citecolor=black}

\newcommand{\lsim}{\stackrel{<}{_\sim}}
\newcommand{\gsim}{\stackrel{>}{_\sim}}

\newcommand{\bea}{\begin{eqnarray}}
\newcommand{\eea}{\end{eqnarray}}
\newcommand{\beq}{\begin{equation}}
\newcommand{\eeq}{\end{equation}}

\counterwithin*{equation}{section}

\def\pv{P}
\def\tpv{p}
\def\pvc{p_{D \!\!\!\!\! \not\,\,\,\,\,}}
\def\pvg{p_G}
\def\vf{\varphi}
\def\vfup{{\mathrel{\raisebox{-1.7pt}{\_}}}\mkern -6.6mu\varphi}
\def\GAUG{\xi}
\def\SdS{S_{\! \mathsmaller{\mathsmaller {\rm dS}}}}
\newcommand{\lambdabar}{{\mkern0.75mu\mathchar '26\mkern -9.75mu\lambda}}
\def\lambb{\lambda}
\def\LAMB{\lambdabar}
\def\lah{\lambda_{\scriptscriptstyle H}}
\def\fun{\omega}
\def\Ai{{\rm Ai}}
\def\Bi{{\rm Bi}}
\def\ett{g_\epsilon}
\def\WAV{\zeta}
\def\xE{x_{\! \mathsmaller E}}
\def\vfupE{\vfup_{\! \mathsmaller E}}

\def\vfupud{\vfup_{\mathsmaller{\mathsmaller \uparrow \downarrow}}}
\def\vfupu{\vfup_{\mathsmaller{\mathsmaller \uparrow}}}
\def\vfupd{\vfup_{\mathsmaller{\mathsmaller  \downarrow}}}

\def\gupm{g_{\mathsmaller{\mathsmaller{\mathsmaller 1}}}^{\mathsmaller{\mathsmaller{\mathsmaller{\mathsmaller{\mathsmaller{\mathsmaller (\pm)}}}}}}}
\def\gdpm{g_{\mathsmaller{\mathsmaller{\mathsmaller 2}}}^{\mathsmaller{\mathsmaller{\mathsmaller{\mathsmaller{\mathsmaller{\mathsmaller (\pm)}}}}}}}
\def\gudpm{g_{\mathsmaller{\mathsmaller{\mathsmaller 1,2}}}^{\mathsmaller{\mathsmaller{\mathsmaller{\mathsmaller{\mathsmaller{\mathsmaller (\pm)}}}}}}}
\def\gudp{g_{\mathsmaller{\mathsmaller{\mathsmaller 1,2}}}^{\mathsmaller{\mathsmaller{\mathsmaller{\mathsmaller{\mathsmaller{\mathsmaller (+)}}}}}}}
\def\gudm{g_{\mathsmaller{\mathsmaller{\mathsmaller 1,2}}}^{\mathsmaller{\mathsmaller{\mathsmaller{\mathsmaller{\mathsmaller{\mathsmaller (-)}}}}}}}
\def\gdum{g_{\mathsmaller{\mathsmaller{\mathsmaller 2,1}}}^{\mathsmaller{\mathsmaller{\mathsmaller{\mathsmaller{\mathsmaller{\mathsmaller (-)}}}}}}}
\def\Phiudp{\Phi_{\mathsmaller{\mathsmaller{\mathsmaller 1,2}}}^{\mathsmaller{\mathsmaller{\mathsmaller{\mathsmaller{\mathsmaller{\mathsmaller (+)}}}}}}}
\def\Phiudm{\Phi_{\mathsmaller{\mathsmaller{\mathsmaller 1,2}}}^{\mathsmaller{\mathsmaller{\mathsmaller{\mathsmaller{\mathsmaller{\mathsmaller (-)}}}}}}}
\def\Phidup{\Phi_{\mathsmaller{\mathsmaller{\mathsmaller 2,1}}}^{\mathsmaller{\mathsmaller{\mathsmaller{\mathsmaller{\mathsmaller{\mathsmaller (+)}}}}}}}
\def\Phidum{\Phi_{\mathsmaller{\mathsmaller{\mathsmaller 2,1}}}^{\mathsmaller{\mathsmaller{\mathsmaller{\mathsmaller{\mathsmaller{\mathsmaller (-)}}}}}}}
\def\Phiupm{\Phi_{\mathsmaller{\mathsmaller{\mathsmaller 1}}}^{\mathsmaller{\mathsmaller{\mathsmaller{\mathsmaller{\mathsmaller{\mathsmaller (\pm)}}}}}}}
\def\Phidpm{\Phi_{\mathsmaller{\mathsmaller{\mathsmaller 2}}}^{\mathsmaller{\mathsmaller{\mathsmaller{\mathsmaller{\mathsmaller{\mathsmaller (\pm)}}}}}}}
\def\Phiudpm{\Phi_{\mathsmaller{\mathsmaller{\mathsmaller 1,2}}}^{\mathsmaller{\mathsmaller{\mathsmaller{\mathsmaller{\mathsmaller{\mathsmaller (\pm)}}}}}}}

\def\eq#1{eq.~(\ref{#1})}
\def\fig#1{fig.~\ref{#1}}
\def\sec#1{sect.~\ref{#1}}

\definecolor{shamrockgreen}{rgb}{0.0,0.62,0.38}

\begin{document}

{\hfill CERN-TH-2021-077}

\vspace{2cm}

\begin{center}
\boldmath

{\textbf{\LARGE Self-Organised Localisation}}

\unboldmath

\bigskip

\vspace{0.4 truecm}

{\bf Gian F. Giudice},$^a$ {\bf Matthew McCullough},$^{a,b^*}$ {\bf Tevong You}$^{a,b,c}$
 \\[5mm]

{$^a$\it CERN, Theoretical Physics Department, Geneva, Switzerland}\\[2mm]
{$^b$\it DAMTP, University of Cambridge, Cambridge, UK}\\[2mm]
{$^c$\it Cavendish Laboratory, University of Cambridge, Cambridge, UK}\\[2mm]

\vspace{2cm}

{\bf Abstract }
\end{center}

\begin{quote}
We describe a new phenomenon in quantum cosmology: self-organised localisation. When the fundamental parameters of a theory are functions of a scalar field subject to large fluctuations during inflation, quantum phase transitions can act as dynamical attractors. As a result, the theory parameters are probabilistically localised around the critical value and the Universe finds itself at the edge of a phase transition. We illustrate how self-organised localisation could account for the observed near-criticality of the Higgs self-coupling, the naturalness of the Higgs mass, or the smallness of the cosmological constant.
\end{quote}

\thispagestyle{empty}
\vfill

{\smaller{$^*$\it On leave.}}

\newpage

\tableofcontents

\newpage

\section{Introduction}
\label{sec:intro}

The paradigm of symmetry underlies the construction of the Standard Model and General Relativity (SM+GR) and thus, arguably, the most successful scientific effective field theory ever created.  It is jarring that, after decades of symmetry-based speculation, the paradigm has thus far failed to uncover the microscopic origins of SM+GR. Once we add the experimental information about the value of the cosmological constant and the lack of new physics at the LHC, it becomes natural to conclude that perhaps it is time for the paradigm to shift.  If so, the questions we seek to answer remain, but the way we hope to answer them must change, and not adiabatically.  The trail beyond symmetry has few signposts and must be forged anew.  There are, however, some breadcrumbs.  The Higgs naturalness problem, the cosmological constant problem, and the peculiar feature of Higgs vacuum metastability all share one feature in common: criticality.  The measurement of the Higgs mass, together with a high-energy extrapolation of the Standard Model, have revealed that the Universe may exist in a near-critical state, with two phases subsisting side-by-side, as if we lived in water while ice is but just a fluctuation away.  The hierarchy problem also exhibits the hallmark of criticality, with a mass parameter seemingly tuned arbitrarily close to the symmetry-breaking point, as if the Universe couldn't decide between an ordered or disordered phase, finally opting to live in a state as close to indecision as possible, with just a smidgen of order.  So too for the cosmological constant, where we find ourselves teetering inexplicably close to the critical boundary between explosion and implosion, apparently opting, by a slither, for the former.  All of these observations suggest criticality, viewed as a coexistence of phases, as a potential signpost towards a new paradigm.

If these examples are not the product of symmetry, then criticality must be arrived at by other means.  If, in some competing paradigm, criticality emerges out of a non-critical state in a form of self-organised criticality~\cite{Bak:1987xua}, then both a timeline and a mechanism for the organisation are required.  It is natural to suppose that cosmological evolution could furnish both.  In this work we suggest that fundamental criticality, as classified by Ehrenfest~\cite{Ehrenfest} in terms of finding a Universe balanced on the precipice between two different phases of existence, may actually be a generic cosmological occurrence.  We will show that, in relatively typical circumstances, the cosmological evolution of empirical parameters is towards criticality, not from it.

The emergence of order out of disorder is not a new concept, rather it has inspired the development of some of the most universal conceptual advancements in theoretical physics.  The prototypical ferromagnet furnishes a familiar and cherished example.  As the temperature of the system is lowered below some critical value,  order spontaneously emerges in the form of a bulk magnetisation, with the prevalence of the order increasing to completeness at zero temperature.  The nature of the classical transition between the two phases is diagnosed by a change in the symmetry of the ground state and also by the analytic properties of the magnetisation as a function of temperature.  A first-order phase transition occurs for a discontinuity in the magnetisation and second-order for a discontinuity in its derivative.

Phase transitions need not be classical.  In a quantum phase transition it is not the temperature which varies, but some background field such as the magnetic field. The system may be held at a constant temperature, but a transition between two distinct phases occurs as the background field $\phi$ is varied across some critical threshold.  In practice, in the scalar potential, $\phi$ is coupled as $V =(\phi -\phi_c ) \, \mathcal{O}$ to some operator $\mathcal{O}$ whose expectation value changes as $\phi$ passes through some critical value $\phi_c$.  If $\langle \mathcal{O} \rangle$ is discontinuous across the critical point, as in a first-order phase transition, then $V' = \partial V/\partial \phi$ is discontinuous.  If, on the other hand, $\langle \mathcal{O} \rangle$ changes continuously from $\langle \mathcal{O} \rangle = 0$ for $\phi>\phi_c$ to $\langle \mathcal{O} \rangle \propto \phi_c -\phi$ for $\phi<\phi_c$, then $V''$ is discontinuous at the critical point.  In particle physics parlay, it is the tadpole (first-order) and the mass (second-order) of the scalar field, respectively, which are discontinuous.  In this work, we seek to study the role of quantum phase transitions in the early Universe, with a particular focus on first-order transitions.

Classical phase transitions have long been studied in cosmology and are known to have been instrumental for the evolution of the Universe, as exemplified by the QCD and the electroweak phase transitions, not to mention the many speculative hypotheses about other critical phenomena related to inflation, grand unification or quantum gravity. While work remains to be done, the theory of classical cosmological phase transitions is well established, with the pattern of symmetry breaking and derivative discontinuities, manifesting itself at a grander scale in the laboratory of the cosmos.

An obvious difference with cosmology is that in the laboratory one has the freedom to increase or decrease temperatures, background fields, or other parameters at will, and sit as close to, or far from, the critical points of classical or quantum phase transitions as desired. On the other hand, the inevitable march of the Hubble rate to lower values means that we may not reverse cosmological evolution to replay, or tune parameters to sit arbitrarily close to any possible cosmological phase transition. In the case of classical cosmological phase transitions, our knowledge about the thermal history of the Universe allows us to reconstruct with reasonable confidence the conditions for their occurrence. However, the situation is more intricate for quantum phase transitions, as their occurrence depends on the cosmological evolution and fate of the $\phi$ background field.

In this work we establish some aspects of the theory of cosmological quantum phase transitions. Unexpectedly we find that, under circumstances which are not atypical, inflation in general relativity, quantum fluctuations of scalar fields seeded by the inflating geometry and the discontinuities present at a quantum critical point conspire in a special interplay that leads to a universal phenomenon. When these features borrowed from the main pillars of modern physics are combined together, they cooperate to localise scalar fields exponentially close to the critical point.  

The feature of a critical point acting as an attractor is closely akin to self-organised criticality~\cite{Bak:1987xua}. For this reason, we will call Self-Organised Localisation (SOL) the phenomenon discussed in this paper, in which a period of inflation induces a dynamical attraction towards the critical point of a quantum phase transition. A very interesting, and related, point of view has been presented in refs.~\cite{Khoury:2019yoo,Khoury:2019ajl,Kartvelishvili:2020thd}. Although the approach followed in refs.~\cite{Khoury:2019yoo,Khoury:2019ajl,Kartvelishvili:2020thd} is different from ours, an important common aspect is the use of methods in quantum statistical mechanics and critical phenomena as central elements for the description of physical processes in eternal inflation.

The prediction of near-criticality based on SOL has a probabilistic nature. This is not surprising, as we are dealing with a quantum phenomenon. Just as physicists a hundred years ago had to resign to the fact that phenomena at the atomic level must be described probabilistically, so too predictions in the domain of quantum cosmology may have a probabilistic nature. The important difference is that, while in atomic physics the statistical sample is provided by many laboratory measurements, in our setting the statistical sample is given by different patches of the Universe not necessarily in causal contact. The information about the statistical ensemble is distributed in the spacetime geography of the full multiverse, even beyond the event horizon of our observable patch. This causes several well-know interpretation problems that will also be discussed in this paper.

The consequences of SOL for predicting physical parameters are profound, as SOL represents a radical change of perspective with respect to the usual intuition based on Effective Field Theories (EFT). When physical parameters in an EFT are functions of more fundamental parameters that belong to an underlying UV theory, they are expected to obey a specific structure dictated by quantum mechanics and symmetry properties. In particular, the smallness of an EFT parameter should be accounted for by the selection rules defined by symmetry or else it is destabilised by large quantum fluctuations. SOL defies this logic without modifying the power counting based on symmetry quantum numbers and without introducing new-physics energy scales. The SOL reasoning is that some EFT parameters, which are functions of scalar fields belonging to an underlying theory, are attracted towards their critical values as the result of the evolution during inflation of the fields governing the dynamics of the corresponding quantum phase transition.

The mechanism of SOL may have various applications in the interface between particle physics and cosmology. In this paper we start the exploration of the phenomenon and discuss its implications for the determination of Standard Model parameters. In \sec{sec:scalar} we give some introductory material about the dynamics of light scalars in an inflationary background. In \sec{sec:SOL} we describe the general features of SOL, studying specific examples of potentials that exhibit critical behaviour. Next, we discuss how SOL could address some of the outstanding open questions in fundamental physics:
near-criticality of the Higgs self-coupling (\sec{sec:nearcrit}), Higgs naturalness (\sec{sec:natur}) and the smallness of the cosmological constant (\sec{sec:cc}). In \sec{sec:concl} we give a comprehensive summary of our study, presenting all the key results. Finally, the appendix, written in a self-contained form, outlines some general properties of the stochastic equation and gives a compendium of analytical solutions.

\section{Scalar Fields in an Inflationary Universe}
\label{sec:scalar}

\subsection{The Stochastic Approach}

During inflation, scalar fields are subject to quantum fluctuations fuelled by the Hubble rate.  In the stochastic approach~\cite{Vilenkin:1983xq,Starobinsky:1986fxa,Starobinsky:1986fx,Rey:1986zk,Goncharov:1987ir,Sasaki:1987gy,Nakao:1988yi,Kandrup:1988sc,Nambu:1987ef,Nambu:1989uf,Mijic:1990qx,Salopek:1990re,Linde:1993nz,Linde:1993xx,Starobinsky:1994bd}, the field evolution is governed by a Langevin-like equation where quantum fluctuations correspond, qualitatively, to a random walk.  Starting from an initial distribution of field values in a patch of the Universe, the distribution changes as the patch grows and the field values fluctuate.  After a sufficiently long time, with respect to a given foliation, the distribution will take a stationary form which is universal, in the sense that is independent of  initial conditions but is specified only by field boundary conditions. This asymptotic distribution can in principle be used to determine the relative likelihood of different values of $\phi$ at the end of inflation. 

To determine this distribution one could calculate the average of a large number of random walks using the Langevin approach. Equivalently, one may instead use a Fokker-Planck (FP) equation which describes the average behaviour of solutions to the Langevin problem \cite{Starobinsky:1986fx,Goncharov:1987ir}.  The relevant FP equation, in proper time and with Ito ordering for the diffusion term~\cite{Winitzki:1995pg, Vilenkin:1999kd}, is
\beq
\frac{\partial}{\partial \phi} \left[   \frac{\hbar}{8 \pi^2} \frac{\partial (H^{3}\, \pv_{\rm FP} )}{\partial\phi} +\frac{V' \, \pv_{\rm FP}}{3 H} \right]  = \frac{\partial \pv_{\rm FP}}{\partial t}  ~,
\label{eq:FP}
\eeq
where $\pv_{\rm FP}(\phi,t)$ is a normalised time-evolving probability distribution, $H$ is the Hubble rate, $V(\phi)$ is the scalar potential and $V'$ its first derivative.  

Instead of considering the probability distribution $\pv_{\rm FP}$, for our purposes we are interested in determining the volume-weighted distribution of field values, found from calculating $\langle e^{3 H t} \rangle$ averaged over the random walks.  This distribution will be referred to as $\pv(\phi,t)$ and, as a function of proper time, is a solution to the volume-weighted Fokker-Planck (FPV) equation~\cite{Nambu:1988je, Nakao:1988yi, Nambu:1989uf, Linde:1993xx}
\beq
\frac{\partial}{\partial \phi} \left[   \frac{\hbar}{8 \pi^2} \frac{\partial (H^{3}\, \pv )}{\partial\phi} +\frac{V' \, \pv}{3 H} \right] +3H\pv = \frac{\partial \pv}{\partial t}  ~.
\label{eq:FPV}
\eeq
Here $\pv(\phi,t)$ is \emph{not} a normalised time-evolving probability distribution, but instead can be thought of as describing the volume fraction distribution for $\phi$ obtained after averaging over a large number of field trajectories. Once initial and boundary conditions for $\pv(\phi,t)$ are set, the evolution of the distribution may thus be determined. 

It is important to be clear on how to interpret the distributions $\pv_{\rm FP}$ and $\pv$.  The function $\pv_{\rm FP}$ describes the probability distribution of measuring the field value $\phi$ after some proper time when $\phi$ has undertaken a random walk, as dictated by the Langevin equation.  The function $\pv$ is essentially reporting the probability distribution multiplied by the volume expansion factor corresponding to the field value $\phi$.  As a result, in the multiverse we may assign a sort of probabilistic interpretation to this quantity.  For simplicity, suppose we begin with a large number of patches, all with the same initial field value, and let them undertake their random walk, which we stop at some specific proper time.  We now add up the volume of all patches, with their associated final field value, chop the entire multiverse up into equal-sized portions and make a histogram of the field value in each portion.  This is the distribution reported by $\pv$.

We may not, as individual observers within our own Hubble patch, measure the distribution of field values in the multiverse and this leads to several open questions about the probabilistic interpretation of $\pv$ as an `observable'. In the following section, we will discuss some of these conceptual difficulties about the probabilistic interpretation of the solutions to the FPV equation.

\subsection{On Time, Eternal Inflation and Quantum Gravity}
\label{sec:time}
We wish to calculate the volume distribution of $\phi$ values and ascribe to it a probabilistic interpretation.  One could take an abstract view of $\pv$, removed from questions of observers, and interpret it probabilistically at any timescale.  Indeed, this is the approach often taken to consider important dynamical questions such as the onset of eternal inflation and the approach to steady states.  On the other hand, as also often noted, if we wish to interpret $\pv$ as representing the likelihood of measurements of fundamental parameters performed by observers we must define observers more carefully.  Importantly, these two subtly different interpretations of $\pv$ may have profoundly different outcomes if inflation is eternal.

Since observers can only exist on patches of the multiverse which have reheated after inflation has terminated, a meaningful statistical sample to compare probabilities across the multiverse is given by patches that live on the 3-volume hypersurface of all reheating events in spacetime, which is called the `reheating surface' (for a discussion see, e.g., chapter 6 of ref.~\cite{Winitzki:2008zz}). 

In particular, this implies that the field $\phi$, which governs the SOL dynamics, in no sense can play the role of the inflaton.  By definition, inflation would terminate at a specific $\phi$ value and thus all `observers' would measure an essentially unique value of $\phi$, determined by the reheating process, instead of being part of a statistical distribution $\pv$ which carries the fingerprints of SOL. However, note that this does not preclude the possibility that $\phi$ could give the dominant contribution to the vacuum energy density in certain field regions.

While the reheating surface is the bridge between the stochastic approach during inflation and probabilistic predictions of observables in inhabitable universes, its calculation requires knowledge of the specific model of inflation. Given a certain model we could, in principle, determine the joint $\phi$ and inflaton distribution, from which we could derive the probabilistic distribution of physical observables. However, given that the true model of inflation is still unknown, in this work we treat inflation as a fixed background hosting the $\phi$ dynamics, with the goal of determining the inflaton-independent properties of the $\phi$ distribution.

The problem of identifying the reheating surface is further exacerbated by the fact that, for the parameters chosen in all the phenomenological applications that we will consider in this work, the SOL dynamics  occurs in a regime where inflation is eternal~\cite{Vilenkin:1983xq,Guth:2007ng,Linde:1986fc, Linde:1986fd}. This does not, however, imply that all applications of SOL necessarily require eternal inflation as a fundamental ingredient. In non-eternal inflation the definition of the reheating surface is unambiguous: all spacetime trajectories eventually reach the reheating stage and inflation ends everywhere within a finite time. The most likely field value measured by an observer corresponds to the patch that occupies the largest volume fraction of the total reheating surface. Instead, in eternal inflation, the reheating surface is infinite because at all times there exist trajectories that are still inflating. This leads to the well-known `measure problem' (as reviewed, for example, in refs.~\cite{Winitzki:2008zz,Freivogel:2011eg}), which afflicts any prediction from eternal inflation.

In eternal inflation, probabilities obtained with any volume-based measure are independent of the initial conditions and instead correspond to a stationary state which expands, in some time foliation, uniformly at a constant rate~\cite{Aryal:1987vn, Nambu:1989uf, Linde:1993xx, Linde:1993nz, GarciaBellido:1994ci, Winitzki:2001np}.  This stationarity follows automatically from the eternal nature of inflation and is independent of choice of cut-off or time foliation (for a discussion see again chapter 6 of ref.~\cite{Winitzki:2008zz}), as it can be understood from the volume-weighting, which exponentially favours observers at the latest possible times.  As a result, in practical terms the fraction of all observers who have a finite proper time in their past light cone is zero, even though their spacetime is past-incomplete (see e.g.~\cite{Borde:1993xh,Vilenkin:1995yd,Borde:2001nh,Mithani:2012ii,Susskind:2012yv,Susskind:2012xf} for discussions on these aspects, including the distinction between technical and practical finiteness).  This renders any question of dynamics from the perspective of observers moot, since only a stationary solution remains at infinite time. However, this needs not be necessarily the case for an interpretation of $\pv$ without observers, where dynamics of the solution can occur at finite times.

Problematically, the specific form of this stationary distribution is not unambiguous.  As mentioned, in practice the `clock' which is stopped when inflation ends is the inflaton itself when it arrives at the reheating point on the potential.  Absent a model of inflation we have opted to consider a time-like variable as a proxy for the inflaton, noting that this ansatz may or may not be valid for different inflationary models. Moreover, the appropriate choice of time parametrisation depends on how the infinite reheating surface is regularised.  As the solution $\pv$ depends on the prescription of time foliation we say it has a residual gauge dependence.  Physically, this can be understood by observing with a specific example that, instead of computing the $\phi$ distribution that enters a steady-state with respect to proper time, one could also consider a distribution whose $\phi$ functional form becomes independent of the number of $e$-foldings that have occurred.

To make this explicit, following refs.~\cite{Winitzki:1995pg,Linde:2010xz}, we define a one-parameter family of gauges $\GAUG$ (with $0\le \GAUG \le 1$) characterised by the time coordinate $t_\xi$ such that $dt_\xi/dt = (H/H_0)^{1-\xi}$, where $H_0$ is a reference constant value of the Hubble rate. In terms of the new time coordinate, \eq{eq:FPV} becomes
\begin{equation}
\frac{\partial}{\partial \phi} \left[   \frac{\hbar}{8 \pi^2} \frac{\partial (H^{2+\xi}\, \pv )}{\partial\phi} +\frac{V' \, \pv}{3 H^{2-\xi}} \right] +3H^\xi \pv = H_0^{\xi -1}\, \frac{\partial \pv}{\partial t_\xi}  ~,
\end{equation}
where we have redefined $\pv \to (H_0/H)^{1-\xi} \pv$. 
Proper-time gauge corresponds to $\xi=1$ and $e$-folding gauge to $\xi=0$, with values in between interpolating between the two. 

After expanding $H$ around a background inflationary value $H_0$ and dropping sub-dominant terms, the only remaining effect of $\xi$ is to multiply the volume term. The residual $\xi$-dependence is therefore a characteristic of the FPV extension of the usual Fokker-Planck treatment. The parameter $\xi$ encodes the ambiguity coming from our inability to determine the reheating surface and, therefore, the statistical sample of universes over which probabilities are computed. As we will show in the following, many of the consequences of SOL are fairly robust against variations of $\GAUG$, with exponential localisation persisting as long as $\GAUG \ne 0$. In particular, in all applications that we will consider, the exponential localisation is lost only if $\GAUG =0$ with an extraordinary accuracy, a case which corresponds to a strict choice of $e$-folding gauge or scale factor measure~\cite{DeSimone:2008bq, Bousso:2008hz, DeSimone:2008if,Bousso:2012tp,Vilenkin:2019mwc,Olum:2021pux}, for which volume-growth effects are absent.  Contrary to the case $\GAUG =0$, in measures for which volume effects can dominate at late times a number of paradoxes arise, particularly the `Youngness Paradox'.  The extent to which this paradox afflicts interpretations of the $\phi$ distributions calculated here will be discussed in detail in \sec{sec:measure}. 

A different concern, detailed in refs.~\cite{ArkaniHamed:2007ky,Creminelli:2008es,Dubovsky:2008rf,Dubovsky:2011uy}, is that an observer in an asymptotically flat region can only measure a limited number of inflationary modes, bounded by the entropy of the surrounding dS space.  This argument limits the amount of dS space outside the horizon whose modes would eventually enter the horizon, giving the following bound on the number of $e$-foldings possible in non-eternal inflation
\beq
N<{\SdS} = \frac{8\pi^2 M_P^2}{\hbar \, H^2}  ~.
\label{entb}
\eeq
In our phenomenological applications of SOL, the bound in \eq{entb} is grossly violated. However, this only reiterates SOL's need for eternal inflation when applied to questions of phenomenological relevance to fundamental physics. In eternal inflation an observer can never access all of the spacetime volume after reheating, thus \eq{entb} is not a reason for concern. 

The presence of eternal inflation raises the question of whether the semi-classical stochastic approach to the dynamics of $\phi$ remains valid eternally.  Indeed, in \cite{Dvali:2013eja,Dvali:2017eba} it was argued that, when the total number of $e$-foldings violates the bound of \eq{entb}, the semi-classical treatment breaks down due to the large occupation number of gravitons.  It is not clear, however, in what way the stochastic treatment should break down.   For instance, in a QFT it is possible for UV effects to have a significant impact on long timescales without the EFT approach ever breaking down.\footnote{An EFT can undergo a change of regime after a sufficiently long timescale, with UV quantum effects dominating over IR classical effects. However, this does not necessarily imply a breakdown of the effective theory, as long as the energies involved are much smaller than the cutoff. To illustrate this point with an example, consider proton scattering processes at nuclear energies $E_{IR}$, which are appropriately described by a low-energy effective theory. In general, the low-energy theory must include higher-dimensional operators (e.g.~generated by GUT interactions at the mass scale $M_{UV}$) that mediate proton decay with a rate $\Gamma \sim  E_{IR}^5/M_{UV}^4$. The effects of UV interactions are negligible for normal scattering experiments, but they become dominant after a time $t\sim 1/\Gamma$ when they completely change the state of the system as protons disappear from the colliding beams. Although UV interactions dominate the outcome of the experiment, there is no breakdown of the effective-theory description. In other words, as long as we scatter protons at low energies, the processes are adequately described by the effective theory even without detailed knowledge of the UV microscopic dynamics.

Of course, it is not guaranteed that these EFT-based considerations should hold for gravity as well, but let us suppose that this is the case.
The FPV solutions show a change of regime after a time $t \sim 1/\Gamma$, where $\Gamma \sim H^3/M_P^2$ is the typical rate of quantum-gravity processes. Just like in the GUT example, beyond this critical time the dynamics is dominated by UV quantum effects. Nonetheless, as long as the curvature ($\sim H^2/M_P^2$) is small, we posit that the GR effective theory gives an appropriate description and no knowledge of the full quantum-gravity dynamics is needed. Exceeding the critical timescale only gives enough time for the system to become dominated by UV quantum effects, but does not necessarily imply a breakdown of the effective-theory validity.} Returning to the inflationary case, recent calculations appear to support the validity of the stochastic approach even at long times (see, for instance, \cite{Creminelli:2008es,Dubovsky:2008rf,Dubovsky:2011uy,Lewandowski:2013aka,Baumgart:2019clc,Cohen:2020php} for related discussions and in particular \cite{Gorbenko:2019rza} for a recent discussion of the validity of the stochastic approach in eternal inflation).  Thus, while the phase transition to eternal inflation does occur when \eq{entb} is satisfied, which is a very nontrivial outcome, it is not clear that the stochastic description of $\phi$ dynamics should be called into question.  As a result, in our study we will take for granted that, even in an eternally inflating background, the dynamics of $\phi$ can be described through the semi-classical stochastic approach.

A final question concerns the Swampland conjectures (see e.g.~ref.~\cite{Palti:2019pca} for a review).  There are two facets to consider.  The first is inflation.  The Swampland de Sitter conjecture is~\cite{Matsui:2018bsy}
\beq
|\nabla V| > c \frac{V}{M_P}~, 
\eeq
where $c$ is an $\mathcal{O}(1)$ parameter. This suggests that the slow-roll parameter $\epsilon$ should also be an $\mathcal{O}(1)$ parameter, in contradiction with slow-roll itself.  Thus a na\"ive contradiction with the Swampland conjectures arises already within the context of generic, but not necessarily all, slow-roll inflationary models.  Furthermore, the condition for eternal inflation, written in terms of the Swampland de Sitter conjecture \cite{Obied:2018sgi}, is
\beq
\frac{H}{M_P} \gtrsim 2 \pi c ~.
\eeq
Thus, unless the Hubble rate is parametrically close to the Planck scale, on the edge of the validity of a semi-classical approach, the eternally inflating backgrounds we will assume throughout are likely in the Swampland.  Similarly, the $\phi$ potential required for SOL will also typically violate both the Swampland de Sitter Conjecture and the Distance Conjecture \cite{Ooguri:2006in,Klaewer:2016kiy}, thus it would seem that both the inflationary sector and SOL sectors are firmly in the Swampland.  Given the incertitude of these conjectures we will not attempt to address this tension here, although it should be kept in mind that the fundamental ingredients for SOL appear, at least at first glance, to be at odds with UV-completion within string theory.  For further discussion on this see, for example, \cite{Matsui:2018bsy,Dimopoulos:2018upl,Kinney:2018kew,Brahma:2019iyy,Rudelius:2019cfh,Wang:2019eym,Blanco-Pillado:2019tdf}.

\subsection{The Measure Problem}
\label{sec:measure}

Many solutions have been proposed to the measure problem. One of the earliest attempts was the proper time cut-off measure\cite{Linde:1993nz,Linde:1993xx}. For finite proper time, the reheating surface is finite. As this time cut-off is taken to infinity the ratio of volume distributions remains constant. Predictions in proper time cut-off measure therefore correspond to the asymptotically stationary volume distribution in proper time. Unfortunately, this measure is in conflict with observation due to an exponential preference for younger vacua over older ones such as ours \cite{Tegmark:2004qd,Guth:2007ng,Linde:1994gy,Freivogel:2011eg} As well as predicting apparent violations of the Copernican principle that our location in the Universe is not special \cite{Linde:1994gy,Linde:1996hg}.\footnote{An additional potential problem relates to `Boltzmann Brains'.  However, since our models all predict reheated patches which must ultimately decay to AdS, either by tunnelling or $\phi$ rolling to AdS, it is not clear to what extent this problem would apply to our setup.}

\subsubsection*{The Youngness Paradox}

To see this Youngness Paradox, consider, for example, a metastable false vacuum for the inflaton $\chi$ at $\chi_A$ whose potential energy density is greater than that of a stable vacuum at $\chi_B$, where we assume reheating occurs relatively soon after tunnelling to vacuum B. The Hubble expansion rate at $\chi_A$ is greater than that at $\chi_B$, $H_A > H_B$, but in equilibrium the ratio of volume distributions $P_{A,B}(t) \equiv P(\chi, t)$ at $A$ and $B$ is constant so they must be expanding at the same rate,
\begin{equation}
	P_A(t) = P_A(0)e^{3H_A t} \quad , \quad P_B(t) = P_B(0)e^{3H_A t} \, .
	\label{eq:PBmeasure}
\end{equation}
This is due to a vacuum decay flux from $A$ to $B$ which maintains the steady state. The pre-existing $B$ vacuum from $t=0$ has a volume $P_B^{\text{old}}(t) = P_B(0) e^{3H_B t}$ after a proper time $t$. The new vacuum created by $A$ decays to $B$ is $P_B^{\text{new}}(t) = P_B(t) - P_B^{\text{old}}(t)$. The volume ratio of new to old vacua is then
\begin{equation}
	\frac{P_B^{\text{new}}(t)}{P_B^{\text{old}}(t)} = e^{3(H_A - H_B)t} - 1 \, .
\end{equation}
We therefore see that, in steady state where everything expands at the same rate, there is an exponential preference for newly created vacua over older ones.  This result generalises to slow-roll inflationary potentials.  In this case, reheating occurs whenever the inflaton reaches the reheating point on the inflaton potential.  However, on the trajectories leading up to that event, the inflaton will typically be much further up the potential with significantly higher Hubble rate.  

In this work, the field $\phi$ is not the inflaton and the $\phi$-distribution is sampled at the moment of reheating, which is unrelated to the $\phi$ dynamics.  There are three important differences between $\phi$ and the inflaton that must be kept in mind regarding questions such as the Youngness Paradox.
\begin{itemize}
\item The $\phi$ potential is extremely shallow and the field range exponentially larger than the Planck scale.  Thus if we repeat the above exercise the furthest up the potential that $\phi$ could have explored relative to the reheated patch is $|\Delta \phi| \lesssim H_B H_A t$.  For $t \sim 1/H_B$, as in our Universe at the present day, this becomes $|\Delta \phi| \lesssim H_A$.  In all applications we will consider the $\phi$-potential is so shallow that the difference in Hubble rate for $|\Delta \phi| \sim H_A$ is significantly smaller than the present day Hubble constant, thus there is no Youngness Paradox pertaining to $\phi$ in this regard.
\item In all of our applications we will also find that the $\phi$ distribution is highly localised in a field range over which the change in Hubble constant changes by an amount smaller than the present day Hubble constant, so no Youngness Paradox is associated with the stationary distribution of $\phi$ which is sampled.
\item  In all of our applications the $\phi$ potential is so shallow that subsequent to reheating $\phi$ rolls by a distance which changes the Hubble constant by an amount smaller than the present day value of the Hubble constant, so no Youngness Paradox is associated with the subsequent dynamics of $\phi$ after inflation, as compared to patches that reheated more recently.
\end{itemize}
Therefore, as regards the dynamics of $\phi$, patches of the Universe which are 13.7 billion years old have effectively the same value of $\phi$ and $V(\phi)$ as patches which have just reheated, hence there is no Youngness Paradox associated with $\phi$, even if we work in proper time gauge. 

This suggests that the Youngness Paradox is decoupled from the detailed form of the stationary $\phi$ distribution. However, it remains to understand if working in a proper-time-like gauge for $\phi$ dynamics could implicitly mean a Youngness Paradox is reintroduced in the inflaton sector by proxy. To see this, we now consider two proposed measures that avoid the Youngness Paradox for the inflaton and discuss their application when including the $\phi$ distribution. 

\subsubsection*{Scale factor measure}	

In our first example, we may employ scale factor cut-off measure in e-folding gauge for the inflationary sector, where e-foldings in the simple false vacuum model above is $N_A = H_A t$. Working with this variable is a constant relabelling of proper time as regards $\phi$ dynamics and the form of the stationary distribution for $\phi$ is unchanged. The inflationary sector in this measure does not suffer from a youngness problem, yet the stationary $\phi$ distribution is the same as in proper-time gauge and there is no Youngness Paradox stemming from the $\phi$ measure. We therefore see that stationarity for the $\phi$ distribution in one gauge does not necessarily imply stationarity for inflaton dynamics in the same gauge, hence questions concerning the Youngness Paradox are relegated to the assumptions made for the measure applied to the inflaton dynamics.

More generally, the decoupling of the measure paradoxes between $\phi$ and inflationary sectors can become more pronounced in asynchronous gauges which weight volumes according to spacetime events.  Such a prescription seems plausible given that the effective past-eternity of eternal inflation for the majority of observers may suggest that the important clock registers events according to the termination of inflation, rather than at some fixed time at which it began.  In this case conditioning on reheating gives the inflationary sector a special prominence as compared to the $\phi$ dynamics.

\subsubsection*{Stationary measure}

In our second example, one may consider the stationary measure for both the inflationary \emph{and} $\phi$ sectors whose prescription circumvents the Youngness Paradox in proper time for the inflationary sector \cite{Linde:2007nm,Linde:2008xf}.  Contrary to the previous discussion, in this case the \emph{same measure} is assumed for both fields. The toy model of eternal inflation in ref.~\cite{Linde:2010xz} is similar to the previous false vacuum setup, but after the inflaton $\chi$ tunnels from $A$ to $B$, it slow-rolls along a shallow potential lasting a fixed $N_B = H_B t_B$ e-foldings before reheating. In addition, there are two such shallow potentials: one in domain $B_-$ and the other in domain $B_+$, both with the same energy density. Reheating occurs at $\chi_-$ for the $B_-$ vacuum and at $\chi_+$ for $B_+$. They are expanding with the same Hubble rate $H_B$, but due to the difference in slow-roll distance to reach the point of reheating one has a smaller number of e-foldings than the other, $N_{B-} < N_{B^+}$. 

In this example setup we wish to compare the volume distributions in the two domains $B_-$ and $B_+$ at reheating (the post-reheating evolution is identical for the two so we neglect this here). Following \cite{Linde:2010xz}, this can be estimated as follows. The volume distribution at $A$ grows by inflationary expansion and shrinks due to decay to either $B_+$ or $B_-$ with equal rate $r < 3H_A/2$ per unit time,  
\begin{equation}
	\frac{d P_A(t)}{dt} = (3H_A - 2r) P_A(t) \quad \Rightarrow \quad P_A(t) = P_A(0) e^{(3H_A - 2r)t} \, .
\end{equation}
The total volume at $B_\pm$ increases by the rate $r$ from $A$ vacuum decay with an initial volume $P_A(t_A)$ determined by the time $t_A$ spent in the metastable vacuum $A$. There is also a fixed expansion factor from the slow-roll phase in $B$ that lasts $N_{B^{\pm}} = H_B t_{B^\pm}$ e-foldings. The total time is then $t = t_A + t_{B^\pm}$. We have
\begin{align}
	\frac{dP_{B^\pm}(t)}{dt} &= r e^{3 H_B t_{B^\pm}} P_A(t - t_{B^\pm}) \,  
	\label{eq:diffPBLinde} \\
	\Rightarrow \, P_{B^\pm}(t) &= \frac{r P_A(0)}{3H_A - 2r} e^{3N_{B^\pm}} \exp\left[\frac{(3H_A - 2r)}{H_B}(N - N_{B^\pm})\right] \, ,
	\label{eq:PBLinde}
\end{align} 
where we have dropped negligible corrections of $\mathcal{O}(H_B/H_A)$ in \eq{eq:diffPBLinde}.  The ratio of volume distributions at $B_+$ and $B_-$ then tends to a constant,
\begin{equation}
	\lim_{t\to\infty} \frac{P_{B^+}}{P_{B^-}} = e^{3(N_{B^+} - N_{B^-})}\exp\left[ -\frac{(3H_A - 2r)}{H_B}(N_{B^+} - N_{B^-}) \right] \, .
\end{equation}
We see that the volume distribution at $B_+$ is exponentially suppressed with respect to $B_-$. Proper time cut-off measure therefore favours the vacuum with a smaller number of e-foldings, $N_{B^-} < N_{B^+}$. The reason is that less e-foldings means more time spent inflating at a higher expansion rate in the metastable vacuum $A$.  

Now, in stationary measure we wish to compare the volume distributions at reheating in $B_\pm$, not from the total time $t$ but synchronised from the point where they were respectively created after tunnelling. This means undoing the volume expansion from sitting in the metastable vacuum $A$, since the time before tunnelling is irrelevant. The physically meaningful comparison is between the two vacua when they were created and expanded at the same rate. The stationary measure volume distribution modifies \eq{eq:PBLinde} to become
\begin{align}
	\left. P_{B^\pm}(t) \right|_{\text{stationary measure}} &=  \exp\left[-\frac{(3H_A - 2r)}{H_B}(N - N_{B^\pm})\right] P_{B^\pm}(t) \\
	&= \frac{r P_A(0)}{3H_A - 2r} e^{3N_{B^\pm}} \, ,
\end{align}
so the volume ratio is 
\begin{equation}
	\lim_{t\to\infty} \frac{P_{B^+}}{P_{B^-}} = e^{3(N_{B^+} - N_{B^-})} \, ,
\end{equation}
which now exponentially favours vacuum $B_+$ with longer e-foldings since there is no longer any gain from inflating in $A$.

If we add a scalar field $\phi$ to the above toy model then the volume distribution becomes a function of both $\phi$ and the inflaton $\chi$, $P(\chi, \phi, t)$. However, just like the post-inflationary evolution is identical at $B_+$ and $B_-$ and so does not affect the stationary measure procedure, so are the dynamics and distribution of $\phi$ the same for both domains. In practise this means we can write $P(\chi, \phi, t) \simeq P(\chi, t)P(\phi,t)$, assuming a negligible dependence of the $\phi$ distribution on the varying inflationary Hubble rate that is treated as approximately constant for the FPV dynamics of $\phi$. Moreover, the stationary measure mandates synchronisation when in the stationary regime with respect to all fields and processes, which implies $P(\phi,t)$ must be the asymptotically stationary distribution for $\phi$, $P(\phi) \equiv \lim_{t\to \infty} P(\phi,t)$. The main point is that the synchronisation procedure of stationary measure concerns only the inflaton sector and is independent of $\phi$. The inflationary volume expansion that is undone by the stationary measure is that of the inflaton which from the point of view of $P(\phi)$ is just an overall rescaling. Similarly, from the inflaton's point of view the $\phi$ distribution and energy density is a constant background that is sampled identically in both domains. 

\hfill

In summary we see that, while the various measure problems remain in eternal inflation, they apply in a decoupled manner to the inflationary and $\phi$ sectors.  Due to the localisation of the stationary $\phi$ distribution and the shallow $\phi$-potential there is no Youngness Paradox associated with $\phi$, even though some additional prescription must presumably be applied to resolve the measure problem paradoxes associated with the inflaton sector. The above two examples illustrate how this may be applied in the context of SOL. As a result of these considerations we expect that when the correct understanding and/or framework is found for eternal inflation resolving the various paradoxes, the qualitative features of SOL will remain.

With all of these independent, but interrelated, aspects in mind we may now proceed to detailed calculations of $\pv$.

\subsection{EFT Potential}
\label{sec:EFT}
To illustrate the nature of the dynamics of the field distribution, and the ultimate stationary configurations, we must solve the FPV equation explicitly. We will consider the following class of potentials 
\beq
V= 3H_0^2 M_P^2 + \ett^2 f^4 \fun (\vf) ~,~~~~
\fun(\vf)=  \sum_{n=1}^\infty \frac{c_n}{n!} \vf^n 
~,~~~~ \vf \equiv \frac{\phi}{f}~,~~~~ \fun(0)=0
~,
\label{potinEFT}
\eeq
which is characterised by three quantities: {\it (i)} a constant background vacuum energy $V_0=3H_0^2 M_P^2$ which drives the inflationary expansion; {\it (ii)} a scale $f$ which defines the field range of $\phi$ where we trust the EFT approach, such that $|\vf|\le 1$; {\it (iii)} a coupling constant $\ett$ which defines an overall size of the $\phi$ potential. The parameters $f$ and $\ett$ encode the physical information about $\phi$ interactions, while the Wilson coefficients $c_n$ are $\mathcal{O}(1)$ numbers with no large hierarchies, implying that $\fun(\vf)$ and all of its derivatives are also $\mathcal{O}(1)$ quantities across the field slice. 

We employ the EFT in the regime 
\beq
V_0= 3H_0^2 M_P^2 \gg \ett^2 f^4 ~.
\label{appinf}
\eeq
This could result from dynamics in which the vacuum energy density associated with $\phi$ gives only a small modulation of the constant background value $V_0$ that is primarily responsible for the inflationary process. But \eq{potinEFT} could also be viewed as an expansion around a generic field point (taken at $\vf =0$ with an appropriate coordinate choice) where $V_0=V(\vf \! = \! 0)$ is the leading constant term. In this case, $f$ is interpreted as the largest field excursion such that $|V(\phi)-V_0|\ll V_0$. For this reason, we will call {\it perturbative range} the field region in which \eq{appinf} is satisfied.

The structure in \eq{potinEFT} encompasses a general class of scalar theories, but in our applications we will be particularly interested in the case in which $\ett \ll 1$ and $f\gg M_P$. The hypothesis $\ett \ll 1$ is naturally realised when $\phi$ is a Goldstone boson coming from an underlying spontaneously broken global symmetry, while $\ett$ measures the amount of explicit symmetry breaking.

For potentials of the form in \eq{potinEFT}, we can use the condition in \eq{appinf} and expand the FPV equation at leading order in $\ett$
\beq
\frac{\alpha}{2} \frac{\partial^2 \pv}{\partial \vf^2}+ \frac{\partial \left(\fun ' \pv\right)}{\partial \vf} +\beta \fun \pv = \frac{\partial \pv}{\partial T}~.
\label{eq:fpv2}
\eeq
The dimensionless parameters $\alpha$, $\beta$ and $T$ are defined as
\beq
\alpha = \frac{3 \hbar H_0^4}{4 \pi^2 \ett^2 f^4} ~,~~~~ \beta = \frac{3\GAUG f^2}{2 M_P^2}~,~~~~ T = \frac{t}{t_R}  ~,~~~~  t_R = \frac{3 H_0}{\ett^2 f^2} ~.
\label{eq:params}
\eeq
We also reabsorbed the constant part of the expansion term through a redefinition
\beq
\pv \to e^{3 H_0 t} \pv ~.
\eeq
By separation of variables, we can write $\pv$ as a linear combination of the stationary distributions $\tpv$, whose field functional dependence does not change form with time,
\beq
\pv(\vf , T) = \sum_\lambb e^{\lambb T} \tpv (\vf ,\lambb) ~~.
\eeq
Since any initial state is built out of the eigenfunctions $\tpv$, the one with the greatest eigenvalue ($\lambb =\lambb_{\rm max}$) will eventually come to dominate the distribution in steady state, for any physical initial condition.

Let us consider the parameters entering \eq{eq:fpv2} and their interpretation. First take $t_R$.  The typical gradient of the EFT potential is $V'\sim \ett^2 f^3$ and the slow-roll velocity is $\dot{\phi}=V'/3 H_0$.  Thus the typical time taken to classically evolve along the potential is $\Delta t \sim 3 H_0 \Delta \phi/V'\sim t_R$.  If the dynamics is dominantly quantum, in the sense that quantum fluctuations exceed the rate of classical evolution, then relaxation is much slower. Thus $t_R$ represents an estimate of the minimum time it takes for relaxation to a stationary state. Indeed, explicit calculations show that the evolution time of the system is $\alpha \beta \, t_S$ for $\alpha \beta \ll 1$, and $\sqrt{\alpha \beta} \, t_S$ for $\alpha \beta \gg 1$, where $t_S={\SdS}/H_0$ is the largest timescale allowed by the entropy bound, see \eq{entb}. Thus, for $\alpha \beta \gg 1$, one is implicitly discussing relaxations on timescales so long that eternal inflation is an inevitable assumption.

The parameter $\alpha$ is proportional to $\hbar$ and thus it describes a quantum effect. 
The smaller $\alpha$ the more classical the evolution, with $\alpha\to0$ furnishing the classical limit. This result can also be understood by estimating, on dimensional grounds, the `thermal' dS corrections to the scalar potential as 
$\Delta V \propto T_{\rm dS}^4$ where $T_{\rm dS}=H_0/2\pi$ is the Gibbons-Hawking temperature \cite{Gibbons:1977mu}. As a result we see that $\alpha \sim T_{\rm dS}^4/V$ parameterises the scale of `thermal' dS corrections to the potential relative to the overall height of the potential.  Thus, if $\alpha \ll 1$, the evolution in the FP equation will be dominantly classical. However, the condition that the $\phi$ potential gives only a small contribution to the background vacuum energy, see \eq{appinf}, gives the lower bound $\alpha \gg 1/{\SdS}$.

Finally, $\beta$ parameterises the field range in Planck units. It also encodes all the information about the $\GAUG$-dependence in the perturbative FPV. In our applications, we will be interested in the case $\beta\gg1$, with a large number of $e$-foldings during which the field makes super-Planckian field excursions. Although the energy density always remains sub-Planckian, the regime $\beta\gg1$ is in conflict with the Swampland `Distance Conjecture' \cite{Ooguri:2006in}.

More relevant for this work, however, is that $\beta$ sets the scale of the volume effects, allowing to identify a third regime of parameter space by treating effects on purely perturbative grounds.  As discussed, if $\alpha\ll1$ then the dynamics is dominantly classical.  However, a quantum fluctuation combined with volume growth will have effects which scale, perturbatively, as $\alpha \beta$.  Thus, if $\alpha \beta \gg 1$ then, despite the fact that quantum fluctuations are small, we should expect to enter a regime in which the combination of quantum and volume effects dominates the dynamics.  It may seem that we are in presence of a strongly-coupled regime, but one can still trust the result because the FPV follows directly from the FP solution, which is entirely calculable also in this case.  Going further, if $\alpha^2 \beta \gg 1$ then NLO quantum effects combined with volume growth will also be greater than classical evolution, suggesting a transition to another regime of solutions.  In \sec{sec:PS} we will show more precisely how these changes of regime occur and what form the solutions take in each.  

\subsection{Boundary Conditions and Eigenvalues}
To calculate the stationary state at large times one must supplement the FPV equation with boundary conditions at the endpoints of the field range. Boundary conditions encode information of the dynamics beyond the EFT field range and their choice requires knowledge of the UV completion. From an EFT perspective, all we can do is to make reasonable guesses. A simple, but relatively generic, assumption is to take absorbing boundary conditions at the field endpoints $\pv (\phi_{\text{min},\text{max}} ) = 0$, which correspond to discarding every path which ventures beyond these boundaries.  Physically, this means that one considers only trajectories which remain within the EFT.  Alternatively, one could choose reflecting boundary conditions, $\pv' (\phi_{\text{min},\text{max}} ) = 0$, implying that UV physics repulses
any trajectory incident on the boundary.

If the potential $V$ and its first derivative are continuous functions of the scalar field then, for any linear combination of absorbing or reflecting boundary conditions, the Sturm-Liouville theorem ensures that the eigenvalue spectrum is uniquely determined. Hence, the asymptotic stationary state is fully determined by the choice of boundary conditions and any information about the initial condition is erased. 

\subsection{Positivity}
\label{sec:PS}

Volumes are positive quantities and probabilities are too, thus any physical volume-weighted field distribution is strictly positive.  In particular, the eigenfunction $\tpv(\lambb_{\rm max})$ must be positive everywhere because, up to a time-dependent (but field-independent) volume expansion factor, it is equal to the asymptotic physical distribution. 
This property offers a powerful criterion for identifying a physical stationary state without solving the full eigenvalue problem because positivity typically singles out $\tpv(\lambb_{\rm max})$ among all possible FPV stationary solutions satisfying given boundary conditions.

Let us consider the FPV equation for a monotonically increasing potential $V$, whose form is completely general and not restricted to the EFT structure nor to the perturbative range. We use the spectral representation
\beq
\pv (\phi , t )= \sum_{\lah} \, e^{3\lah^\GAUG H_0t } \, \tpv (\phi ,\lah) ~,
\label{declah}
\eeq
where $H_0$ is a constant reference value of the Hubble rate. The parametrisation of the eigenvalues has been chosen to conform with the overall expansion rate of each steady-state solution, which is given by $3H_0 (H/H_0)^\GAUG$. The FPV for the spectral modes is 
\beq
 \frac{\hbar \, H^2}{8\pi^2}\,   \tpv^{\prime \prime} +\frac{M_P^2V'}V \, \tpv^\prime +
 \left[ 3 - 3\Big( \frac{\lah H_0}{H}\Big)^\GAUG   +(\GAUG -2) \epsilon +\eta \right] \tpv =0~,
\label{slowrollfpv2}
\eeq
\beq
\epsilon =\frac{M_P^2 \, V'^{ 2}}{2\, V^2}~,~~~
\eta= \frac{M_P^2\, V''}{V} ~,
\label{slowrollpar}
\eeq
where $\epsilon$ and $\eta$ must be smaller than one in slow-roll regime and therefore can be neglected in \eq{slowrollfpv2} anywhere outside the field region where $H$ is approximately equal to $\lah H_0$.

Suppose that the distribution $\tpv$ is peaked around a field value ${\bar \phi}$ (such that $\tpv' ({\bar \phi})=0$) with width $\sigma^2 \equiv -\tpv ({\bar \phi})/\tpv'' ({\bar \phi})$. Then \eq{slowrollfpv2} in slow-roll regime yields the relation
\beq
\Big( \frac{\lah H_0}{\bar H}\Big)^\GAUG =1-\frac{\hbar \, {\bar H}^2}{24\pi^2\sigma^2}~,~~~~{\bar H}\equiv H({\bar \phi})~.
\label{relazzz}
\eeq
As long as $\sigma \gg {\bar H}$, we find that $\lah$, which measures the expansion rate of the stationary solution, is proportional to the Hubble rate evaluated at the peak and $\lah = {\bar H}/{H_0}$.

In analogy with an algebraic quadratic equation, we can define a `discriminant' of \eq{slowrollfpv2}
\beq
D=2\epsilon + \frac{3\hbar \, H^2}{2\pi^2 M_P^2} \bigg[ \left( \frac{\lah H_0}{H}\right)^\GAUG -1\bigg] ~,
\label{discrim}
\eeq
working at leading order in slow-roll parameters.
The `discriminant' $D$ provides a useful means for identifying when a stationary solution becomes negative since the FPV solutions become oscillatory in the vicinity of a field point where $D$ turns negative.  This can be understood by comparing the second-order linear differential equation to an algebraic quadratic equation and using the criterion for determining when the roots are real or imaginary. Since negative discriminant implies oscillatory, hence negative, solutions we infer that the solution with the largest eigenvalue is the one which remains positive-definite over the greatest field range. 

Equation~(\ref{discrim}) shows that positivity is always satisfied below the peak ($H \ll \lah H_0$). However, well above the peak ($H \gg \lah H_0$), the solution remains positive only if 
\beq
V'> H^3 ~.
\eeq
We recognise this inequality as the familiar Classical-beats-Quantum (CbQ) condition. We stress that positivity requires CbQ only above the peak and not in the full field range.

If CbQ does not hold, then \eq{discrim} gives us information about the location of possible peaks in the stationary solution. Indeed, positivity requires
\beq
V({\bar \phi}) > V_f\,  \bigg( 1-\frac{2\pi^2M_P^6 V_f^{\prime 2}}{\hbar \, V_f^3}\bigg)^{\frac{2}{\GAUG}} ~~\Rightarrow ~~
{\bar \phi} > f -\frac{4\pi^2M_P^2 V_f^{\prime }}{\hbar\, 9\GAUG\, H_f^4} ~,
\label{locatpee}
\eeq
where $f$ is the upper endpoint of the field range and $V_f =V(f)$ is the maximum of the monotonic potential. Equation (\ref{locatpee}) shows that, when CbQ does not hold, any peak (if it exists) must be close to the maximum of the potential.

The extreme case is when $D$ becomes negative before $\phi$ reaches $f$, but the solution remains positive because the peak is given by the first oscillation of $\tpv$. In this case, the solution of \eq{slowrollfpv2} develops an imaginary part in the argument of the exponential and, in the limit $\phi \to f$ and for absorbing boundary condition at the upper endpoint, we find
\beq
\tpv =e^{\frac{S_fV^\prime_f (f-\phi)}{2V_f}}\, \sin \left[\frac{S_f \sqrt{-D} (f-\phi)}{2M_P} \right] ~,
~~~
D = - \frac{12\GAUG H_f' (f-{\bar \phi})}{S_fH_f} ~,
\eeq
\beq
S_f =\frac{8\pi^2M_P^2}{\hbar \, H_f^2}~,~~~~H_f\equiv H(f) ~.
\eeq
The peak is located at the point where the argument of the sine is $\pi/2$ and therefore
\beq
{\bar \phi} = f - \bigg( \frac{\hbar \, M_P^2 H_f^4}{16\, \GAUG V_f'}\bigg)^{\frac{1}{3}} ~.
\eeq
In this case, the peak has a distance from the endpoint of the same size as its width.

In conclusion, just based on the positivity requirement, we can identify three regimes for possible distribution peaks, which depend on the behaviour of the monotonic potential near the top of the field range.

\begin{itemize}
\item \textbf{Classical (C): $H_f^3<V_f^\prime <M_PH_f^2$}. In this regime, the CbQ and slow-roll conditions are satisfied near the top of the potential. The peak, if it exists, is generically far from the upper endpoint with a location determined by the eigenvalue $\lah$.
\item  \textbf{Quantum+Volume (QV): $\sqrt{\GAUG}\, H_f^4/M_P <V_f^\prime <H_f^3$}. In this case, if allowed by boundary conditions, the peak must be close to the upper endpoint, at a distance $M_P^2 V_f^\prime /\GAUG H_f^4$. The peak is well separated from the endpoint, since its displacement is greater than its width.
\item  \textbf{Quantum$^2$+Volume (Q$^2$V): $V_f^\prime < \sqrt{\GAUG}\, H_f^4/M_P$}. In this case the peak, if it exists, is close to the upper endpoint, at a distance  $(M_P^2H_f^4/\GAUG V_f')^{1/3}$ with a separation comparable to the width.
\end{itemize}

\subsection{Peak Properties}
\label{sec:peak}
We can further investigate the nature of local peaks in stationary volume-weighted field distributions by studying the case of EFT potentials within a perturbative field range.
When expressed in terms of the modes $\tpv(\vf, \lambb)$, the FPV in \eq{eq:fpv2} becomes
\beq
\frac{\alpha}{2}\, \tpv''+ \fun' \tpv' +\left(\fun''+\beta \fun - \lambb \right) \tpv = 0 ~,
\label{eq:peakpos}
\eeq
where $\fun (\vf)$ is a monotonically increasing function such that $\fun (0)=0$ and $\fun (\pm 1)=\pm 1$. The corresponding `discriminant' is
\beq
D=\fun'^2 +2\alpha (\lambb -\beta \fun -\fun'' )~.
\label{eq:det}
\eeq

 Let us consider absorbing boundary conditions at the endpoints $\vf =\pm 1$.  The ratio of the coefficients of the two general solutions of \eq{eq:peakpos} can be determined by the boundary condition at $\vf =-1$, while the boundary condition at $\vf =1$ can be used to determine the eigenvalue spectrum. In order to cross zero at the upper endpoint and be positive elsewhere, the solution corresponding to $\lambb_{\rm max}$ must enter the oscillating regime at $\vf =1$. As a result, $\lambb_{\rm max}$ can be estimated as the solution of $D=0$ at $\vf =1$.  Using the expression of the discriminant in \eq{eq:det} together with $\alpha \ll 1$ and $\beta \gg 1$, we find 
\beq
 \lambb_{\rm max} = \beta - \frac{\fun_1^{\prime \,2}}{2\alpha}~,
\label{eq:alpei}
\eeq
where $\fun_1^{\prime}\equiv \fun'(1)$ is a number of order unity.

Now consider a hypothetical localised peak at $\vf \!=\! {\bar \vf}$ with width $\sigma$ defined as $\sigma^2 = -\tpv ({\bar \vf})/\tpv'' ({\bar \vf})$, in analogy with a Gaussian distribution. From \eq{eq:peakpos}, we find
\beq
\lambb = \beta \fun({\bar \vf})  + \fun''({\bar \vf})-\frac{\alpha}{2\sigma^2}~.
\label{pikpik}
\eeq
This shows that, for $\alpha \ll 1$ and $\beta \gg 1$, the eigenvalue $\lambb$ measures the height of the potential at the peak location. Assuming an absorbing boundary condition at $\vf =1$, we can combine this result with \eq{eq:alpei}, finding that the peak location is
\beq
\fun({\bar \vf}) =1 -\frac{\fun_1^{\prime \,2}}{2\alpha \beta} ~.
\eeq
  This gives a simple, and general, illustration of the two different parameter regimes.  For a classically dominated solution ($\alpha \beta \ll1$) the implied peak slides down the potential, as expected for classical evolution. The classical tendency to minimise the potential energy can only be prevented by imposing appropriate boundary conditions at $\vf =-1$.

On the other hand, if $\alpha \beta \gg 1$ the peak is located near the top of the potential, with the distribution settling as high as possible consistently with boundary conditions.  This neatly demonstrates the general condition under which the volume distribution will preferentially be peaked atop a potential, with $1/\alpha \beta$ governing the distance from the true maximum.  This confirms the expectation that departures from classical evolution occur when $\alpha \beta \gg 1$. 

We can also estimate the width $\sigma$ of the peak by taking the derivative of \eq{eq:peakpos} evaluated at ${\bar \vf}$. In the limit $\alpha \ll1$ and $\beta \gg 1$, we find
\beq
\sigma = \sqrt{\frac{1}{\beta}}~.
\label{eq:sigpla}
\eeq
This corresponds to a width of Planckian size for the dimensionful field $\phi$.

The result in \eq{eq:sigpla} breaks down if the distance between the position of the peak and the top of the potential ($\sim 1/\alpha \beta$) is smaller than the implied width ($\sim 1/\sqrt{\beta}$), which occurs whenever $\alpha^2 \beta \gg 1$.
Note that, from a perturbative perspective, this simply reflects the fact that NLO quantum effects combined with volume growth now dominate over classical evolution, and thus one is in a highly quantum regime of the theory.  In this case the width can be calculated directly from the discriminant condition, since the peak will be the first oscillation after entering the oscillatory regime, in order to satisfy the boundary condition.  From the frequency arising as the square root of the discriminant we find that in the oscillatory regime the solution assuming absorbing boundary conditions must be  of the form
\beq
\tpv =e^{\frac{\fun_1^{\prime}}{\alpha}(1-\vf)} \, \sin \Big[\sqrt{\frac{2\beta \Delta \fun}{\alpha}} \,(1-\vf) \Big] ~,~~~\Delta \fun \equiv 1-\fun({\bar \vf})= \fun_1^{\prime}(1-{\bar \vf )}
\eeq
and the peak is such that
\beq
1-{\bar \vf} \approx \Big( \frac{\alpha}{\beta} \Big)^{1/3}~,~~~~\sigma \approx \Big( \frac{\alpha}{\beta} \Big)^{1/3} ~. 
\eeq
Thus we see that the nature of the peak in this case arises solely as a result of entering the oscillatory regime.

\begin{figure}[t]
\begin{center}
\includegraphics[width=0.7\columnwidth]{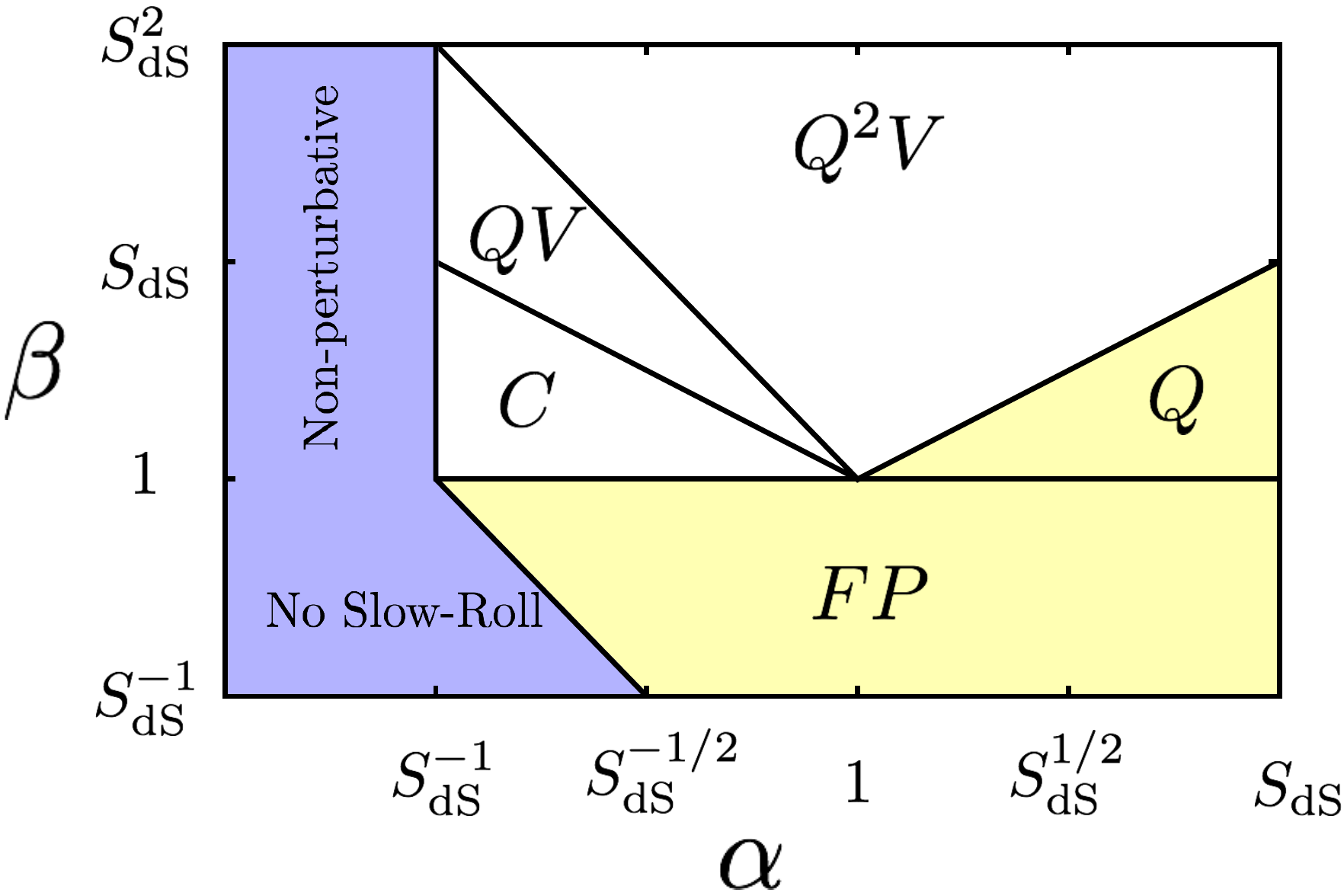} 
\end{center}
\caption{The different regimes of the FPV dynamics for an EFT potential parametrised by $\alpha$ and $\beta$. 
For $\alpha \ll 1/ {\SdS}$ the field range is non-perturbative and for $\alpha^2 \beta \ll \GAUG/ {\SdS^2}$ the dynamics is no longer in slow-roll regime (shown for $\xi =1$). The Q and FP regimes are relatively uninteresting for SOL because the volume term is only a subleading effect.}
\label{fig:ab}
\end{figure}

Armed with these estimates we may now sketch the parameter space for EFT potentials of the kind in \eq{potinEFT}.

\begin{itemize}
\item \textbf{Classical (C): $\alpha \beta \ll 1$}.  In this case the FPV distribution naturally evolves down the potential, as much as it is allowed by boundary conditions.
\item  \textbf{Quantum+Volume (QV): $\alpha \beta \gg 1$, $\alpha^2 \beta \ll 1$}.  In this case the FPV distribution naturally climbs the potential.  The resulting asymptotic peak will settle at a distance $1/\alpha \beta$ from the top of the potential with a width $\sigma \approx (1/\beta)^{1/2} $.
\item  \textbf{Quantum$^2$+Volume (Q$^2$V): $\alpha \beta \gg 1$, $\alpha^2 \beta \gg 1$}.  Also in this case the FPV distribution naturally climbs the potential.  Since NLO quantum effects are large, the asymptotic peak will be as close to the top of the potential as possible, with a displacement of  comparable magnitude to the width of the peak.  The width follows from the discriminant and is $\sigma \approx (\alpha/\beta)^{1/3}$.
\end{itemize}

The different regimes are summarised graphically in \fig{fig:ab}. For $\alpha \ll 1/\SdS$, the perturbative expansion of the FPV breaks down. When the slow-roll parameter $\epsilon = 3\GAUG \fun^{\prime \, 2} /(\SdS^2\alpha^2 \beta)$ is larger than one, the FPV description can no longer be trusted as the contribution from kinetic energy becomes important. When $\beta <1$, the FPV solutions are approximately given by the corresponding FP solutions up to a time-dependent, but nearly field-independent, factor. In the Q regime ($1<\beta <\alpha$), the dynamics is dominated by quantum fluctuations. In both the FP and Q regimes, the width of the distribution grows in time, eventually occupying the full field range. Since the asymptotic distribution is essentially uniform in field space, the FPV dynamics does not lead to interesting effects for our applications. In this work, we focus instead on C, QV and Q$^2$V, which are the three regimes where the volume term characterises the distinguishing features of FPV dynamics.

We can now show how the three regimes of the EFT coincide with the corresponding regimes defined for a general potential in \sec{sec:PS}. The EFT potential can be written as
\beq
V=V_0 \left[ 1+\frac{2 \fun(\vf )}{{\SdS} \alpha}\right] ~.
\eeq
Replacing this expression in \eq{discrim}, we find
\beq
D= \frac{3\GAUG}{ \SdS^2 \, \alpha^2\beta}\, \Big( \fun'^2(\vf) +2\alpha \beta \left[\fun({\bar \vf}) -\fun(\vf)\right] \Big)~,
\eeq
which, in slow-roll approximation, agrees with \eq{eq:det} up to an arbitrary constant. From this, it is immediate to see that the regimes defined in \sec{sec:PS} from positivity arguments are in perfect correspondence with those defined in the EFT directly from the FPV. We note that simply translating the CbQ condition in terms of the parameters $\alpha$ and $\beta$ would lead to an incorrect answer because it does not properly take into account the perturbative expansion of the Hubble rate.

\subsection{Linear Potential}
\label{sec:linear}
To elucidate the estimates and the different regimes found in \sec{sec:peak} we compare them here with the case of a linear potential $\fun(\vf) = \vf$, in which \eq{eq:peakpos} allows for the following analytic solution (see appendix)
\beq
\tpv (\vf, \lambb ) =e^{-\frac{\vf}{\alpha}} \left[ a(\lambb)\, \Ai (x) +b(\lambb)\,  \Bi (x) \right] ~,~~~x=\frac{1+2\alpha (\lambb -\beta \vf)}{(2\alpha^2 \beta)^{2/3}}~,
\label{sollinuf}
\eeq
where $\Ai (x)$ and $\Bi (x)$ are the Airy functions, while $a$ and $b$ are two arbitrary coefficients. Since we are interested in cases where the volume term is important for the dynamics, we take $\beta > 1$ and $\beta >\alpha$. 

The first solution in \eq{sollinuf} has a peak at $\vf ={\bar \vf}$ with width $\sigma^2 =- \tpv ({\bar \vf})/\tpv'' ({\bar \vf})$,
\beq
{\bar \vf} =\left\{
\begin{array}{c} 
\frac{\lambb}{\beta} \\ \frac{\lambb}{\beta}-a_1'\big( \frac{\alpha}{2\beta} \big)^{1/3} 
\end{array} \right.
~,~~~
\sigma =\left\{
\begin{array}{c} 
\frac{1}{\sqrt{\beta}} \\ \frac{1}{\sqrt{-a_1'}}\big( \frac{\alpha}{2\beta} \big)^{1/3} 
\end{array} \right.
~~~
\begin{array}{c} 
({\rm C~or~QV~regime}) \\[7pt] ({\rm Q}^2{\rm V~regime})
\end{array}
\eeq
where $a_1'=-1.02$ is the first zero of $\Ai^{\, \prime} (x)$. The location of the peak grows with $\lambb$, but it always remains below the upper perturbative endpoint (${\bar \vf}\le 1$) because of the bound on the maximum eigenvalue $\lambb_{\rm max} \le \beta$~(see \eq{lmax} in the appendix). However, for $\lambda <-\beta$, the peak disappears below the lower endpoint of the perturbative field range. 
In the C and QV regimes, the local behaviour of the solution around the peak is well approximated by a Gaussian function with a variance $\sigma$ which is universal, in the sense that is independent of the eigenvalue $\lambb$ and the coupling constant of the linear potential. It has a typical Planckian size in physical units (as long as $\xi \ne 0$). In the Q$^2$V regime, the peak is non-Gaussian, with a characteristic width $\sigma$ which is independent of $\lambb$ and grows as the coupling constant decreases. 

The first solution in \eq{sollinuf}  turns negative, entering a subsequent oscillatory phase, for $\vf > \vf_{\rm pos}$ with
\beq
\vf_{\rm pos} = \frac{1+2\alpha \lambb -a_1 (2\alpha^2 \beta)^{2/3}}{2\alpha \beta} ~,
\eeq
where $a_1= -2.34$ is the first zero of $\Ai (x)$. In the C regime, the oscillatory phase exists only when the peak has vanished below the lower endpoint of the field range. In the quantum regimes,
the separation between the peak and the end of the positivity range is small and  independent of the eigenvalue $\lambb$, 
\beq
\vf_{\rm pos} - {\bar \vf} =
\left\{
\begin{array}{c} 
\frac{1}{2\alpha\beta} \\ (a_1'-a_1)\big( \frac{\alpha}{2\beta} \big)^{1/3} 
\end{array} \right.
~~~
\begin{array}{c} 
({\rm QV~regime}) \\[7pt] ({\rm Q}^2{\rm V~regime})
\end{array}
\eeq
In the QV regime, the peak is well within the positivity region ($\vf_{\rm pos} - {\bar \vf} \gg \sigma$) while, in the Q$^2$V regime, is at the edge ($\vf_{\rm pos} - {\bar \vf} \approx \sigma$). Indeed, the peak in the Q$^2$V regime is not an isolated Gaussian, as in the case of C and QV, but corresponds to the first oscillation of $\Ai (x)$ before the function turns negative. Positivity of the asymptotic solution, together with the upper bound on $\lambb_{\rm max}$, imply that the asymptotic location of the peak (${\bar \vf}_{\infty}$) must be in the range
\bea
\beta \Big( 1- \frac{1}{2\alpha \beta}\Big) < \lambb_{\rm max} < \beta &\Rightarrow &
1-\frac{1}{2\alpha \beta} < {\bar \vf}_{\infty} <1  ~~~ ({\rm QV}) \nonumber \\
\beta \Big[ 1+ a_1 \Big(\frac{\alpha}{2\beta}\Big)^{1/3}\Big] < \lambb_{\rm max} < \beta &\Rightarrow &
1-(a_1'-a_1)\Big(\frac{\alpha}{2\beta}\Big)^{1/3} < {\bar \vf}_{\infty} <1  ~~~({\rm Q}^2{\rm V})
\eea
This means that, both in the QV and Q$^2$V regimes, the location of the peak is parametrically close to the upper endpoint, independently of the choice of boundary conditions. On the other hand, in the C regime, the peak location is entirely a matter of boundary conditions.

The second solution in \eq{sollinuf} is monotonically decreasing until it turns negative, entering a subsequent oscillatory phase, for 
\beq
\vf > \vf_{\rm pos}-(b_1-a_1) \Big( \frac{\alpha}{2\beta} \Big)^{\frac13} ~,
\eeq
where $b_1= -1.17$ is the first zero of $\Bi (x)$.

The previous considerations were independent of boundary conditions. Let us now focus on the case of absorbing boundary conditions at the endpoints of the field range, $\tpv(\vf \! = \! \pm 1) =0$. The solution in \eq{sollinuf} becomes, up to an overall constant, 
\beq
\tpv = e^{-\frac{\vf}{\alpha}} \left[ \frac{\Ai (x_n)}{\Ai (\xE)} - \frac{\Bi (x_n)}{\Bi (\xE)}   \right] ~,~~~~
\lambb_n = \beta -\frac{1}{2\alpha}+ a_n\, \Big( \frac{\alpha\beta^2}{2} \Big)^{1/3} ~,
\label{soluzecco}
\eeq
\beq
x_n=a_n +\Big( \frac{2\beta}{\alpha} \Big)^{1/3} (1-\vf) ~,~~~~
\xE = x_n (\vf \! = \! -1)~,
\eeq
where $a_n$ are the zeros of $\Ai (x)$. The asymptotic distribution at large times corresponds to the largest eigenvalue, which is obtained for $n=1$. The asymptotic distribution in \eq{soluzecco} has the following behaviours in the three different regimes. 

\subsubsection*{C regime}

For $\alpha \beta \ll 1$, we find 
\beq
{\bar \vf } = -1 +\alpha ~,~~~~  \sigma = \alpha ~.
\eeq
In the C regime, the peak of the distribution slides down the potential and, being supported only by the absorbing boundary condition, it settles at a distance $\alpha$ from the lower endpoint of the field range. The steeper the potential, the closer to the endpoint is the peak, but it always remains 1-$\sigma$ away from it. Note, however, that $\sigma$ measures only the local property of the distribution at the maximum. Since the peak in the C regime is fairly asymmetric, its spread away from the endpoint is actually larger than $\alpha$.

\subsubsection*{QV regime}

For $\alpha \beta \gg 1$ and $\alpha^2 \beta \ll 1$, we find
\beq
{\bar \vf} = 1-\frac{1}{2\alpha \beta}~,~~~~\sigma = \sqrt{\frac{1}{\beta}}~.
\label{QVreglin}
\eeq
The width of the peak is independent of the coupling constant in the potential. The peak is located a short distance away from the upper endpoint, but slides down as the potential gets steeper. Although being relatively near to the upper endpoint, the peak is always separated from it by a distance much larger than the width.

\subsubsection*{Q\boldmath$^2$V regime}

For $\alpha \beta \gg 1$ and $\alpha^2 \beta \gg 1$, we find
\beq
{\bar \vf} = 1 -(a_1^\prime -a_1)\, \Big( \frac{\alpha}{2\beta}\Big)^{1/3}
~,~~~~\sigma = \frac{1}{\sqrt{-a_1^\prime}}\, \Big( \frac{\alpha}{2\beta}\Big)^{1/3}~.
\eeq
The peak is located near the upper endpoint, but it moves away from it the shallower the potential. However, its distance from the endpoint remains constant in units of width, being always equal to about 1.3 $\sigma$.

\bigskip

\begin{figure}[t]
\begin{center}
\includegraphics[width=0.45\columnwidth]{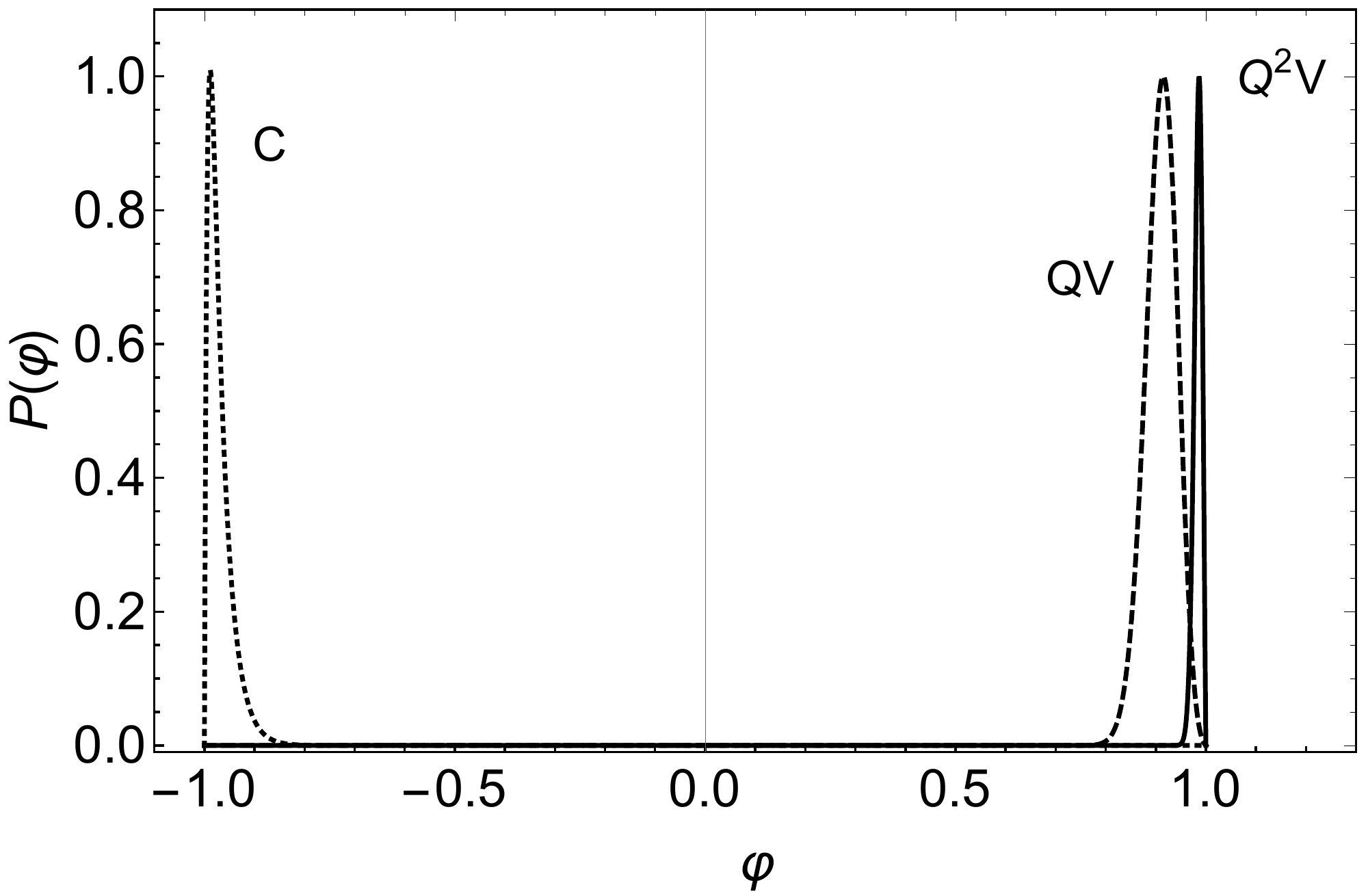} 
\includegraphics[width=0.45\columnwidth]{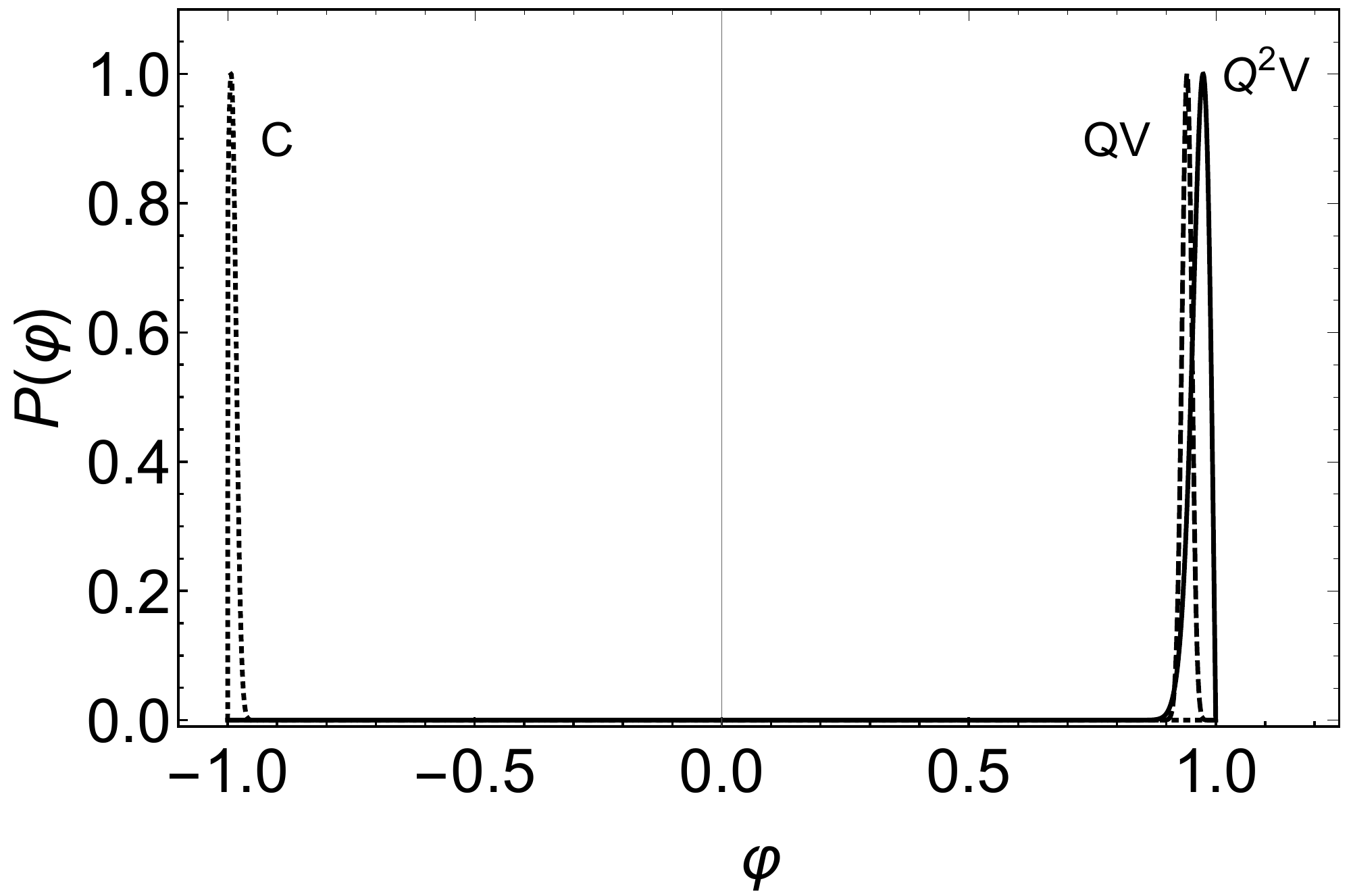} 
\end{center}
\caption{The asymptotic volume-weighted field distribution for a linear potential with absorbing boundary conditions at $\vf =\pm 1$, for fixed $\alpha=10^{-2}$ (left panel) with $\beta=10$ (C regime), $10^3$ (QV regime), $2\times 10^4$ (Q$^2$V regime), and for fixed $\beta=10^4$ (right panel) with $\alpha=2.5\times 10^{-5}$ (C regime), $10^{-3}$ (QV regime), $10^{-1}$ (Q$^2$V regime).}
\label{fig:linearFPVsolutions}
\end{figure}

These analytic results for the linear potential confirm the general EFT expectations for the three parameter regimes introduced in \sec{sec:peak}. The FPV distribution in \eq{soluzecco} is plotted in \fig{fig:linearFPVsolutions} for parameter values corresponding to the three regimes discussed, varying $\beta$ keeping $\alpha$ fixed on the left and vice versa on the right. 

\subsection{The Classical Regime}
\label{sec:determ}

We now want to show that the classical regime allows for a full characterisation of the FPV solutions for a general potential, beyond EFT or perturbative field range.

Consider a monotonically increasing potential $V(\phi)$ and define the following special field points
\beq
V(\phi_{\rm AdS} ) =0 ~,~~~ V(0) =V_0 ~,~~~ V(\phi_P) =2 V_0~,~~~ V(f )=M^4/g_*^2~.
\eeq
The point $\phi_{\rm AdS}$ is where the potential crosses zero. It is the absolute lower endpoint of the field range that we can consider, since the FPV is defined only in dS space. The point $\phi =0$ is defined, after an appropriate coordinate shift, to be  a generic field value around which we will expand perturbatively in a field region such that $|V(\phi) -V_0|\ll V_0$. 
Therefore, $\phi_P$ gives the maximum field value where we can trust a perturbative expansion, which is valid in the range
\beq
\phi_{\rm AdS} \ll \phi \ll \phi_P~.
\eeq
The point $\phi =f$ defines the absolute upper endpoint of the field range that we can consider, since $M$ describes the energy cutoff at which the EFT description breaks down, while $g_*$ is a typical coupling constant of the UV theory. Because of the onset of quantum gravity at Planckian energies, $M/g_*$ must be smaller than $M_P$.

\subsubsection*{The diffusionless solution}

Consider the diffusionless limit ($\hbar \to 0$) of the FPV in \eq{slowrollfpv2}. Up to an overall normalisation, the solution is
\beq
\pvc = \frac{V}{V'} \Big( \frac{V_0}{V}\Big)^{\frac{\GAUG}{2}} \, \exp \frac{3}{M_P^2}\int d V\, \frac{V}{V'^2}\left[ \left( \lah \sqrt{\frac{V_0}{V}}\right)^\GAUG -1\right]
 ~.
\label{classol}
\eeq
The solution $\pvc$ in \eq{classol} has a peak at ${\bar \phi}$ with width $\sigma$ such that
\beq
{\bar H}\equiv H({\bar \phi}) =H_0\,  \lah~,~~~\sigma = \sqrt{\frac{2}{3\xi}}\, M_P
\label{peakpvc}
\eeq
at leading order in the slow-roll parameters $\epsilon $ and $\eta $.

By replacing \eq{classol} into \eq{slowrollfpv2}, we can compare the relative size of the diffusion ($Q$) and drift ($C$) terms. At leading order in slow-roll parameters, their ratio is
\beq
\frac{Q}{C} = \frac{\hbar \, 27 }{8\pi^2}\left( \frac{H^3}{V'}\right)^2 \,\left[ \Big( \frac{\bar H}{H}\Big)^\GAUG -1 \right] ~.
\eeq
This shows that the diffusionless approximation is valid when the potential satisfies the CbQ condition in the field region above the peak and the stronger condition $V'>H^{3-\GAUG /2}{\bar H}^{\GAUG /2}$ below. The diffusion term can be important at the initial stages of the time evolution to create a spread of the distribution. However, in the classical regime, it becomes irrelevant once the width has reached Planckian size since the subsequent evolution is largely determined by drift and volume terms alone.

The behaviour of $\pvc$ well above the peak is obtained by taking $ V/V_0 \gg \lah^2$ in \eq{classol} and we find, at leading order in slow-roll parameters,
\beq
\frac{\pvc'}{\pvc} =-\frac{3V}{M_P^2V'} ~~~~({\rm for~large}~\phi ) ~.
\label{behabov}
\eeq
As $\phi \to \phi_{\rm AdS}$, \eq{classol} shows that the diffusionless solution goes like $\pvc \propto V^{1-{\GAUG}/{2}}$ and therefore it vanishes exactly at the AdS boundary. The behaviour below the peak ($ V/V_0 \ll \lah^2$)  is
\beq
\frac{\pvc'}{\pvc} =\frac{3V}{M_P^2V'} \Big( \frac{\bar V}{V}\Big)^{\frac{\GAUG}{2}} +
\Big( 1-\frac{\xi}{2}\Big) \frac{V'}{V} ~~~~({\rm for~small}~\phi ) ~,
\eeq
where the first term dominates in the slow-roll regime.

For a linear potential $V = V_0+V'\phi$ with constant $V'$, we can perform the integral in \eq{classol} and obtain
\beq
\pvc = \frac{V}{V'} \Big( \frac{V_0}{V}\Big)^{\frac{\GAUG}{2}} \, \exp \frac{3V^2}{2M_P^2V'^2}
\left[ \frac{4}{4-\GAUG}\left( \lah \sqrt{\frac{V_0}{V}}\right)^\GAUG -1\right]
 ~.
\eeq
It can be verified that, in a perturbative field range around $\phi =0$, the solution $\pvc$ coincides with the first of the two solutions in \eq{sollinuf}, $\tpv = \exp(-\vf /\alpha) \Ai (x)$.

Just for illustration consider, as a second example, an exponential potential
\beq
V = V_0 \left[ 1+ a\, \big(e^{\frac{\kappa \phi}{M_P}} -1\big)\right] ~,~~~ a= \Big(1- e^{\frac{\kappa \phi_{\rm AdS}}{M_P}}\Big)^{-1}~.
\eeq
In this case, the integral in \eq{classol} gives, for $\xi =1$,
\beq
\pvc = {\scriptstyle \sqrt{V} \, \exp \big[  { \frac{3}{\kappa^2} \big(  
\frac{\lah}{\sqrt{a-1}}\arctan \frac{r}{\sqrt{a-1}}-\frac{a-1+\lah r}{a}\, e^{-\frac{\kappa \phi}{M_P}}
 \big) -\big( \kappa +\frac{3}{ \kappa}\big) \, \frac{\phi}{M_P} }\big]~,~~~r=\sqrt{\frac{V}{V_0}}}~,
\eeq 
while, for a general $\xi$, $\pvc$ can be expressed in terms of hypergeometric functions. 

\subsubsection*{The Gibbs solution}

Consider the equation
\beq
 \frac{\hbar \, H^2}{8\pi^2}\,   \tpv ^{\prime \prime} +\frac{M_P^2V'}V \, \tpv^\prime +
(\eta -4 \epsilon ) \tpv =0~,
\label{eqacaso}
 \eeq
 which can be conveniently rewritten as
 \beq
 \tpv ^{\prime \prime} = ( {\mathcal V}' \tpv )' ~,~~~~ {\mathcal V}=\frac{24\pi^2M_P^4}{\hbar \, V} ~.
 \eeq
 One of its two solutions is
 \beq
 \pvg =\exp \left( \frac{24\pi^2M_P^4}{\hbar \, V} \right) ~,
 \label{gibbs}
 \eeq
 which is the stationary Gibbs distribution corresponding to the potential ${\mathcal V}$.

Equation (\ref{eqacaso}) properly accounts for the first and second derivative terms in the FPV and differs from \eq{slowrollfpv2} by terms which are small when evaluated on $\pvg$ if the following conditions are satisfied
\bea
H^3 < V' < V/M_P ~~&~~{\rm for}~V> \lah^2 V_0 ~,\nonumber \\
H^{3-\GAUG/2} (\lah H_0 )^{\GAUG /2}< V' < V/M_P ~~&~~{\rm for}~V< \lah^2 V_0 ~.
\label{classrangesr}
\eea
These are exactly the same classical slow-roll conditions under which the diffusionless approximation gives a valid solution to the FPV equation.

The Gibbs distribution in \eq{gibbs} is monotonically decreasing, it explodes in the region of small $V$ and satisfies
\beq
\frac{\pvg^\prime}{\pvg} = -\frac{V_0 {\SdS} V' }{V^2} ~.
\eeq
Therefore its relative rate of decrease at large $\phi$ is faster than for $\pvc$ because 
\beq
\frac{|{\pvg^\prime}/{\pvg}|}{|{\pvc'}/{\pvc}|} \sim \left( \frac{V'}{H^3}\right)^2,
\eeq 
which is larger than one in the CbQ regime. For a linear potential in a perturbative field range around $\phi =0$, the solution $\pvg$ coincides with the second solution in \eq{sollinuf}, $\tpv = \exp(-\vf /\alpha) \Bi (x)$.

\subsubsection*{Summary}

The FPV solutions can be fully characterised in the classical slow-roll regime, which is defined by \eq{classrangesr}. One solution, called $\pvc$ and given in \eq{classol}, corresponds to the case in which the diffusion term is negligible. The drift and volume terms in the FPV balance each other and create a peak in the stationary solutions with Planckian width and location at the field point where the potential height matches the global expansion rate measured by the eigenvalue $\lah$. 

The second solution, called $\pvg$ and given in \eq{gibbs}, is monotonically decreasing and gives an exponential preference towards the lowest possible value of the potential, up to small quantum fluctuations. The proper combination of $\pvc$ and $\pvg$ which gives the physical FPV solution can be determined only with knowledge of the boundary conditions.

\subsection{Junction Conditions}
\label{sec:junc}
Consider the cosmological fate of the scalar field $\phi$ in the case that, at some critical value $\phi_c$, it triggers a first-order quantum phase transition. Since the potential is continuous, the FPV solution $\pv$ and its time derivative are also continuous. However, when the gradient of the potential is discontinuous at $\phi_c$, the gradient of the solution $\partial \pv/\partial \phi$ will not be continuous. We may understand the dynamics at the critical point by integrating the FPV in \eq{eq:FPV} across the critical point, with continuity of the equation leading to the constraint
\beq
\lim_{\epsilon \to 0}\int_{\phi_c -\epsilon}^{\phi_c +\epsilon} d\phi \, \frac{\partial}{\partial \phi} \left[ \frac{V' \pv}{3 H}  + \frac{\hbar}{8 \pi^2} \frac{\partial}{\partial\phi} (H^{3} \pv) \right] =0  ~.
\label{limlimlim}
\eeq
Since this is a total derivative, the integral is directly given by the boundary terms and we find
\beq
\frac{\Delta \pv^\prime}{\pv (\phi_c)} = - \frac{24 \pi^2 M_P^4\, \Delta V' }{\hbar \,V^2(\phi_c)}~,
\label{eq:const}
\eeq
where the discontinuities for any quantity $Q$ is defined by
\beq
\Delta Q  = \lim_{\epsilon \to 0}\left[ Q (\phi_c+\epsilon) - Q (\phi_c-\epsilon) \right] ~.
\label{defdisc}
\eeq
In \eq{eq:const} we have dropped a correction proportional to $V/M_P^4$ which is always negligible as long as the dynamics is not quantum-gravity dominated.

The junction condition in \eq{eq:const} does not rely on any assumption on the form of the potential nor any approximations beyond those implicit in the use of the stochastic approach. It is also valid for all times and therefore it holds for each $\tpv (\lambb)$, mode by mode. In the approximation of \eq{appinf} and for the EFT potential, the junction condition in \eq{eq:const} becomes
\beq
\frac{\Delta \pv'}{\pv(\vf_c )} = -\frac{2  \Delta \omega' }{\alpha}~.
\label{eq:jung}
\eeq

Since potential gradient discontinuities with $\Delta \omega' =\mathcal{O}(1)$ are the hallmark of a first-order phase transition, we see that the phase transition imprints itself upon the field distribution as a discontinuity with the opposite sign.  The sign of the discontinuity $\Delta \omega'$ is negative for phase transitions, thus $\Delta \pv'$ must be positive.  The negativity of the gradient discontinuity follows from the fact that, in thermal equilibrium, the lower of the two available states is occupied.  Thus, the potential of the equilibrium state as a function of $\vf$ will track the lowest of any branches of vacua, stable or not, that may exist at any $\vf$ value in different quantum phases.  

Another example of junction conditions relevant to our applications is the case of a multivalued potential with two branches corresponding to two different phases. The potentials $V_{a,b}(\phi)$ on the two branches are both continuous and differentiable, but branch $b$ has a termination point at $\phi_c $, where $V_b (\phi_c ) > V_a (\phi_c )$, and therefore exists only for $\phi \ge \phi_c $. If the field undergoes a sudden phase transition from $b$ to $a$ at the critical point  $\phi_c $, then the junction condition on the $b$ branch must be
\beq
\pv_b (\phi_c ) =0
~.
\label{eqquno}
\eeq 

The junction conditions on the $a$ branch are obtained by considering the sum of the two FPV on the branches. By integrating this equation in a neighbourhood of the critical point, following the same procedure as in \eq{limlimlim}, and exploiting the continuity of $\pv_a$ and $\pv_a + \pv_b$, we find the junction conditions
\beq
\Delta \pv_a^\prime = -\pv_b^\prime(\phi_c )  ~,~~~~
\Delta \pv_a =0~,
\label{eqqdue}
\eeq
where the discontinuities across the critical point are defined as in \eq{defdisc}. Equation~(\ref{eqqdue}) describes flux conservation. Although the FPV does not have a conserved current, there is an effective conservation law because the volume term does not enter the discontinuity.

\section{Self-Organised Localisation}
\label{sec:SOL}

An essential element of SOL is the localisation of the volume-weighted field distribution triggered by a quantum phase transition. We will illustrate the mechanism using some simple examples that exhibit different features of the SOL phenomenon.

\subsection{Localisation in the Pyramid Scheme}
\label{sec:pyra}

As a first prototype example, let us consider the `pyramid' potential defined by $\fun(\vf ) = -|\vf |$.
The discontinuity of the potential gradient at $\vf_c =0$ reflects an underlying first-order phase transition. Taking absorbing boundary conditions at the endpoints of the field range, $\tpv(\vf \!=\! \pm 1)\!=\! 0$, and the junction condition in \eq{eq:jung}, the solution of \eq{eq:peakpos} for the mode corresponding to the largest eigenvalue is (for $\beta \gg \alpha$ and up to an overall constant)
\beq
\tpv = e^{\frac{|\vf|}{\alpha}} \left[ 2\Ai (x) - e^{-\frac43 \xE^{3/2}} \Bi (x) \right] ~,~~~~
\lambb_{\rm max} =  a\, \Big( \frac{\alpha\beta^2}{2} \Big)^{1/3}-\frac{1}{2\alpha} ~,
\label{distrpyr}
\eeq
\beq
x=a +\Big( \frac{2\beta}{\alpha} \Big)^{1/3} |\vf | ~,~~~~
\xE = x(|\vf | \! =\! 1)~,
\eeq
where $a$ is the largest solution of the equation
\beq
\frac{\Ai^{\, \prime}(a)}{\Ai (a)}=(2\alpha^2 \beta)^{-1/3} ~~~\Rightarrow ~~~
a=\left\{ \begin{array}{cc}
a_1 & ({\rm for~}\alpha^2 \beta \ll 1) \\
a_1^\prime & ({\rm for~}\alpha^2 \beta \gg 1) 
\end{array} \right. ~.
\eeq
Therefore, $a$ always lies in the range $-2.34\! < \! a\! < \!  -1.02$. The distribution in \eq{distrpyr} exhibits a pair of symmetric peaks whose locations ${\bar \vf }_\pm$ and width $\sigma$ have different behaviours in the three following regimes.

\subsubsection*{C regime}

For $\alpha \beta \ll 1$, we find 
\beq
{\bar \vf }_\pm = \pm (1 -\alpha) ~,~~~~  \sigma = \alpha ~.
\eeq
The asymptotic distribution is localised in the proximity of the field endpoints because the boundary conditions prevent any further descent.

\subsubsection*{QV regime}

For $\alpha \beta \gg 1$ and $\alpha^2 \beta \ll 1$, we find
\beq
{\bar \vf}_\pm = \pm \frac{1}{2\alpha \beta}~,~~~~\sigma = \sqrt{\frac{1}{\beta}}~.
\eeq
The peaks are located 
relatively close to the critical point, although they are always well separated in units of width since $|{\bar \vf}_+\! -\! {\bar \vf}_- |/\sigma \gg 1$.

\begin{figure}[t]
\begin{center}
\includegraphics[width=0.47\columnwidth]{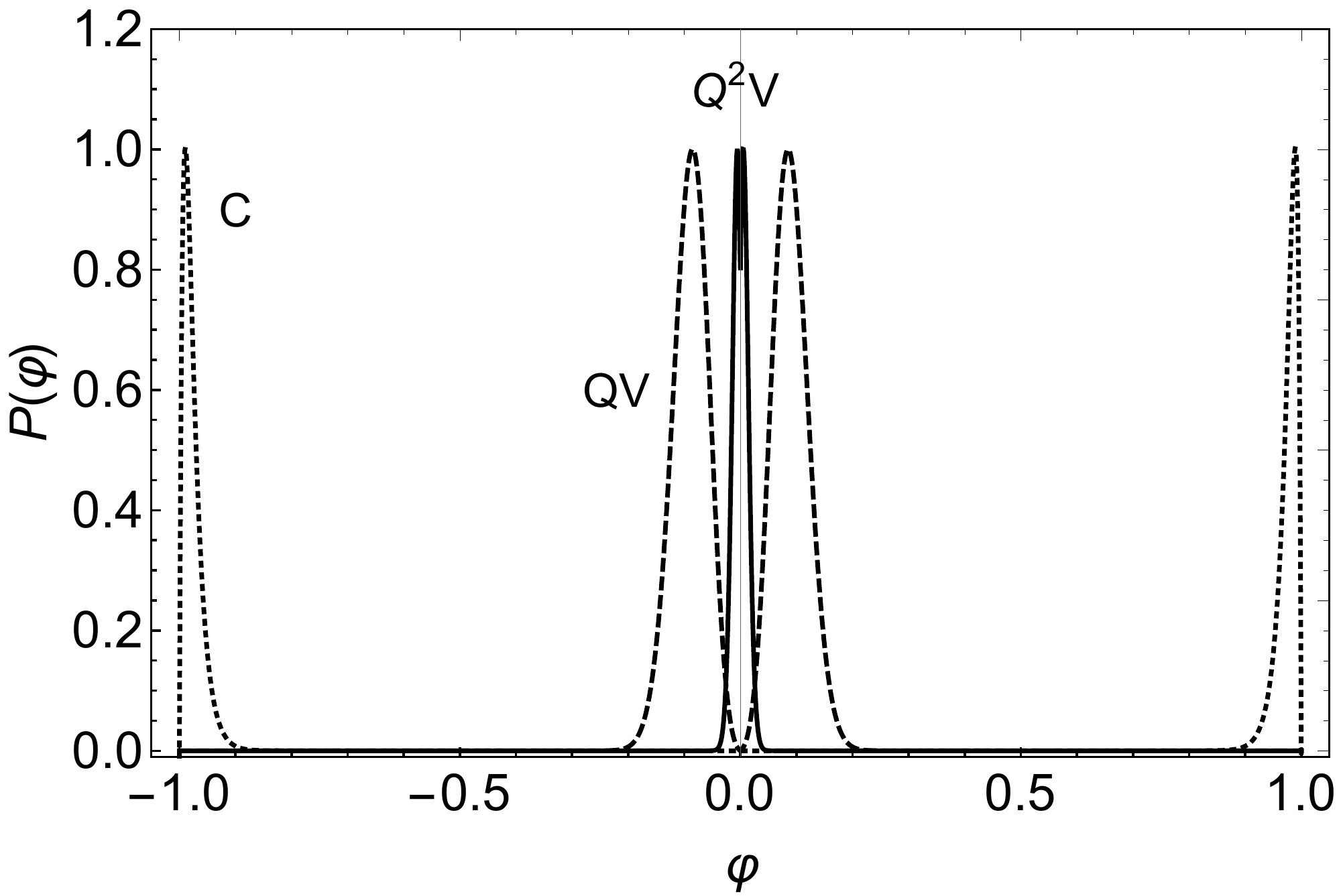} ~~ \includegraphics[width=0.47\columnwidth]{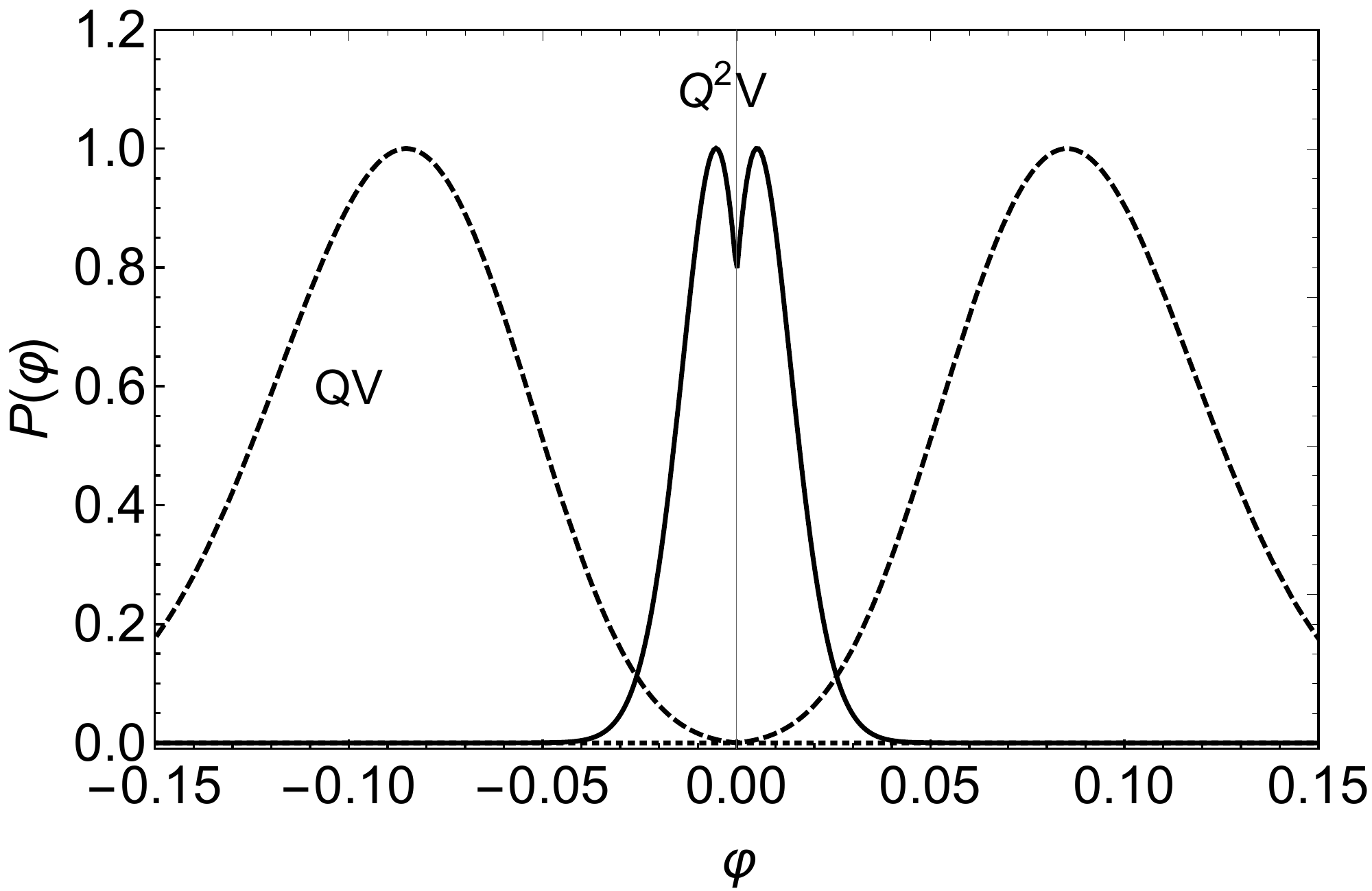} 
\end{center}
\caption{The asymptotic volume-weighted field distribution for a pyramid potential with absorbing boundary conditions at $\vf =\pm 1$, for $\alpha=10^{-2}$ and $\beta=10$ (C regime), $10^3$ (QV regime), $2\times 10^4$ (Q$^2$V regime). The figure shows the occurrence of SOL and of a peak at or near the critical point for the latter two regimes, zoomed in on the right.}
\label{fig:1stvaca}
\end{figure}

\subsubsection*{Q\boldmath$^2$V regime}

For $\alpha \beta \gg 1$ and $\alpha^2 \beta \gg 1$, we find
\beq
{\bar \vf}_\pm = \mp \frac{1}{a_1^\prime} \, \Big( \frac{2}{\alpha \beta^2}\Big)^{1/3}~,~~~~\sigma = \frac{1}{\sqrt{-a_1^\prime}}\, \Big( \frac{\alpha}{2\beta}\Big)^{1/3}~.
\eeq
The two peaks are much broader than their separation, since $|{\bar \vf}_+\! -\! {\bar \vf}_- |/\sigma \ll 1$, and therefore in practice they form a single peak centred at the critical point. 

\bigskip

The
asymptotic FPV distribution in \eq{distrpyr} is plotted in \fig{fig:1stvaca} for a representative choice of parameters, showing the narrow peak located at the critical point in the $Q^2V$ regime, as well as solutions in the $C$ and $QV$ regimes (left panel). Zooming in closely on the critical point (right panel) we 
see that the $Q^2 V$ distribution has a local minimum, as predicted by the junction condition, even though this behaviour persists only for a short field range. Globally, the distribution is peaked around the critical point, as expected.  

\bigskip

This calculation illustrates a simple realisation of SOL, where the volume-weighted field distribution becomes localised at the critical point of a first-order quantum phase transition during a long period of inflation. The pyramid example has been chosen because it allows for a simple analytical treatment. However, the SOL mechanism remains qualitatively the same for a general class of potentials where a negative $\Delta g'$ creates a cusp, corresponding to a local maximum of the potential at the critical point, and even for discontinuous potentials at the maximum.

\subsection{Microscopic Features of the Phase Transition}
\label{sec:micro}

The example of SOL presented in the previous section relies only on the nature of the dynamical regimes on either side of the critical point. Implicitly we have assumed that the system immediately undergoes the phase transition to the true minimum, neglecting the microscopic details of the transition and any associated timescale. In order to illustrate how these details can matter, we will study a toy model inspired by Landau theory.

\begin{figure}[t]
\begin{center}
\includegraphics[width=0.95\columnwidth]{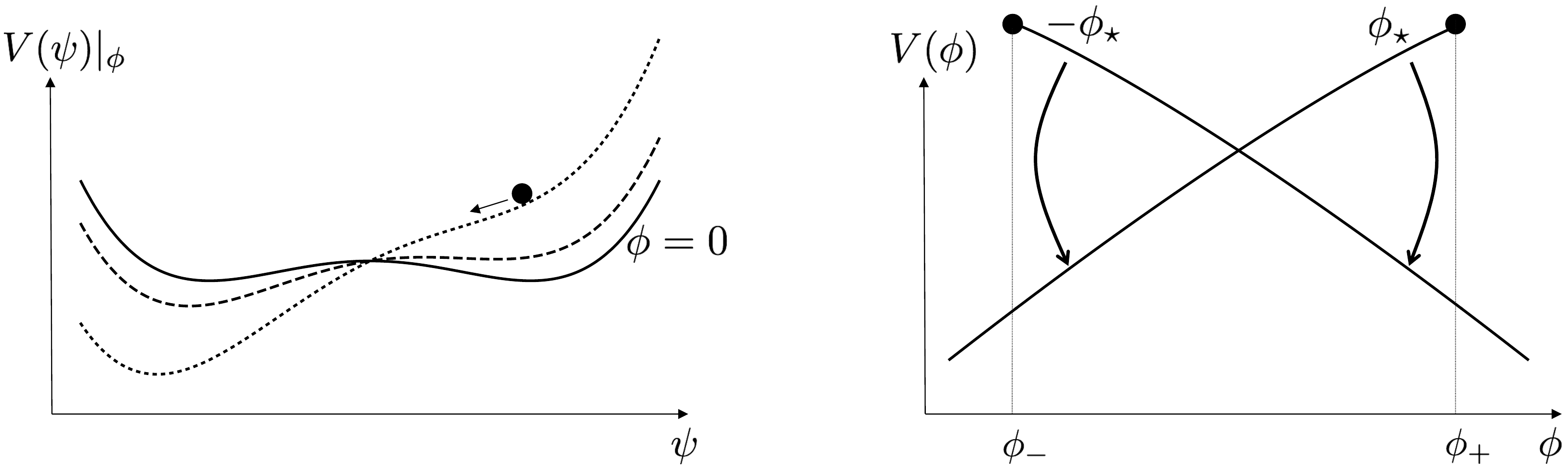} 
\end{center}
\caption{Vacuum structure in the Landau model, where $\phi$ plays an analogous role to an external magnetic field. Left panel: $V$ as a function of $\psi$ for different values of $\phi$. Right panel: $V$ as a function of $\phi$ with $\psi$ fixed at its two possible minima.}
\label{fig:1stvac}
\end{figure}

Consider the scalar potential
\beq
V=\frac{\lambda}{4} \left( \psi^2-\rho^2 \right)^2 + \kappa \phi \psi ~~,
\label{eq:Landau}
\eeq
where $\psi$ is the microscopic scalar field and $\phi$ plays a role analogous to the $Z_2$-breaking external magnetic field in the Ising model.  
The $\psi$ vacuum, as a function of $\phi$, has two configurations
\beq
\langle \psi \rangle_\pm =\pm \rho \,  C ( \phi / \phi_\pm ) ~~~~~{\rm for}~\phi \, \,
\stackanchor[0.5pt]{\scalebox{0.8}{$<$}}{\scalebox{0.8}{$>$}} \,\, \phi_\pm
\eeq
\beq
\phi_\pm =\pm \frac{2\lambda \rho^3}{3\sqrt{3} \kappa} ~,~~~
C(x) = \left\{ 
\begin{array}{ll}
\frac{2}{\sqrt{3}} \, \cos \left[ \frac{\arccos (-x )}{3}\right] & {\rm for}~|x|<1  \vspace{0.2cm}\\
\frac{2}{\sqrt{3}} \, \cosh \left[  \frac{{\rm arcosh} (-x )}{3}\right] & {\rm for}~x < -1 
\end{array}
\right. ~.
\eeq
The two branches of the $\phi$ potential on these $\psi$ configurations are 
\beq
V(\phi, \langle \psi \rangle_\pm )=
\frac{\lambda \rho^4}{4} \left[ 1+2 C^2 (\phi /\phi_\pm )-3C^4 (\phi /\phi_\pm )  \right] ~~~~~{\rm for}~\phi \, \,
\stackanchor[0.5pt]{\scalebox{0.8}{$<$}}{\scalebox{0.8}{$>$}} \,\, \phi_\pm ~,
\eeq
which are shown in \fig{fig:1stvac}. For $\phi$ large and negative, the true minimum of $\psi$ is at $\psi_+$. As $\phi$ increases, the minimum is lifted and, at $\phi =0$, it becomes degenerate with the configuration $\psi_-$. However, $\psi_+$ persists as a local minimum in a `supercooled' phase beyond the critical point $\phi_c =0$ until it becomes classically unstable at $\phi = \phi_+$.

In the previous section we have used a simplified treatment assuming immediate tunnelling to the true vacuum, with a sudden transition at $\phi_c$ generating the first-order discontinuity. In practice, this means assuming that the dynamical evolution of the field $\phi$ strictly follows the lower, true, vacuum.  However, in reality the tunnelling rate to this vacuum at a given $\phi$-point depends on the microscopic physics.  Parametrically, at a generic value for $\phi$ it scales as $\Gamma \sim e^{-\zeta/\hbar \lambda}$, where $\zeta$ is a numerical coefficient which must be evaluated on a case-by-case basis.  If the timescale for this tunnelling is greater than the other cosmological timescales, the assumption of immediate tunnelling will break down.

As shown in \fig{fig:1stvac}, when $\phi$ increases beyond $\phi_+$, the metastable vacuum eventually disappears and $\psi$ rolls directly to the true vacuum, with no tunnelling required.  If the typical gradient of the $\psi$ potential exceeds the slow roll condition, this rolling is essentially instantaneous from a cosmological perspective. Depending on the dynamical timescales, $\psi$ may tunnel to the true vacuum long before $\phi_+$ has been reached, or the tunnelling process may be so slow that $\phi$ evolves along the supercooled branch all the way up to $\phi_+$  before proceeding to the true vacuum.  A complete modelling of this dynamical process would require a random walk which can explore metastable branches of the vacuum structure and take into account the probability of tunnelling to the true vacuum from the excited state.  However, we may provide a simplified treatment by noting that, since the tunnelling rate depends exponentially on the vacuum energy difference, there will be a fairly abrupt change of behaviour at a given field point $\phi_\star$ (with $0<\phi_\star < \phi_+$) such that tunnelling to the true vacuum is essentially forbidden for $\phi < \phi_\star$, whereas it becomes inevitable for $\phi > \phi_\star$.  In the supercooled state, this corresponds to an absorbing boundary condition at $\phi_\star$ because once $\phi$ has ventured beyond this point it essentially vanishes from this branch, appearing on the lower one.  On the lower branch the boundary condition must conserve flux, including the flux from the upper branch, and also be continuous.  As a result, at $\phi_\pm$ one must impose an absorbing boundary condition on the upper branch, and a boundary condition on the lower branch for which the derivatives of the distribution have zero sum, as discussed in \sec{sec:junc}.

One can now picture a random walk on a vacuum structure such as in \fig{fig:1stvac}.  In the classical regime the field will roll down the potential.  However in the quantum regimes it will climb the potential, drop off the end of a branch, and then start climbing again.  The final steady state will reflect the preference for high altitudes, but also the fact that one has essentially absorbing boundary conditions at the top of a branch.  With this in mind we see that the field will be found in proximity to the highest values of the potential, with a distribution given by the standard widths in the QV or Q$^2$V regimes. Explicit calculations confirm this expectation. If the width is greater than the length of the supercooled branch, the presence of the metastable states is of little significance, since the distribution will encompass the entire region.  However, if the width is smaller, then  the distribution will be found localised at a point on the supercooled branch away from the degenerate point and will remain stable relative to the timescales relevant for the dynamics.  Typically, since quantum tunnelling will often not lead to a phase transition which can complete through percolation, as in old inflation, the supercooled branch can be effectively stable.  In the applications to follow, these aspects will be important.

\subsection{Localisation in the Waterfall Scheme}
\label{sec:water}

Let us consider a multivalued potential describing two phases of the underlying system. In phase $v$ (to be identified with the vacuum in the `visible' configuration of the system), we take a generic, but monotonically increasing, potential defined within a field range $\vf_E^- < \vf < \vf_E^+$  in which the scalar field gives only a small modulation of the background vacuum energy. In phase $h$ (corresponding to a `hidden' configuration), we take a flat potential and we choose coordinates such that the two phases are degenerate in energy at the point {$\vf = \vf_c$. Nonetheless, a phase transition occurs only at the point $\vf = \vf_T$, as illustrated in \fig{fig:ccfig} (left panel). We will refer to this setup as the `waterfall potential'.

The question at hand is to determine the final stationary volume distribution of $\vf$ as it fluctuates and explores the vacuum structure of the waterfall potential.  In particular, we are interested in whether the field can become exponentially localised in phase $v$ at the point of energy degeneracy between the two vacua. This would constitute a self-organisation of the vacuum energy in phase $v$ to be as close to the one in phase $h$ as the width of the distribution will tolerate.

Since the crossing point of the two phases is not a local maximum and in both quantum regimes the field prefers to localise at maxima, we must consider a classical regime if we are to have localisation at the point of degeneracy.  Furthermore, as found in \sec{sec:PS}, the greatest separation one can find between the location of a peak and the point at which the solution no longer satisfies positivity is $\Delta \vf \sim 1/(\alpha \beta)$. Thus if the asymptotic solution in phase $v$ is to be positive everywhere and localised at the degenerate point, we must have $\alpha \beta \lesssim 1$.  This is the classical regime.  

\begin{figure}[t]
\begin{center}
\includegraphics[width=0.45\columnwidth]{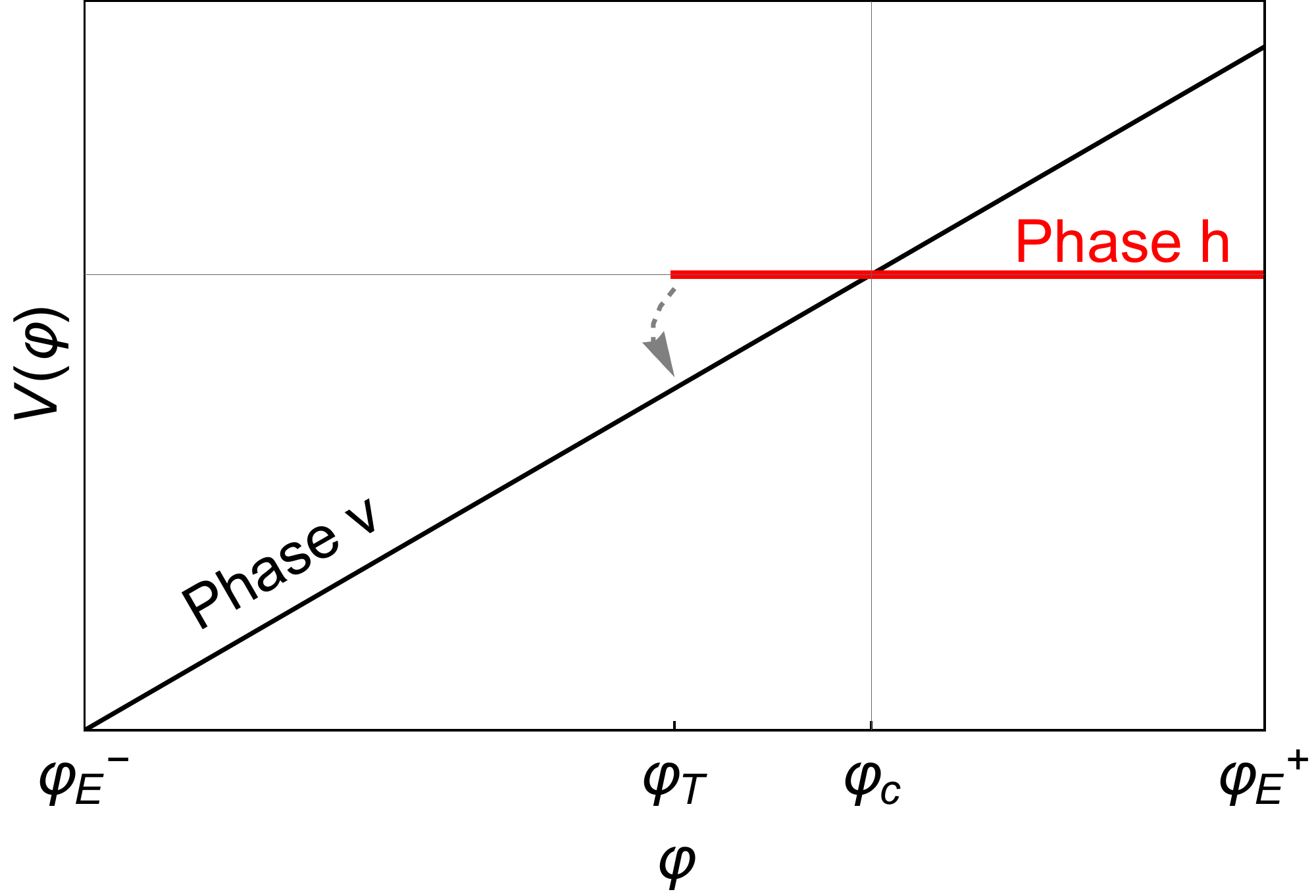} ~~ \includegraphics[width=0.485\columnwidth]{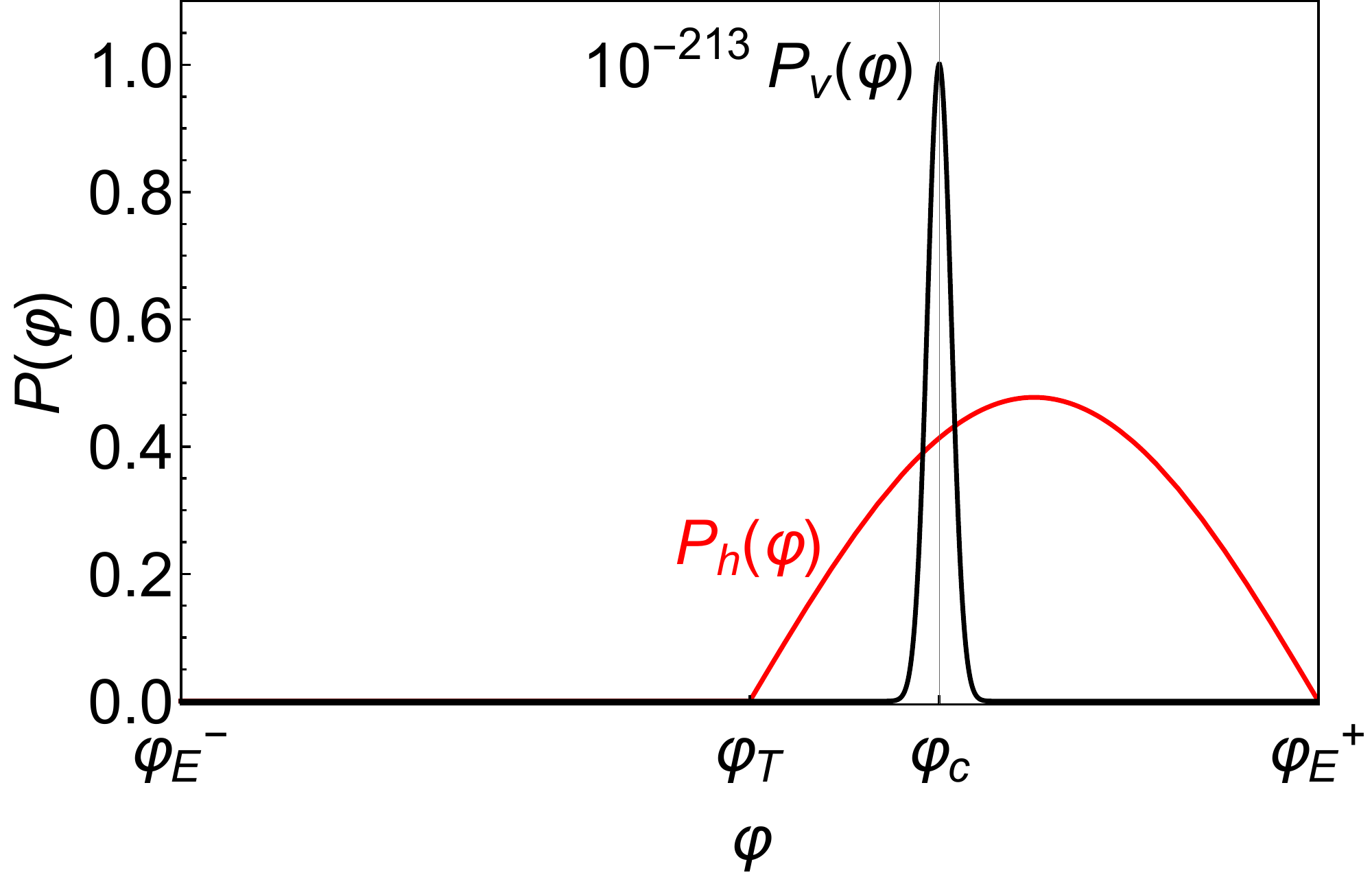}
\end{center}
\caption{{\it Left panel.} A sketch of the scalar potential of the waterfall scheme. The two branches $v$ and $h$ are degenerate in energy at a generic field value $\vf_c$, but transitions between them occur only at $\vf =\vf_T$. {\it Right panel.} The asymptotic volume-weighted distribution in phase $v$ (in black and multiplied by a reduction factor of $10^{-213}$) and phase $h$ (in red) for $\alpha=10^{-5}$, $\beta=10^{3}$, $\vf_T =-0.5$, $\vf_c =0$, $\vf_E^+=1$ and $k_v=k_h=1$ (the value of $\vf_E^-$ is inconsequential for the result). The distribution is dominantly found in phase $v$, peaked near $\vf =\vf_c$.}
\label{fig:ccfig}
\end{figure}

It remains to specify the boundary conditions. In the following, with no loss of generality, we make a coordinate choice such that $\vf_c =0$ and $\vf_E^\pm = \pm 1$ (although the result is insensitive to the location of the lower endpoint). For phase $h$ we choose an absorbing boundary condition at the upper endpoint,  $\pv_h(1)=0$, but this is unimportant for our conclusions and analogous results are obtained for a reflecting boundary condition. For
 phase $v$ we also choose an absorbing boundary condition at the lower endpoint, $\pv_v(-1) = 0$. However, at the upper endpoint, we set a non-vanishing boundary condition for the derivative of the distribution, $\pv_v^\prime (1)\neq0$.  We assume this is due to a UV sector in the landscape which provides a source of flux, possibly due to tunnelling.  This will not in general be constant, however once steady state is reached this will become a constant, up to the overall inflationary contribution common to all sectors.  Factoring this out we assume a value of the derivative equal to $-k_v$, with $k_v$ being a positive constant that parametrises these UV properties beyond our EFT knowledge. In practice, the chosen boundary condition is a way to enforce that, in the far UV, the distribution asymptotically approaches the diffusionless solution rather than the Gibbs one discussed in \sec{sec:determ}.

Since boundary conditions are determined by the UV completion, they are incalculable within the EFT and we have no way to assign any meaningful measure of their likelihood in theory space. However it is important to assess whether fine-tuning of boundary conditions was required to realise the diffusionless solution on the visible branch. This solution is realised for $k_v$ above a certain
value $k_v^{{D \!\!\!\!\! \not\,\,\,\,\,}}$, defined such that for $k_v<k_v^{{D \!\!\!\!\! \not\,\,\,\,\,}}$ the solution is predominantly Gibbs-like and no peak is generated. Since the absolute normalisation of the combined distribution is unphysical (there is a degeneracy in rescaling simultaneously $k_v$ and the parameter $k_h$ that normalises the solution in the $h$ phase, defined below) one cannot attach any meaning to the absolute numerical values of $k_v$.  Furthermore, on phase $v$ the solutions behave exponentially whereas on phase $h$ they do not, so a numerical comparison between $k_v$ and $k_h$ is unlikely to provide any insight.  More practically, once $k_v$ is in the diffusionless window, we may consider the logarithmic derivative of the position or width of the peak with respect to $k_v$ as a fine-tuning criterion. Our calculation shows that, once in this basin of attraction, fine-tuning of $k_v$ is not necessary.
 
A complementary perspective comes from considering the inflationary physics.  In slow-roll the dominant contribution of a field to the vacuum energy density, and hence inflationary rate, is given by the corresponding height of the scalar potential.  As a result, on physical grounds, for a given eigenvalue (i.e.~inflationary rate) we expect the field to be located at the corresponding position on the scalar potential.  This is precisely the case for the diffusionless solution, giving some physical motivation for the family of boundary conditions that realise this solution.  On the other hand, for Gibbs-like solutions the field is not located at the corresponding height on the potential, but very far from it.  This means that the inflationary rate is not being supported by the vacuum energy of a slowly-rolling scalar, but instead by a large diffusion term, even though the dynamics is in the C regime.  This would seem to imply that for the Gibbs-like class of solutions individual trajectories must be outside of the slow-roll regime and so we wish to avoid Gibbs-like solutions in the case that they correspond to implausible or fine-tuned boundary conditions.  It would, however, be valuable to further investigate the possible UV motivations for the different ranges of IR boundary conditions and how they delineate the corresponding classes of solutions.

The junction conditions at $\vf =\vf_T$ are dictated by the structure of the phase transition and follow from eqs.~(\ref{eqquno})--(\ref{eqqdue}) 
\beq
\pv_h (\vf_T) = 0~,~~~\Delta \pv_v (\vf_T) =0~,~~~\Delta \pv_v^\prime (\vf_T) = -\pv_h^\prime (\vf_T) ~.
\eeq

With these boundary and junction conditions in mind, we can easily determine the properties of the stationary solutions. The mode that satisfies positivity in the $h$ phase must be of the form
\beq
\tpv_h \propto \sin \left[\sqrt{\frac{-2 \lambb}{\alpha}}\, (\vf-\vf_T) \right] ~,~~~~
\lambb = -\frac{\pi^2 \alpha}{2(1-\vf_T)^2} ~.
\eeq
Now consider what this implies for the system in the $v$ phase.  For $\beta \gg 1$, \eq{pikpik} shows that, if a peak of the distribution exists, its location ${\bar \vf }$ must correspond to a potential height
\beq
\fun_v ({\bar \vf }) = -\frac{\pi^2 \alpha}{2(1-\vf_T)^2 \beta} ~,
\label{height}
\eeq
which can be very small in the regime we are considering ($\alpha \beta \ll 1$, $\beta \gg 1$).

This is a highly non-trivial result.  In isolation, the field value corresponding to $\fun \approx 0$ is not special, thus there is no reason for the stationary solution in phase $v$ to be peaked at this point.  Yet, as a result of a steady state being reached in phase $h$, where the vacuum may decay to vacuum $v$, a peak is formed in phase $v$ arbitrarily close to the point where the two vacua are degenerate, even though the tunnelling between the two vacua at this specific point is exponentially suppressed.

These considerations are confirmed by the analytical solution of the FPV, which can be obtained for an exactly linear potential in phase $v$, taking $\fun_v= \vf $. The asymptotic solution in the perturbative C regime is
\beq
\left\{ \begin{array}{lc}
{\displaystyle \pv_v  =  \alpha \, e^{\frac{\vf_T-\vf}{\alpha}} \left( a_\pm \, { A}_\vf + b_\pm \, { B}_\vf \right)} & ({\rm for~}\vf \, \stackanchor[0pt]{$>$}{$<$} \,\vf_T) \\[5pt]
{\displaystyle \pv_h  = \frac{(1-\vf_T)k_h}{\pi}\,  \sin \Big[ \frac{\pi  (\vf -\vf_T)}{(1-\vf_T)} \Big]} & ({\rm for~}\vf >\vf_T )
\end{array}
\right. 
 \label{apbp}
\eeq
\beq
{A}_\vf =\frac{\Ai (x)}{\Ai (x_T)}~,~~~{ B}_\vf =\frac{\Bi (x)}{\Bi (x_T)}~,
\eeq
\beq
\begin{array}{llll}
a_- = \frac{k_h}{2} + a_+  ~,& b_- = - z \, a_- ~,&
a_+=  \frac{k_v z_v-k_h z_h}{\alpha \beta}  ~,& b_+ = \frac{k_h}{2} + b_- ~,
 \end{array}
 \label{coefjun}
\eeq
\beq
z=\frac{{ A}_-}{{ B}_-}\approx e^{-\frac{2(1+\vf_T)}{\alpha}}~,~~~
z_v= \frac{e^{\frac{1-\vf_T}{\alpha}}}{{A}_+}\,  \approx e^{\frac{\beta}{2} (1-\vf_T^2)}~,~~~
z_h = \frac{{ B}_+}{{ A}_+}\approx e^{-\frac{2(1-\vf_T)}{\alpha}} ~,
 \label{coefjun2}
\eeq
where $x$ is defined in \eq{sollinuf}, $x_T=x(\vf_T)$, and ${ A}_\pm$ or ${B}_\pm$ are the functions ${ A}_\vf$ or ${ B}_\vf$ evaluated at $\vf =\pm 1$. Finally, $k_h$ is a normalisation constant determined by initial conditions. In \eq{coefjun}, corrections ${\mathcal O}(\alpha \beta )$ have been neglected, consistently with the approximation of the C regime. The expressions in \eq{coefjun2} after the symbols $\approx$ are given only to show the size of the corresponding quantities, but should not be used in evaluating the distribution since they neglect sizeable corrections.

Whenever $k_v$ is sufficiently larger than $k_h$, the term proportional to $a_+$ dominates the behaviour of $\pv_v$ and the distribution shows a peak with location and width given by 
\beq
{\bar \vf} =-\frac{\pi^2 \alpha}{2(1-\vf_T)^2 \beta} ~,~~~~\sigma = \sqrt{\frac{1}{\beta}} ~.
\eeq
This result is in agreement with \eq{height}. The solution given in \eq{apbp} is shown in \fig{fig:ccfig} (right panel) for a representative choice of parameters.

In a neighbourhood of the peak location ${\bar \vf}$, we can expand the solution and obtain
\beq
\pv_v  = \alpha\,  a_+ \, e^{\frac{\beta}{2}[\vf_T^2- (\vf -  {\bar \vf})^2] }
\approx \frac{k_v}{\beta} \, e^{\frac{\beta}{2}[1- (\vf -  {\bar \vf})^2] }
~,
\eeq
where the last expression is valid as long as $k_v$ is not exceedingly small.
Therefore, the relative probability for the system to reside asymptotically in the two phases is
\beq
\frac{\int d\vf \, \pv_v}{\int d\vf \, \pv_h}= \sqrt{\frac{\pi^5}{2\beta^3}}\, \frac{k_v \, e^{\frac{\beta }{2}}}{ k_h(1-\vf_T)^2}
 ~.
\label{intrap}
\eeq
The relative probability depends on the ratio $k_v/k_h$. If this ratio is large enough to ensure that $a_+>b_+$, then phase $v$ is favoured by an exponentially large factor, since $\int \pv_v /\int \pv_h \gsim \sqrt{\alpha^2/ \beta}\, \exp (\beta \vf_T^2 /2 )$. If the term proportional to $a_+$ dominates the behaviour of $\pv_v$ only up to the crossing point $\vf =0$, then phase $h$ is exponentially favoured since $\int \pv_v /\int \pv_h \sim  \sqrt{\alpha^2/ \beta}\, \exp (\vf_T/\alpha)$, where $\vf_T$ is negative.

Note that stationarity in this instance is a form of equilibrium.  For any non-vanishing stationary configuration, as enforced by the boundary conditions, the solutions on both branches must inflate at precisely the same rate, and this is the fundamental reason for SOL in this case.  There is a form of equilibrium between the two branches, much as two thermally-coupled boxes of gas will reach equilibrium even if their microphysics is radically different.  Indeed, since the Hubble scale can be interpreted as a form of temperature of de Sitter space this equality of constant Hubble scales is very much analogous to thermal equilibrium.

The waterfall scheme illustrates a striking form of SOL. Thanks to the interplay between two different phases, the vacuum energy is dynamically self-tuned to the right value to guarantee degeneracy. Localisation in phase $a$ occurs at a point which is not an extremum of the potential nor special for an observer confined to this phase. The hidden property of the localisation point is its degeneracy in energy with a different phase of the theory, which communicates with phase $a$ only through transitions that occur in a field region far from the localisation point. 

\subsection{Determining Fundamental Parameters with SOL}
\label{sec:detpar}

One of the most interesting applications of SOL is the determination through cosmological evolution of fundamental parameters of a microscopic theory, which we will identify here with the SM, but which could equally be any of its field-theoretical extensions. The speculation that coupling constants could be determined by multiple point criticality, which corresponds to the condition for coexistence of different phases, has already been proposed in ref.~\cite{Bennett:1993pj}, although without offering any concrete theoretical realisation. Conceptual connections between criticality and cosmology have also been suggested in broader contexts \cite{Smolin:1995ug}. In the case of SOL, the basic assumption is that the SM parameters can be promoted to dynamical variables scanned by the cosmological evolution of a new scalar field $\phi$, which will be called {\it apeiron}\footnote{{\it Apeiron}, a Greek word for ``boundless," is the central element of Anaximander's vision of cosmology: it is the origin of cosmic order and balance of forces. Similarly, the apeiron is central in our theory to determine fundamental parameters through cosmological self-organisation. The word {\it apeiron} is not new in physics, but was previously introduced in 1944 as a concept in statistical mechanics by Max Born {\it et al.}~\cite{Born}.} and which has a nearly-flat potential because of an underlying approximate shift symmetry. 

While scanning SM parameters, the apeiron explores a field range $f$.  The SM sector has an energy cutoff $M$, where new physics appears and whose interactions are generically described by a coupling constant $g_*$ which, by na\"ive dimensional analysis, must be smaller than $4\pi$. Based on its symmetry properties, the general form of the EFT potential is
\beq
V  = \frac{M^4}{g_*^2} \,\fun (\vf ) -\vf \, {\mathcal O}_{\rm SM} +V_{\rm SM}
~,
\label{poteft}
\eeq
where $\vf =\phi /f$ varies in the range $|\vf| \le 1$ and $\fun$ is a generic function of order unity, assumed to be monotonic with $\fun' >0$. 

It is natural to interpret the apeiron as a Goldstone boson emerging from a spontaneously broken global symmetry in a hidden sector, with a small amount of explicit symmetry breaking which is responsible for the $\phi$ potential. In this case, $f$ is not to be necessarily identified with the Goldstone  periodicity because we are interested in an EFT field range where $\fun(\vf)$ is a monotonic function. The scale $f$ has to be interpreted as the field excursion for the apeiron to scan some physical parameter. As a result, the Goldstone decay constant must be larger than, or at best comparable with, the scale $f$. The shallowness of the potential in practise requires $f\gg M_P$, which implies that the Goldstone decay constant must be super-Planckian. This is not necessarily in conflict with a field-theoretical treatment as long as the energy densities involved are sub-Planckian but, as discussed in Sec.~\ref{sec:time}, it raises concerns about the embedding of the theory in a quantum-gravity completion. This problem about super-Planckian field excursions, which is common in many cosmological setups, can be circumvented by assuming that the largeness of the scale $f$ is only a mirage created by an underlying non-trivial monodromy~\cite{Silverstein:2008sg} or clockwork mechanism~\cite{Choi:2015fiu,Kaplan:2015fuy,Giudice:2016yja}. Alternatively, one could interpret $\phi$ as a nearly flat direction of a non-compact modulus, protected by an effective supersymmetry residing in the hidden sector. For our applications, we will simply use the EFT expansion in \eq{poteft}, but we will not need to specify the microscopic nature of $\phi$. 

In \eq{poteft}, $V_{\rm SM}$ describes the SM potential and ${\mathcal O}_{\rm SM}$ is an operator made of SM fields with a coupling constant scanned by the apeiron in a range of order one. We have made a field redefinition such that the apeiron coupling to the SM operator is linear in $\vf$, but the generalisation to couplings with other functional dependences and multiple scanning parameters can be contemplated.

 Note that the overall normalisation of the $\vf$ potential in \eq{poteft} is dictated by the SM cutoff $M$. This is consistent with naturalness arguments based on stability under quantum corrections. Loops amount to replacing ${\mathcal O}_{\rm SM}$ with $M^4/(4\pi g_*)^2$ and do not affect the structure of the EFT potential in \eq{poteft}. In the notations of \eq{potinEFT}, the symmetry-breaking coupling $\ett$ is given by
\beq
\ett = \frac{M^2}{g_* f^2}~,
\eeq
and the EFT parameters $\alpha$ and $\beta$ measure  the Hubble rate in units of the SM cutoff ($\alpha \sim H_0^4/M^4$) and the field range in Planckian units ($\beta \sim f^2/M_P^2$), respectively. 

During inflation, the value of $\phi$ undergoes large fluctuations, causing the SM parameters to vary widely. At certain critical values, the SM vacuum structure may suddenly change, shifting from a low-field to a high-field vacuum.  In other words, the $\phi$-space can exhibit a critical point with a first-order quantum phase transition. Much as a cooled ferromagnet spontaneously and discontinuously flips internal spins as an external magnetic field is varied to change orientation, so the SM can spontaneously transition between two vacua once the field $\phi$ evolves past the critical point. 

The phase diagram of the theory is such that $\vf$ triggers a phase transition in the SM sector at a critical point which can be chosen to be $\vf =0$ with an appropriate field shift. The corresponding order parameter is
\beq
\langle {\mathcal O}_{\rm SM} \rangle = \left\{ 
\begin{array}{llc}
v_{\mathsmaller {\rm IR}} & {\rm for}~ \vf <0 &{\rm (IR~phase)} \\
v_{\mathsmaller {\rm UV}} & {\rm for}~ \vf >0 & {\rm (UV~phase)}
\end{array} \right.
\eeq
with $v_{\mathsmaller {\rm IR}} \ll v_{\mathsmaller {\rm UV}}\sim M^4/g_*^2$ describing the two different quantum phases. 

As $\vf$ crosses the critical point, the SM sector backreacts generating a new contribution to the apeiron effective potential. 
This process dynamically drives fundamental parameters towards critical surfaces, providing predictions of physical quantities that defy the EFT logic, since their values do not correspond to enhanced symmetries. The role of the apeiron is to drive the SM to conditions under which two different vacua coexist and are on the verge of a phase transition. This situation of vacua critical co-existence is the smoking gun of the SOL mechanism.

\subsection{SOL Post-Inflationary Dynamics}
\label{sec:post}

SOL can only make a prediction about the SM parameters at the end of inflation, but their actual physical values depend also on post-inflationary dynamics. The late motion of the apeiron induces a time-dependence of the scanned SM parameters and masses. The time-variation of the vacuum energy provides the strongest constraint and also the most universal, since it is independent of the specific model implementation. 

SOL predicts that the FPV distribution of $\phi$ is sharply peaked around a given value ${\bar \phi}$ at the end of a long period of inflation. After reheating, the apeiron will slow roll down the shallow potential according to its equation of motion
\beq
{\dot \phi} \approx - \frac{V'( {\bar \phi} )}{3H} ~~~,~
V'( {\bar \phi} ) \approx \frac{\hbar \, H_0^4}{\pi^2 M_P \sqrt{\alpha^2 \beta}}
~,
\eeq
where $H$ is the Hubble rate of the radiation-dominated Universe and, for simplicity, we set $\GAUG =1$. The apeiron potential can be approximated as
\beq
V(\phi ) \approx \Lambda^4 + V'( {\bar \phi} ) (\phi -  {\bar \phi} ) ~,
\eeq
where we have tuned the vacuum energy such that it is equal to the observed value of the cosmological constant ($\Lambda = 2.46\times 10^{-3}$~eV) in the proximity of the configuration $ {\bar \phi} $. 

The dark-energy behaviour of $\phi$ is conveniently described in terms of the ratio between the pressure $P_\phi$ and energy density $\rho_\phi$ of the apeiron
\beq
w = \frac{P_\phi}{\rho_\phi}= \frac{{\dot \phi}^2/2 -V(\phi)}{{\dot \phi}^2/2 +V(\phi)}\approx -1 +\frac{{\dot \phi}^2}{V({\bar \phi}) } ~.
\label{wdark}
\eeq
Since present observations require that dark energy resemble a cosmological constant with a discrepancy from $w=-1$ of less then percent, we must impose
\beq
\alpha^2 \beta > \left( \frac{\hbar \, H_0^4}{M_P H_{\rm now} \Lambda^2}\right)^2
= \left( \frac{H_0}{2\times 10^{-3}~{\rm eV}}\right)^8
~,
\label{dees2}
\eeq
where $H_{\rm now}=1.37\times 10^{-42}$~GeV is the Hubble rate today.  

It is possible to circumvent the bound in \eq{dees2} at the price of introducing new dynamics into the model, preventing the field $\phi$ to evolve in the thermal environment. One could imagine a trapping mechanism based on the existence of a hidden-sector interaction whose non-perturbative effects break the global shift symmetry, generating a periodic potential that stops $\phi$ from further evolution during the thermal history of the Universe. An even more economic solution is to interpret $\phi$ as the familiar axion and use the non-perturbative QCD interactions to trap the field $\phi$. 

Assuming that the trapping mechanism occurs when the temperature of the Universe is $T_T$, the post-inflationary slow roll gives a relative shift of the field $\phi$ from its initial location ${\bar \phi}$ 
\beq
|\delta_\phi | \approx \frac{|V'({\bar \phi})|\, M_P^2}{{\bar \phi}\, T_T^4}
\approx \frac{H_0^4}{\alpha \beta\, {\bar \vf}\, T_T^4} ~.
\label{eq:deltaphi}
\eeq
The quantity $\delta_\phi$ measures also the relative shift of the physical parameter scanned by the apeiron.
As long as $|\delta_\phi |\! \lsim \! 1$, the post-inflationary evolution does not modify significantly the original SOL prediction. This implies a lower bound on the trapping temperature $T_T$.

\subsection{EFT Parameters and SOL}
\label{sec:range}

It may be useful to summarise here some conditions on EFT parameters that could be of relevance for SOL. We can express the conditions in terms of the following three physical parameters: the inflationary background Hubble rate $H_0$, the SM energy cutoff $M$ where new physics is expected to take place, and the apeiron field excursion $f$ needed to scan the SM parameters in their full range. Equivalently, we can use the parameters $H_0$, $\alpha$ and $\beta$.  

Two conditions we may wish to consider in the EFT framework are the request that {\it (i)} the $\phi$ modulation of the energy density is only a perturbative correction to the background value, see
 \eq{appinf}, and  {\it (ii)} the dark-energy equation of state is not significantly modified by the apeiron slow-roll after inflation, see \eq{dees2}:
 \beq
\! \! \!
\left\{ \begin{array}{c}
{\displaystyle
\alpha > \left( \frac{H_0}{2\times 10^{19}~{\rm GeV}}\right)^2}
\\
{\displaystyle
\alpha^2 \beta > \left( \frac{H_0}{2\times 10^{-3}~{\rm eV}}\right)^8} 
\end{array} \right.
~~{\rm or}~~
\left\{ \begin{array}{cc}{\displaystyle
\frac{H_0}{\rm GeV} > \left( \frac{M}{10^8~{\rm GeV}}\right)^2  10^{-3}} & 
\! \! \Big(\!\!  \begin{array}{c}
{\rm perturbative} \\ {\rm domain}
\end{array} \! \! \Big) \!\! 
\vspace{0.1cm} \\  {\displaystyle
\frac{f}{M_P} > \left( \frac{M}{10^8~{\rm GeV}}\right)^4 10^{79} }& 
\! \! \Big(\!\! \begin{array}{c}
{\rm dark~energy} \\ {\rm EoS}
\end{array} \! \! \Big)\!\! 
\end{array} \right.
\label{SOLregion}
\eeq
where we have taken $g_*={\mathcal O}(1)$. Both conditions are not strictly necessary for SOL. The first condition can be evaded by going beyond the perturbative expansion and considering a wide field range in which the energy variation due to the $\phi$ excursion is large. 
The second condition does not apply in presence of a post-inflation trapping mechanism. Nevertheless, these conditions give an indication for the parameter range of simple SOL models. 

When the two constraints in \eq{SOLregion} are taken together, we find that SOL in the perturbative region and without trapping {\it (i)} can be in the C regime only if $M\lsim 10$~MeV and\footnote{The C regime must also satisfy $H_0 \gsim (M/10~{\rm MeV})^2\, 10^{-14}$~eV, but this bound at best saturates the stronger requirement that the reheating temperature must be large enough to allow for nucleosynthesis which imposes $H_0 \gsim 10^{-14}$~eV.}
 $H_0 \lsim (10~{\rm MeV}/M)\, 10^{-3}$~eV; {\it (ii)} can be in the QV regime only if $M\lsim $~TeV and $H_0\lsim 10^{-3}$~eV; {\it (iii)} must be in the Q$^2$V regime if $M\gsim $~TeV.

The value of $H_0$ is related to the reheating temperature after inflation, which is
\beq
T_{\rm RH} =\left( \frac{90\, H_0^2\, M_P^2}{\pi^2\, g_R}\right)^{1/4} = \sqrt{\frac{H_0}{10^{8}~{\rm GeV}}}~ 9\times 10^{12}~{\rm GeV} ~,
\label{trh}
\eeq
where $g_R$ is the number of relativistic degrees of freedom in the thermal bath. Moreover, $H_0$ is directly related to the ratio $r$ of tensor-to-scalar primordial perturbations, according to $H_0= \sqrt{r}\, 2.58 \times 10^{14}$~GeV. Hence, the non-observation of primordial tensor modes implies $M\lsim 10^{16}$~GeV for the perturbative expansion to be applicable. 

If the system is in the Q$^2$V regime, the number of inflationary $e$-folds necessary to reach the asymptotic state is
\beq
N \sim \frac{\pi g_* f M_P}{\sqrt{\hbar}\, M^2} > \left( \frac{M}{10^8~{\rm GeV}}\right)^2 10^{100} ~.
\label{googol}
\eeq
The colossal super-Planckianity of $f$, see \eq{SOLregion}, and the order-googol number of required $e$-folds, see \eq{googol}, manifestly display the issues about living in the swampland and about the validity of the semi-classical approach discussed in \sec{sec:time}. 

\section{Near-Criticality of the Higgs Self-Coupling from SOL}
\label{sec:nearcrit}

The discovery of the Higgs boson~\cite{Aad:2012tfa, Chatrchyan:2012ufa} has revealed the surprising coincidence that the SM parameters (most notably the Higgs quartic and top Yukawa couplings) lie critically at the edge of a metastability region, where the SM vacuum is close to a phase transition into a high-field vacuum~\cite{Degrassi:2012ry,Buttazzo:2013uya,Bednyakov:2015sca,Andreassen:2017rzq}. Such an intriguing feature of near-criticality could be a hint that the SOL mechanism is operating in the early Universe. 

\subsection{Phase Diagram}

Following the general approach described in \sec{sec:detpar}, we consider a setup in which all SM couplings are $\phi$-dependent scanning parameters although, to simplify the problem, we fix the weak scale and the cosmological constant at their observed values. We will describe how one could address these additional questions in sects.~\ref{sec:natur} and \ref{sec:cc}.

The EFT potential is a general function of $\vf\equiv \phi/f$ and $g_*^2 h^2/M^2$, where $h=({2H_h^\dagger H_h})^{1/2}$ is the real and positive field describing the Higgs boson in unitary gauge, contained in the Higgs doublet $H_h$. For our purposes, the potential  
can be written in the form of \eq{poteft} as\footnote{It may seem that our parametrisation is unnatural since we are assuming a precise correlation between the way the quartic and quadratic Higgs terms scan with $\vf$. In reality, this choice is made only for the sake of simplifying the presentation and it has no impact on our results, since the relevant energy scales are much larger than $v$ and we could equally well set $v=0$. Also, as discussed in \sec{sec:post} the strongest bound on $f$ comes from the time variation of the vacuum energy and not from time variations of the Fermi constant or fermion masses which, unlike the vacuum energy, are sensitive to the scanning procedure of the Higgs parameters.} 
\beq
V (\vf , h) = \frac{M^4}{g_*^2} \,\fun ( \vf ) +\frac{\lambda (\vf , h)}{4} \left( h^2-v^2\right)^2~,
\label{potefthc}
\eeq
where $v=246$~GeV. 
The effective quartic coupling $\lambda$ is scanned by the apeiron variation in the range $|\vf | \le 1$ and, with no loss of generality, we can make a field redefinition such that
\beq
\lambda (\vf , M/g_*)= - g_*^2\, \vf~,~~~\frac{d\, \lambda  (\vf , h)}{d \ln h^2} =\beta_\lambda (h) ~,
\eeq
where $\beta_\lambda$ is the SM beta-function describing the RG evolution for $h<M/g_*$. In practice, each value of $\vf$ identifies one RG trajectory of the quartic coupling. The dependence on all other SM couplings (which, in turn, are also $\vf$-dependent) is encoded in the beta-function. 

To simplify the discussion, let us focus on the case in which $\beta_\lambda$ is negative at all scales, as $\vf$ scans the SM couplings. Then, the potential in \eq{potefthc} has two possible Higgs minima. For values of $\vf$ such that $\lambda (v)$ is positive, we find the usual SM vacuum $\langle h\rangle =v$. For $\vf >0$, the Higgs potential develops an unstable direction  at large field values. We assume that the UV completion gives a positive contribution to $\beta_\lambda$, sufficient to stabilise the potential at $\langle h\rangle  = c_{\mathsmaller {\rm UV}} M$, where $c_{\mathsmaller {\rm UV}}$ is a coefficient of order unity. When $\lambda(h)$ changes sign at an intermediate scale, it is possible to have coexistence of the two vacua for the same value of $\vf$ with a potential barrier separating the two. 

Expanding for small $\vf$ and integrating out the Higgs, the two branches of the $\vf$ potential in the IR and UV phases around the critical point are
\beq
\frac{V (\vf , \langle h\rangle )}{M^4} = 
\left\{
\begin{array}{rll}
\kappa_{\mathsmaller {\rm IR}}  \vf +\dots
 &~ {\rm for}~\vf<\vf_+~&~${\textrm (IR~phase:}$~\langle h\rangle =v)
\vspace{0.2cm}\\
-\kappa_{\mathsmaller {\rm UV}}\vf +\dots
 &~ {\rm for}~\vf>0~&~${\textrm (UV~phase:}$~\langle h\rangle =c_{\mathsmaller {\rm UV}} M)
\end{array}
\right.
\label{pothh}
\eeq
\beq
\kappa_{\mathsmaller {\rm IR}} = \frac{\fun' (0)}{g_*^2}~,~~~~ 
\kappa_{\mathsmaller {\rm UV}} = \frac{g_*^2 c_{\mathsmaller {\rm UV}}^4}{4} -\kappa_{\mathsmaller {\rm IR}} ~,
\eeq
where the parameters $\kappa_{\mathsmaller {\rm IR,UV}}$ are generically of order one and will be taken to be positive. In \eq{pothh} we have not shown explicitly a constant vacuum energy, eventually tuned to reproduce today's cosmological constant. 

\begin{figure}[t]
\begin{center}
\includegraphics[width=0.55\columnwidth]{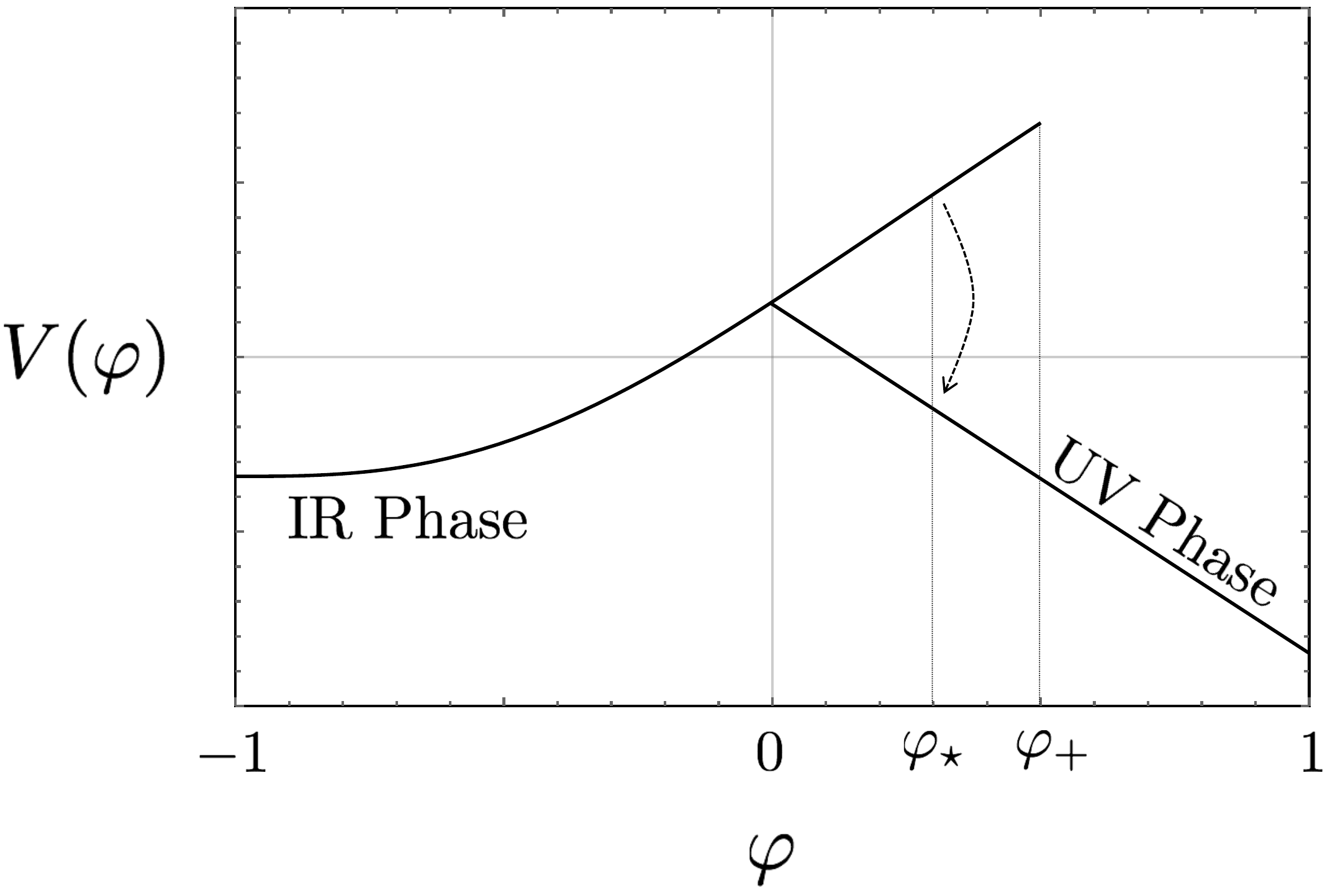} 
\end{center}
\caption{A sketch of the scalar potential $V(\vf , \langle h\rangle )$ with the two branches corresponding to the IR ($\langle h\rangle =v$) and UV ($\langle h\rangle \sim M/g_*$) Higgs phases. The two phases coexist for $0 <\vf <\vf_+$. During inflation, the supercooled IR phase terminates at $\vf_\star$ because of Hubble-induced transitions into the UV phase. }
\label{fig:H2}
\end{figure}

The form of the potential in \eq{pothh} is sketched in \fig{fig:H2}. The IR branch is degenerate with the UV branch at $\vf =0$ (the point corresponding to the RG trajectory such that $\lambda(M/g_*) =0$) and terminates at $\vf=\vf_+$ (corresponding to the RG trajectory such that $\lambda (v)=0$).  Between these two points, the IR branch is in a supercooled phase characterised by $\lambda(\Lambda_I)=0$ with $\Lambda_I$ varying from $M/g_*$ to  $v$, as $\vf$ varies from $0$ to $\vf_+$. In \fig{fig:H2}, the UV branch terminates at $\vf =0$ since we make the plausible assumption that the UV vacuum disappears immediately beyond the critical point, although the precise location of the endpoint depends on the features of the UV completion. The potential in \eq{pothh} is pyramid-like, as the example studied in \sec{sec:pyra}, with microscopic features at the cusp analogous to those discussed in \sec{sec:micro}.

\subsection{SOL Predictions}

To solve the FPV corresponding to the potential in \eq{pothh} we need to determine the boundary conditions that define the matching between the two phases. We will restrict our considerations to the case in which
the Higgs fluctuations around the UV vacuum are damped, so that the Higgs cannot have large excursions away from its UV minimum. Since the typical Higgs mass at the UV vacuum is $M$, this condition implies 
$M>{3H_0}/{2}$. However, even when this condition holds, the potential is relatively shallow for low Higgs fields. Large Higgs fluctuations around the IR minimum are expected during inflation, and these can bring $h$ to explore high field values, possibly getting trapped into the deeper UV minimum. This means that, during the inflationary era, the system cannot access the full supercooled branch that extends up to $\vf_+$ because the Higgs field prematurely drops from the IR to the UV phase for $\vf > \vf_\star$. Consequently, we must impose the boundary condition that, on the IR branch, the FPV distribution $\pv$ vanishes at $\vf_\star$. 

To estimate the value of $\vf_\star$ we must calculate the condition for Hubble fluctuations to drive efficiently the Higgs from the IR to the UV vacuum. Quantum tunnelling is inefficient because the bubbles of true vacuum do not percolate as they are swamped by the fast expansion of the surrounding space. On the other hand, the Higgs field can overcome the potential barrier by means of de-Sitter `thermal' effects proportional to the Gibbons-Hawking temperature $T_H =H_0/2 \pi$~\cite{Gibbons:1977mu}. Quantitatively, the condition for vacuum transition corresponds to the requirement that the rate for processes mediated by Hawking-Moss instantons~\cite{Hawking:1981fz} is unsuppressed, namely
\beq
\Delta V_{\rm max} < \frac{3 \hbar \, H_0^4}{8\pi^2} ~,
\label{gibhawk}
\eeq
where $\Delta V_{\rm max}$ is the potential barrier that the Higgs field has to climb to reach the UV minimum.

Expressing the running coupling constants in terms of SM physical observables, we find that, in the proximity of the SM values for the strong coupling constant and for the Higgs and top-quark masses, the value of $H_0$ that saturates \eq{gibhawk} is 
\beq
\log_{10} \frac{H_0}{\rm GeV} = 9.8 +0.7 \Big(\frac{m_h}{\rm GeV}-125.10\Big) -1.0 \Big( \frac{m_t}{\rm GeV} -173.34 \Big) +0.3  \frac{\alpha_s(m_Z)-0.1184}{0.0007}~.
\label{lastar}
\eeq
This result corresponds to a full NNLO calculation of the barrier of the effective potential in the SM as performed in ref.~\cite{Buttazzo:2013uya}. Since $\Delta V_{\rm max}$ is the difference between two extrema of the potential, the result is scheme and gauge independent. Note that the result is independent of $M$ and it is therefore robust against unknown features of the UV completion. Equation~(\ref{lastar}) gives, for any given value of $H_0$, the SM parameters that correspond to the RG trajectory identified by $\vf_\star$ and such that the Higgs quartic at the UV scale is equal to $\lambda_\star \equiv -g_*^2 \vf_\star $.

As shown in \sec{sec:linear}, the solution of the FPV equation for a linear potential in the Q$^2$V regime with absorbing boundary conditions at $\vf_\star$ is peaked, at asymptotically large times, in the interval between $\vf_\star -\Delta \vf$ and $\vf_\star$ where 
\beq
 \Delta \vf \approx 
 \left( \frac{\hbar \, H_0^4 M_P^2}{4\pi^2 \kappa_{\mathsmaller {\rm IR}} M^4f^2}\right)^{1/3} ~.
\label{incertphi}
\eeq
In other words, the UV value of the Higgs quartic is determined to be equal to $\lambda_\star$ with an uncertainty $\Delta \lambda \approx g_*^2 \Delta \vf$, which is completely negligible  since the bound on the dark-energy equation of state in \eq{SOLregion} implies
\beq
\Delta \lambda < \left( \frac{H_0}{10^{10}~{\rm GeV}} \right)^{\frac43} \left( \frac{10^{12}~{\rm GeV}}{M}\right)^4 10^{-66}~.
\eeq

However, this estimate omits the effects of Hubble-induced Higgs fluctuations, which are unavoidable and must be accounted for to effectively capture the SOL prediction.  A complete treatment would require the computation of the two-field FPV solution $P(\varphi,h)$, from which the maximum likelihood and statistical uncertainty on $\lambda$ could be extracted, subject to the boundary conditions imposed.  For the sake of brevity we may instead estimate the scale of the effect by assuming that $h^2$ is a random variable with RMS fluctuations of $\mathcal{O}(H_0^2)$.  In this case $\varphi$ will see a smeared background value of $h^2$, rather than a value pinned to a local minimum.  Consequently, this will smear the apparent endpoint $\varphi_+$ by an amount proportional to $\Delta {\vf_+} \sim H_0^2/M^2$. Hence the final uncertainty in the prediction of the Higgs self-coupling is
\beq
\Delta \lambda \approx \frac{g_*^2 H_0^2}{M^2} ~,
\eeq
which is still parametrically small but, in the quantum regime we are considering, is much larger than what was estimated from \eq{incertphi}.

In conclusion, SOL predicts that the SM parameters lie in the proximity of the critical point where two Higgs phases coexist and the ordinary Higgs vacuum is at the verge of collapsing into a new high-field configuration. The proper combination of SM parameters is uniquely determined by the Hubble rate $H_0$, which identifies an RG trajectory for the Higgs quartic that crosses zero at a field value of order $H_0$. It is also expected that, at the end of inflation, the Higgs field lies in the IR phase. Although near-criticality of the SM is a robust consequence of SOL and the related statistical uncertainty is small, the precise prediction for the SM parameters has theoretical uncertainties related to the determination of $\vf_\star$. Here we have calculated $\vf_\star$ on the basis of the Hawking-Moss rate but the criterion we have used has an intrinsic order-one uncertainty on the determination of $H_0$, which is associated with the detailed features of the transition between IR and UV vacua.

\begin{figure}[t]
\begin{center}
\includegraphics[width=0.7\columnwidth]{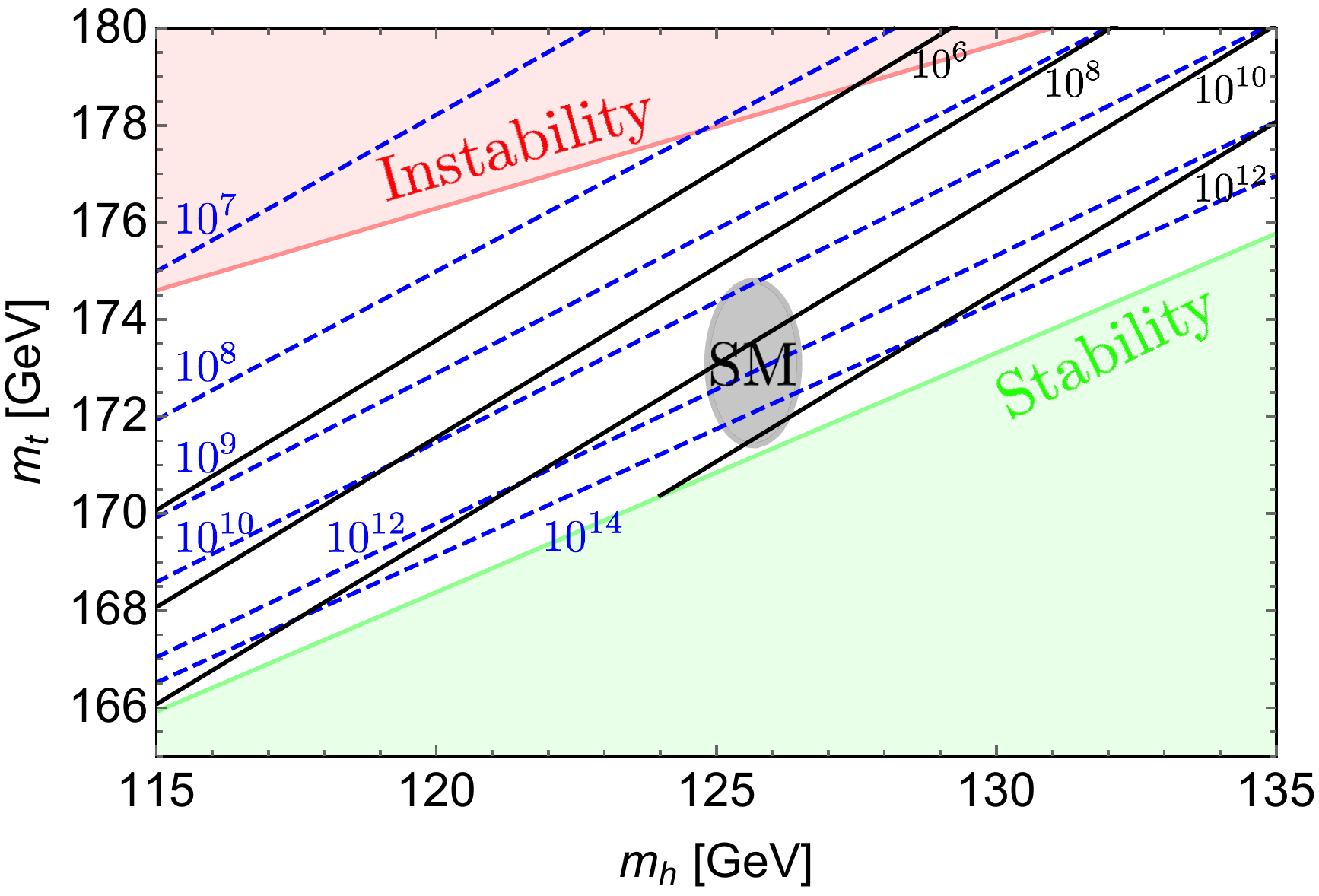} 
\end{center}
\caption{The SOL prediction as a function of the Higgs and top-quark masses, for given values of $H_0$ in GeV (black lines), alongside the absolute stability (green) and instability (red) regions, as calculated in ref.~\cite{Buttazzo:2013uya}. Also shown is the 95\% CL ellipse corresponding to the SM values. The dashed blue lines show the Higgs field value (in GeV) at which the effective potential vanishes and therefore, for $M$ equal to the same value, the IR and UV vacua are exactly degenerate.}
\label{fig:H5}
\end{figure}

\begin{figure}[t]
\begin{center}
\includegraphics[width=0.5\columnwidth]{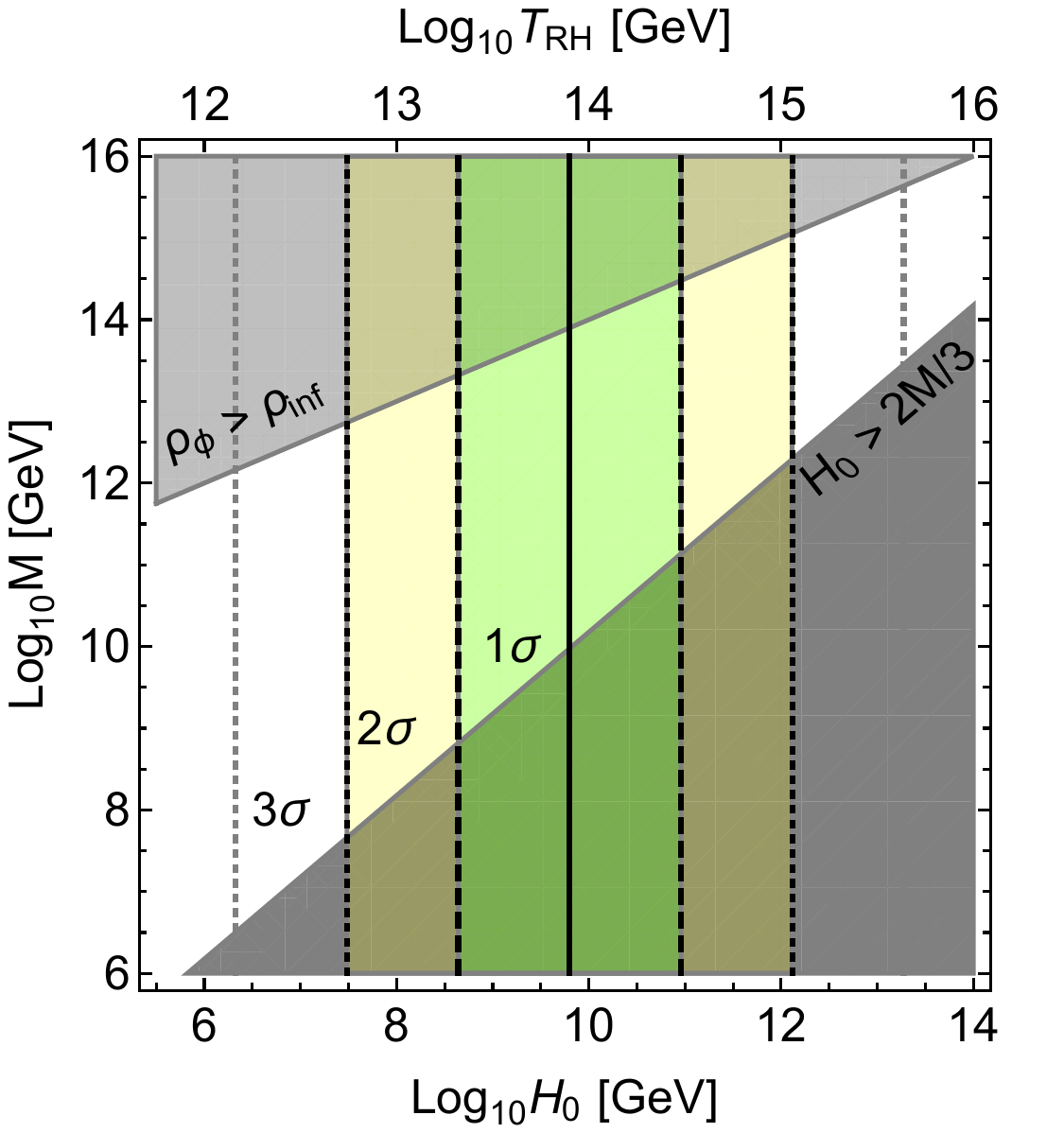} 
\end{center}
\caption{The region of $H_0$ (the Hubble rate during inflation) and $M$ (the energy cutoff of the Higgs sector) in which the SOL prediction is consistent with the SM values of the Higgs and top-quark masses at the 1-$\sigma$ to 3-$\sigma$ levels. Also shown are the corresponding values of the reheating temperature $T_{\rm RH}$. In the shaded regions, the apeiron energy density exceeds the inflationary background ($\rho_\phi > \rho_{\rm inf}$) or quantum fluctuations destabilise the UV phase ($H_0 > 2M/3$).  }
\label{fig:H4}
\end{figure}

In \fig{fig:H5} we show the SOL prediction as a function of the Higgs and top-quark masses, for certain values of $H_0$. We also show the lines corresponding to the critical point at which the IR and UV vacua are degenerate. This is calculated by imposing that the SM effective potential at NNLO, as calculated in ref.~\cite{Buttazzo:2013uya}, vanishes at $h=M/g_*$. These critical lines are uniquely identified by the value of $M$ but, unlike the lines corresponding to $\vf_\star$, contain a residual scheme and gauge dependence because the point $h=M/g_*$ does not correspond to an extremum of the SM potential and are also sensitive to the unknown features of the UV completion. 

In \fig{fig:H4} we show the region of $M$ and $H_0$ where the SOL prediction is consistent with measurements of the Higgs and top-quark masses, allowing for experimental uncertainties and taking $m_h=125.10\pm0.14$~GeV~\cite{Zyla:2020zbs} and $m_t =173.34 \pm 0.76_{\rm exp}\pm 0.30_{\rm th}$~GeV~\cite{ATLAS:2014wva}, where we have added a theoretical error of 300~MeV to account for non-perturbative effects in the top mass. The $M$--$H_0$ space is limited by the previously-mentioned requirement that de-Sitter fluctuations do not affect the Higgs in its UV minimum and by the condition that the inflaton energy density dominates over the typical $\vf$ contribution. These two conditions, shown in \fig{fig:H4}, correspond to
\beq
 \frac{M^2}{\sqrt{3}\, g_*M_P} < H_0 < \frac{2M}{3}~.
\label{cond2}
\eeq
The condition of inflaton domination is not strictly a physical requirement, but it allows us to simplify the calculation by treating the apeiron energy density as a background perturbation. 

The predicted values of $H_0$ are safely below the limit from non-observation of primordial tensor modes but also lead to values of $r$ which are beyond the reach of next-generation CMB experiments. Also shown in \fig{fig:H4} are the corresponding values of the reheating temperature after inflation, given by \eq{trh}. Since the inflaton-domination condition  automatically implies $T_{\rm RH}\gsim M$, the final fate of the Higgs field is insensitive to whether $h$ finds itself on the IR or UV branch at the end of inflation. Once reheating starts, temperature fluctuations dominate the evolution and the Higgs field will be adiabatically driven towards the symmetric configuration as the Universe cools down. It is also interesting that, in the full region shown in \fig{fig:H4}, the reheating temperature is compatible with the lower bound $T_{\rm RH} >2 \times 10^9$~GeV coming from the requirement that the right-handed neutrino explains the cosmic baryon asymmetry through leptogenesis~\cite{Giudice:2003jh}. 

\section{Higgs Naturalness from SOL}
\label{sec:natur}

The SM Higgs bilinear parameter can be interpreted as near-critical with respect to the electroweak phase transition because it happens to be right at the edge of the separation between broken and unbroken phase~\cite{Giudice:2006sn}.  Generically it could take any value between $\pm M^2$, where $M$ is the cutoff of the theory, associated with the scale at which the Higgs mass becomes calculable in terms of more fundamental parameters.  The near-criticality is particularly baffling, as it bluntly violates EFT logic, leading to the well-known Higgs naturalness problem. More than a decade ago, it was suggested that self-organised criticality could provide an explanation of the peculiar value of the Higgs mass~\cite{Giudice:2008bi}, although no concrete realisation was put forward. By now, several ideas have been proposed to explain the Higgs mass in terms of the cosmological evolution~\cite{Dvali:2003br,Dvali:2004tma,Graham:2015cka,Arkani-Hamed:2016rle,Arvanitaki:2016xds,Herraez:2016dxn,Geller:2018xvz,Cheung:2018xnu,Giudice:2019iwl,Kaloper:2019xfj,Dvali:2019mhn,Strumia:2020bdy,Csaki:2020zqz,Arkani-Hamed:2020yna}. 

The EW phase transition between the symmetric and symmetry-breaking vacua  is not immediately suited for SOL, as its quantum nature is second-order. 
However, SOL may address Higgs naturalness by considering the running self-coupling because the underlying quantum phase transition between IR and UV phases, following from Higgs vacuum metastability, is first-order and generates the required discontinuity between the two phases.   Furthermore, the metastability scale is naturally exponentially separated from the scale of UV completion since it arises through the renormalisation group evolution of a marginal parameter.

\subsection{Phase Diagram}
Consider a Higgs mass term which is scanned by the apeiron with a potential given by
\beq
V (\vf , h) = \frac{M^4}{g_*^2} \,\fun (\vf ) -\frac{\vf M^2   h^2 }{2} +\frac{\lambda (h)\, h^4}{4} ~.
\eeq
For simplicity, let us focus on the case in which $\lambda_{\mathsmaller {\rm UV}}\equiv \lambda (M/g_*)$ is negative and $\vf$-independent, while $\beta_\lambda$ is negative at all scales so that $\lambda(h)$ vanishes at an intermediate scale $\Lambda_I$. The Higgs vacuum can be in three different phases and the corresponding apeiron potential, expanded in powers of $\vf$, is
\beq
\! \frac{V (\vf , \langle h\rangle )}{M^4}=
 \left\{ \! \!
\begin{array}{rll}
\kappa_{\mathsmaller {\rm EW}} \vf +\kappa_2\vf^2 +
 \dots
 & {\rm for}~\vf \!<\! 0 &\!\!  ${\textrm (unbroken EW:}$~\langle h\rangle \!=\! 0)
\vspace{0.2cm}\\
 \kappa_{\mathsmaller {\rm EW}}\vf +\kappa_{\mathsmaller {\rm IR}} \vf^2 +
 \dots
 &{\rm for}~0  \!<\! \vf  \!<\! \vf_+ & \!\! ${\textrm (IR~phase:}$~\langle h\rangle \!=\! v )
\vspace{0.2cm}\\
 -\kappa_0 + \kappa_{\mathsmaller {\rm UV}} \vf +\kappa_2\vf^2 +
 \dots
 & {\rm for~any}~\vf  & \!\! ${\textrm (UV~phase:}$~\langle h\rangle \!=\!  c_{\mathsmaller {\rm UV}} M)
\end{array}
\right.
\eeq
\beq
\kappa_{\mathsmaller {\rm EW}} = \frac{\fun' (0)}{g_*^2}~,~~~\kappa_2= \frac{\fun'' (0)}{2g_*^2}
~,~~~\kappa_{\mathsmaller {\rm IR}}=\kappa_2 -\Delta \kappa
~,~~~\kappa_0 = \frac{-\lambda_{\mathsmaller {\rm UV}}c_{\mathsmaller {\rm UV}}^4}{4}
~,~~~\kappa_{\mathsmaller {\rm UV}}=\kappa_{\mathsmaller {\rm EW}}-\frac{c_{\mathsmaller {\rm UV}}^2 }{2} ~.
\label{kapc}
\eeq

First consider the region $\vf \ll (g_* \Lambda_I /M)^2$ and $h \ll \Lambda_I$. A phase transition occurs at $\vf =0$, which corresponds to the critical point for EW symmetry breaking. The IR phase is characterised by a Higgs vacuum $\langle h \rangle =v$, a Higgs mass $m_h$ and an additional contribution to the apeiron potential $\Delta \kappa$ with
\beq
v^2 = \frac{ \vf M^2}{\lambda(v)} ~,~~~~m_h^2 =2\vf M^2
~,~~~~\Delta \kappa = -\frac{1}{4 \lambda (v)}~.
\eeq
The contribution $\Delta \kappa$ gives a discontinuity in $\partial^2 V/\partial \vf^2$, characteristic of a second-order phase transition. This can create a local maximum in the apeiron potential at $\vf = 2 \lambda \omega'(\vf)/g_*^2$.  However, unless one assumes $|\omega'(0)| \ll 1$, contrary to the EFT power counting, the maximum is at a generic field point on the scalar potential, not exponentially close to the critical point $\vf =0$.  Thus the only gain in naturalness that could be found is by assuming a non-generic potential, with the result of pushing the SM UV cutoff only one or possibly two loop factors above the weak scale, as noted in \cite{Cheung:2018xnu,Strumia:2020bdy}. Indeed, the model in ref.~\cite{Strumia:2020bdy} ultimately relies on anthropic selection. Alternatively, one could appeal to a periodic potential from a new axion-like particle in order to create an additional backreaction at the critical point, as proposed in ref.~\cite{Geller:2018xvz}. Because of these limitations, the phase transition between broken and unbroken EW is not suited for SOL and we will follow a different route.

The IR phase is separated from the UV phase, characterised by a high-field Higgs vacuum $\langle h\rangle =c_{\mathsmaller {\rm UV}} M$, by a barrier located at $h_{\rm max}$ and of height $V_{\rm max}$ with
\beq
h_{\rm max}^2= \frac{ \Lambda_I^2}{\sqrt e} ~,~~~~
V_{\rm max} ={\displaystyle \frac{-\beta_I \Lambda_I^4}{8 e}} ~.
\eeq
Here we have approximated the running coupling $\lambda$,
in the neighbourhood of the instability scale $\Lambda_I$, as
\beq
\lambda (h) \approx \beta_I \ln \frac{h^2}{\Lambda_I^2} ~,~~~ \beta_I\equiv \beta_\lambda (\Lambda_I) ~.
\eeq

When $\vf$ approaches $(g_* \Lambda_I /M)^2$ and the mass parameter in the Higgs potential becomes comparable to $g_*\Lambda_I$, the Higgs vacuum structure in the IR phase changes into
\bea
v^2 =  \frac{x_1\Lambda_I^2}{\sqrt{e}} ~,&~~~~
{\displaystyle m_h^2 =2\vf M^2+\frac{2\beta_I x_1 \Lambda_I^2}{\sqrt{e}}}
~,
\\
h_{\rm max}^2= \frac{x_2\Lambda_I^2}{\sqrt{e}} ~,&~~~~
{\displaystyle V_{\rm max} =\frac{-\beta_I x_2^2 \Lambda_I^4}{8 e}-\frac{ x_2 \vf M^2\Lambda_I^2}{4\sqrt{e}}}
 ~,~
\eea
where $x_{1,2}$ (with $x_1<x_2$) are the two solutions of the equation
\beq
x \ln x = \frac{\sqrt{e}\vf M^2}{\beta_I \Lambda_I^2} ~.
\eeq
In this case, we find $\Delta \kappa = {\mathcal O}(1/\beta_I )$.
The IR phase terminates at the value $\vf_+$
\beq
\vf_+= \frac{-\beta_I \, e^{-\frac32}\Lambda_I^2}{M^2} ~,
\eeq
where $x_1=x_2=1/e$. Beyond $\vf_+$, the potential is a monotonically decreasing function of $h$ and the IR vacuum no longer exists. 

\begin{figure}[t]
\begin{center}
\includegraphics[width=0.55\columnwidth]{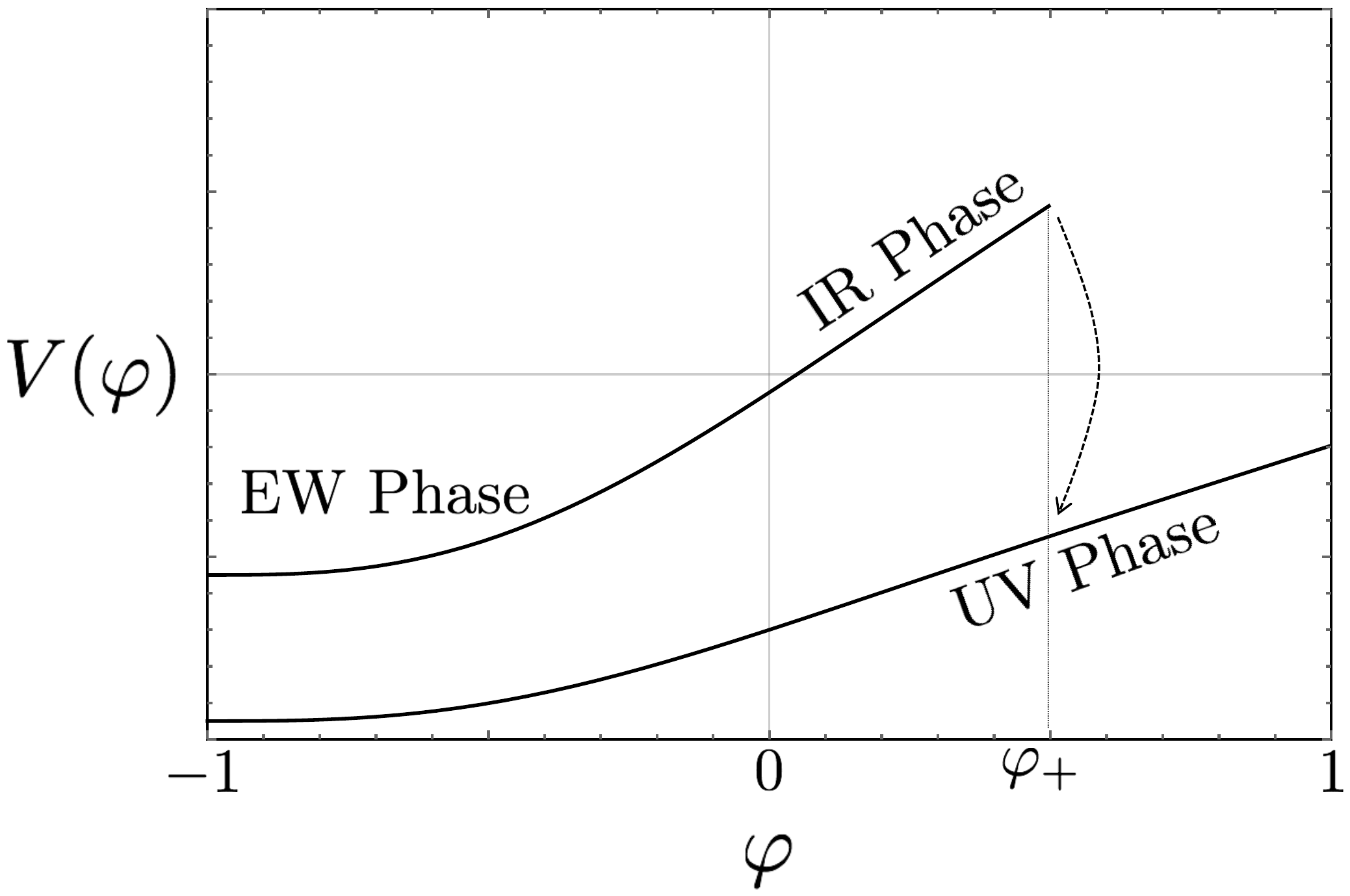} 
\end{center}
\caption{A sketch of the scalar potential $V(\vf , \langle h\rangle )$ with the three branches corresponding to the EW, IR and UV Higgs phases. For $\vf<\vf_+$, two Higgs phases coexist. }
\label{fig:Hpd}
\end{figure}

The coefficients $\kappa$ in \eq{kapc} are of order one and we assume $\kappa_{\mathsmaller {\rm EW}}>0$ and $\kappa_0 > |\kappa_{\mathsmaller {\rm UV}}|, |\kappa_2|$. A sketch of the apeiron potential in the three different phases is shown in \fig{fig:Hpd}.

\subsection{SOL Prediction}

Assuming the apeiron to be in the Q$^2$V regime,
from the results presented in \sec{sec:linear} and in the absence of Hubble-induced Higgs fluctuations, we would conclude that the asymptotic FPV distribution is localised close to $\vf_+$, leading to the following prediction for the Higgs vacuum and mass\footnote{More precisely, the FPV distribution is peaked at ${\bar \vf} =\vf_+ - \delta$ with a width $\sigma = \delta $,
where
$$
\delta \equiv \Big( \frac{\alpha}{2\beta}\Big)^{\frac 13} = \Big( \frac{\sqrt{\hbar}\, g_*H_0^2 M_P}{2\pi M^2 f}\Big)^{\frac 23} ~.
$$
Hence, for small $\delta$, the SOL predictions and their corresponding statistical uncertainties are
$$
v= e^{-\frac 34}  \Lambda_I - \Big( \frac{\delta}{-2 \beta_I}\Big)^{\frac 12}  M~,~~~
\frac{\Delta v}{v} =\Big( \frac{e^{\frac 32}\delta}{- 2\beta_I}\Big)^{\frac 12}  \frac{M}{\Lambda_I}~,~~~
m_h = \Big(\frac{-8 \beta_I \delta}{e^{\frac 32}} \Big)^{\frac 14} \sqrt{\Lambda_I M} 
~,~~~
\frac{\Delta m_h}{m_h} =1
~.
$$
However, the bound in \eq{SOLregion} is so strong that in practice we can set $\delta =0$.
}
\beq
v = e^{-\frac34} \, \Lambda_I ~,~~~~ m_h =0~.
\label{SOLpreved}
\eeq
However, as discussed in \sec{sec:range}, the Hubble-induced Higgs fluctuations must be accounted for.  They will smear the endpoint by an amount proportional to $H_0^2/M^2$ and hence the final prediction for $v^2$ at the end of inflation will be subject to an additional uncertainty of $\mathcal{O} (H_0^2/\lambda )$.  As long as $H_0^2 \ll \lambda \Lambda_I^2$ the quantitative change in the prediction for $v$ is negligible, although the change in the statistical uncertainty is significant with respect to the fixed-Higgs assumption, leading to
\beq
\Delta v \approx \frac{H_0^2}{\lambda v}~.
\label{SOLprevedinc}
\eeq

The vanishing of $m_h$ in \eq{SOLpreved} is signalling that the theory is at the verge of an instability, with the vacuum about to disappear.
The most remarkable aspect of the result in \eq{SOLpreved} is that SOL leads to a natural hierarchy between $v$ and the cutoff energy $M$ as a consequence of the dimensional transmutation which provides the exponentially small ratio $v/M \sim \exp(-\lambda_{\mathsmaller {\rm UV}}/2\beta_\lambda)$. The relevant near-criticality is not between the unbroken and broken EW vacua, which are not separated by any barrier, but between two EW breaking vacua, one in the IR and one in the UV. SOL drives the underlying parameters to a critical situation in which the coexistence of two vacua is about to break down. For an appropriate and generic function $\fun$ and coefficient $c_{\mathsmaller {\rm UV}}$, the IR vacuum is preferred. Just like in the case of the near-criticality of the Higgs quartic discussed in \sec{sec:nearcrit}, the smoking gun of SOL is that SM parameters (in this case the mass term in the Higgs potential) are dynamically driven to a point where the vacuum of the theory is at a critical stage, near to collapse. 

\subsection{Reconciling the Prediction with the SM Higgs}

The SOL prediction for the Higgs parameters in \eq{SOLpreved} is obviously untenable, since the instability scale in the SM is $\Lambda_I \sim 10^{10}$--$10^{12}$~GeV. A way to bring the prediction closer to reality is to introduce new matter that makes $\beta_\lambda$ more negative, leading to an earlier instability. 

As simple illustrative examples consider two kinds of
new vector-like fermions with mass $M_{VL}$:  a weak doublet $\chi$ and a SM singlet $\psi$. We study two possible Yukawa interactions with the Higgs
\beq
(a)~~ \mathcal{L} = - y_{VL} \bar{\psi}\chi H_h + \text{h.c.} ~,~~~~
(b)~~ \mathcal{L} = - y_{VL} \bar{\psi }L H_h + \text{h.c.} ~,
\eeq
where $L$ and $H_h$ are the lepton and Higgs doublets respectively. Case $(b)$ corresponds to what is known as inverse seesaw~\cite{Wyler:1982dd, Mohapatra:1986bd, Ma:1987zm}, although this model does not generate any neutrino mass. Direct searches for vector-like fermions have led to mass limits of about 800~GeV for $\chi$ and 300~GeV for $\psi$~\cite{Sirunyan:2019ofn, Aad:2015dha, Bissmann:2020lge}. Figure~\ref{fig:VLleptons} shows the two-loop running of the Higgs quartic coupling for two choices of masses and couplings for the two cases: $(a)$ $M_{VL} = 1.5$ TeV, $y_{VL}=1.5$, and $(b)$ $M_{VL}=400$ GeV, $y_{VL}=2$. We see that the instability scale can be lowered to $\Lambda_I \sim 2$ TeV in case $(a)$, or 500 GeV in case $(b)$, and that the Higgs potential barrier separating the IR and UV vacua is reduced accordingly, as the apeiron scans the Higgs mass. 

\begin{figure}[h] 
\centering
\includegraphics[width=0.44\textwidth]{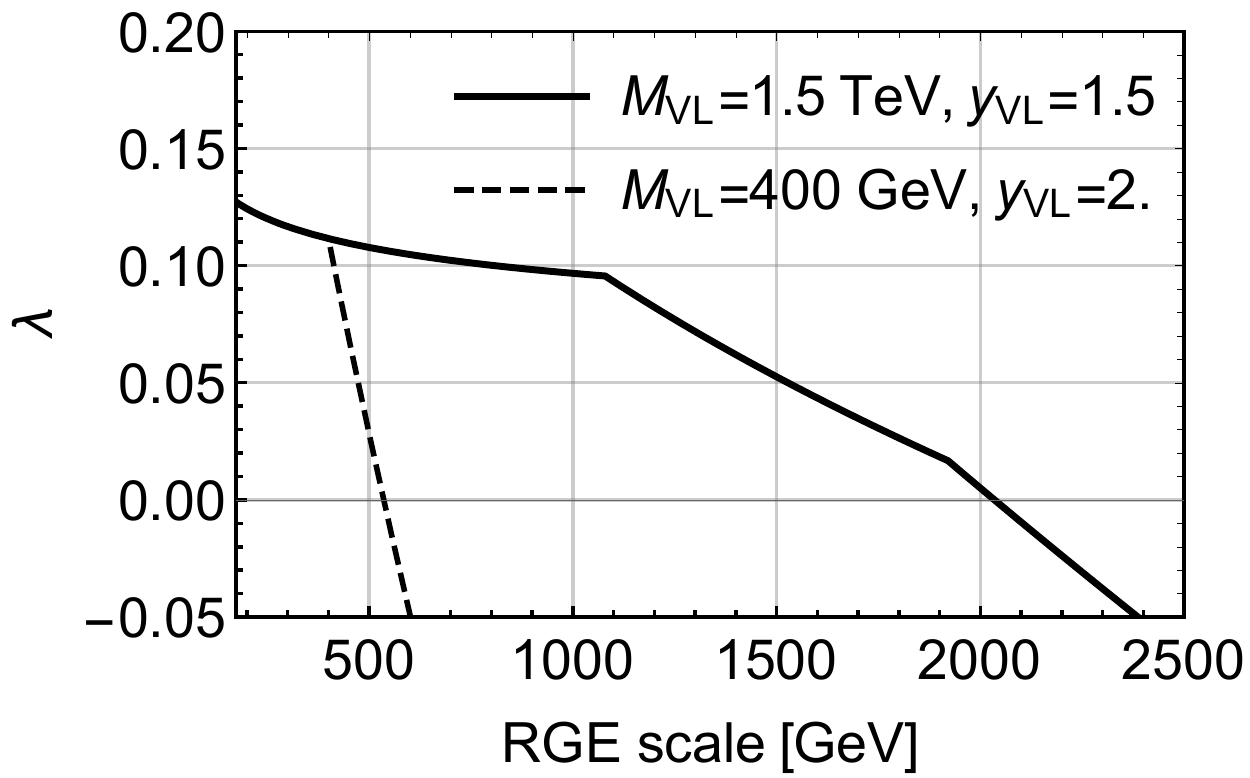}
\includegraphics[width=0.43\textwidth]{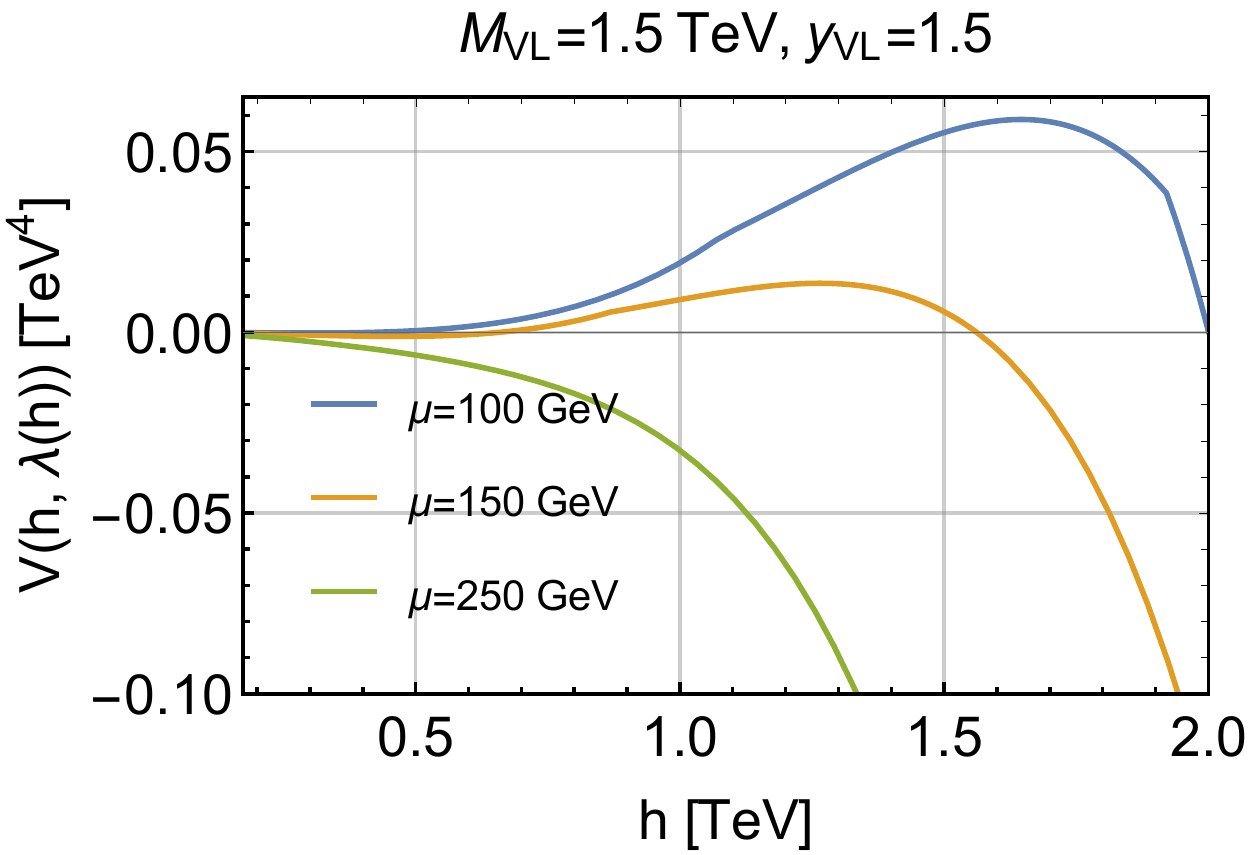}
\caption{\it \label{fig:VLleptons} Two-loop RGE evolution of the Higgs quartic $\lambda$ (left) and the Higgs potential for varying mass $\mu$ in the Higgs bilinear term (right) with the addition of vector-like (VL) doublet and singlet leptons (solid lines) or including only a VL singlet lepton (dashed line).}
\end{figure}

\begin{figure}[h] 
\centering
\includegraphics[width=0.4\textwidth]{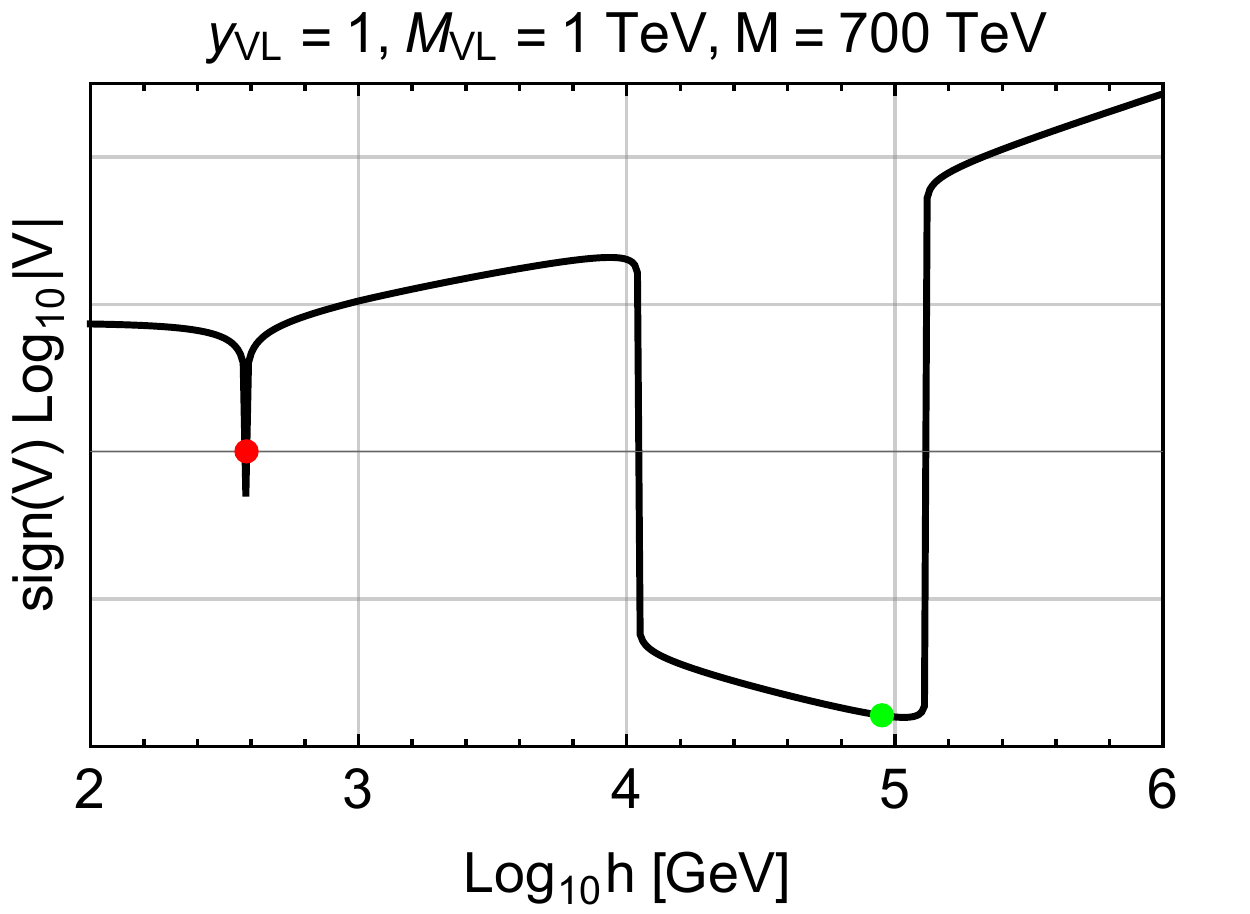}
\includegraphics[width=0.43\textwidth]{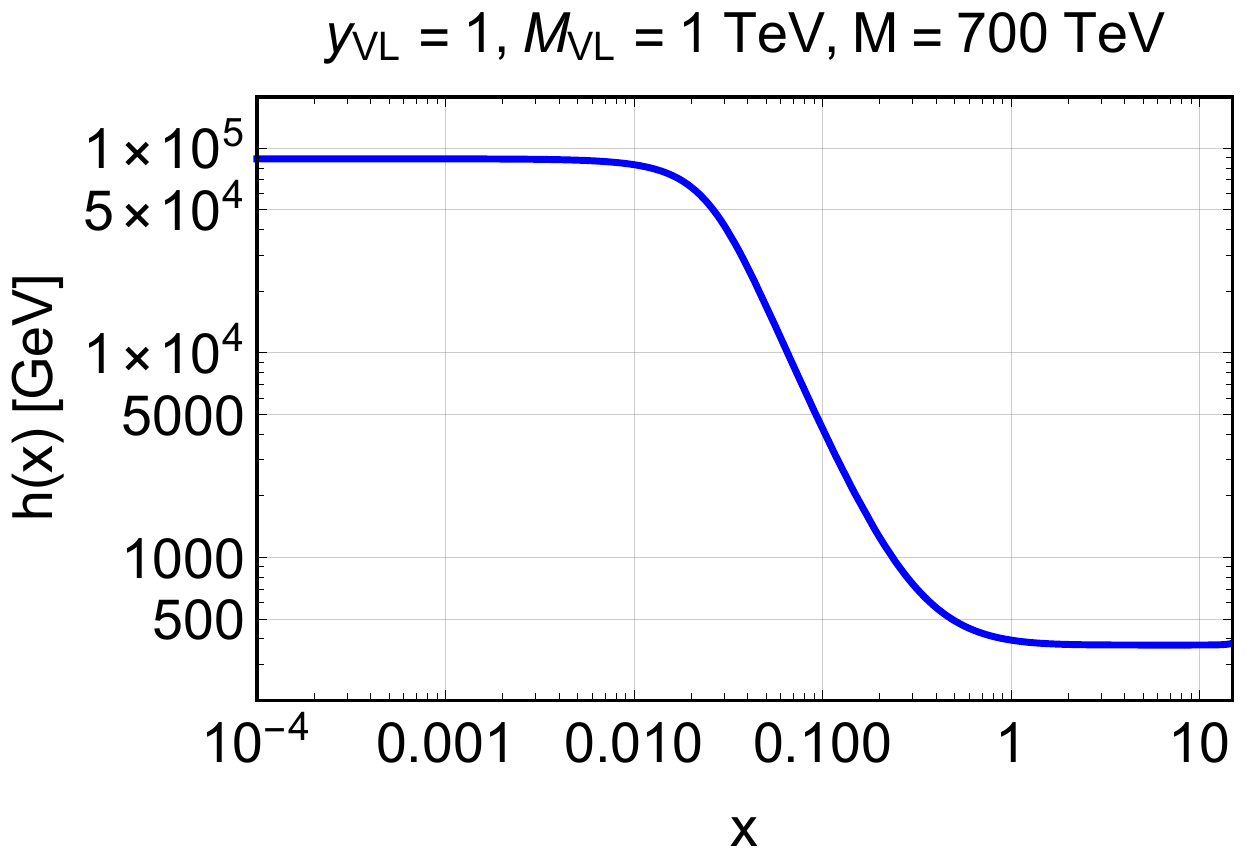}
\caption{\it \label{fig:bounce} Potential (left) with UV vacuum restored by a $|H_h|^6/M^2$ dimension-6 operator at a cut-off $M = 700$ TeV and the corresponding bounce solution (right) between the green and red points, calculated by solving numerically the Euclidean equation of motion for the Higgs field $h$ as a function of the radial coordinate $x$, as described, e.g., in ref.~\cite{Gopalakrishna:2018uxn}. The bounce action for this solution is $S_B=435$, corresponding to an exponentially suppressed probability for vacuum decay, $P \sim e^{404 - S_B}$.}
\end{figure}

The strongest obstruction to lowering $\Lambda_I$ comes from the constraint that quantum tunnelling does not destabilise the Higgs vacuum today. This sets a lower bound on the Higgs quartic $\lambda  \gsim -0.06$~\cite{Isidori:2001bm}, which is quickly violated in the multi-TeV scale for the running corresponding to the parameter choices shown in \fig{fig:VLleptons}. Nonetheless, the bound can be considerably relaxed by adding a dimension-6 operator $|H_h|^6/M^2$, which allows for a sufficiently long-lived Universe even for a cutoff $M$ as large as about $10^6$ GeV, see \fig{fig:bounce}. Or, even more effectively, the bound can be completely evaded if new physics is added above $\Lambda_I$ such that $\beta_\lambda$ is sufficiently small to keep the Higgs quartic in the range $-0.06<\lambda <0$ for energies between $\Lambda_I$ and $M$. This situation, which may appear artificial, could have a justification in setups where the coupling has a quasi-fixed point in the UV and a quasi-conformal running down to the instability scale. 

In conclusion, it is conceivable that new matter in the TeV domain can lower the SM instability scale, bringing the SOL prediction closer to the real world. However, even in the presence of these new particles, the theory is left with an unsatisfactory little-hierarchy problem, since the Higgs mass can be at best in the hundreds of GeV or TeV range. This little-hierarchy problem has a degree of severity comparable to that encountered today in low-energy supersymmetry and is particularly reminiscent of what happens in composite Higgs models, where the weak scale naturally wants to be of the order of the Goldstone scale $f$, with the scale separation $v/f$ requiring some additional dynamics. Similarly, some new ingredients have to be added to our theoretical setup to cure the little hierarchy, and this comes at the price of a certain degree of cancellation among parameters.

One can envisage several approaches to shift the SOL prediction in \eq{SOLpreved} to values slightly below the critical point, from inflaton couplings, curvature couplings, or even trapping in a periodic potential after inflation.  Perhaps, the conceptually simplest solution is to appeal to a certain degree of numerical coincidence (which does not imply any technical fine-tuning) between the Hubble rate during inflation $H_0$ and the instability scale $\Lambda_I$.  In this case, as apparent from eqs.~(\ref{SOLpreved}) and (\ref{SOLprevedinc}), the width of the distribution can cover the breadth of the weak scale and a little hierarchy is accommodated.  Note that this does mean that we can make a sharp prediction of the Higgs vacuum expectation value, but only that the result can be compatible with observation, accepting a parameter coincidence at the level of $\sim v^2/\Lambda_I^2$. 

To summarise, the natural SOL prediction for the weak scale is $v\sim \Lambda_I$, where $\Lambda_I$ is the instability scale in a theory where the SM is augmented with new matter. Since $\Lambda_I$ is generated by dimensional transmutation, the weak scale may be exponentially smaller than the cutoff of the theory, thus offering the setup for a natural explanation of the lightness of the Higgs.  Our existence requires that $\Lambda_I$ is separated from the weak scale by a small hierarchy, which is the SOL incarnation of the little hierarchy problem.  This discrepancy may be accommodated by a coincidence between (finite) dS-induced Higgs fluctuations and the instability scale.

\section{A Small Cosmological Constant from SOL}
\label{sec:cc}
The smallness of the cosmological constant is the greatest fine-tuning puzzle in nature (see e.g.~\cite{Zeldovich:1967gd,Zeldovich:1968ehl,Weinberg:1988cp,Nobbenhuis:2004wn,Polchinski:2006gy,Bousso:2007gp,Martin:2012bt,Burgess:2013ara,Padilla:2015aaa}) and attempts to explain it have rarely been successful.  Arguably the most successful explanation is Weinberg's anthropic approach~\cite{Weinberg:1987dv} which requires a vast landscape of dS and AdS vacua in order to render an explanation of the smallness of the cosmological constant in our own multiverse allotment.  Recently, alternative cosmological approaches have been investigated \cite{Arvanitaki:2016xds,Graham:2019bfu,Bloch:2019bvc}.  In this section of the paper we present an alternative approach, which exploits a landscape of cosmological constant values, but is based on a purely dynamical selection process governed by SOL and does not appeal to any anthropic criteria.

\subsection{Theoretical Setup}

We assume that the fundamental theory, in which the Standard Model is embedded as a low-energy limit, is supersymmetric with some of its parameters being scanned by the apeiron field. The theory
possesses at least two different vacua, called $v$ and $h$, each supporting a different quantum phase. Vacuum $h$ preserves supersymmetry and an R-symmetry. In this phase of the theory, the cosmological constant is zero because all auxiliary fields vanish (by supersymmetry) and so does the superpotential (by R-symmetry). Obviously such a configuration does not resemble our universe, and we will refer to it as the `hidden' phase of the theory. We live instead on vacuum $v$, the `visible' phase where both supersymmetry and R-symmetry are broken. 

On vacuum $h$, the apeiron enjoys a shift symmetry that controls its potential, making it at most exponentially shallow. This is not the case on vacuum $v$, where the apeiron energy density changes as the field scans the parameters of the theory. The different apeiron behaviour in the two phases can be justified by the fact that a pseudo-Goldstone potential is very sensitive to the vacuum structure of the QFT from which it emerges.  To illustrate this, consider a QFT not dissimilar from the SM in which the Higgs potential has, due to RG effects, two different vacua.  In the IR vacuum the Higgs vev could be small, or even vanishing, yet in the UV vacuum it is large.  The pion mass depends on the Higgs vev, thus if the Higgs vev is parametrically different in the two vacua so, too, is the pion mass.  This simple example serves as some justification of our underlying assumption, but it does not imply that only a Higgs-like sector would have this required vacuum structure.  Moreover, we assume that the non-perturbative corrections to the apeiron potential on vacuum $h$, as arising from worldsheet, brane, gravitational or additional gauge instantons are small enough, potentially leveraging the additional protection from supersymmetry, as to be a negligible correction (see \cite{Svrcek:2006yi} for a discussion of the typical instanton action for these contributions, in support of our assumption). 

Next we add a second sector of the theory, which is responsible for driving an inflationary process. We assume that this inflationary sector is sequestered, in the sense that it is not coupled directly to other fields in the K\"{a}hler potential or superpotential  and therefore its interactions with the rest of the theory are purely (super)gravitational. As a result, the inflationary sector preserves the apeiron shift symmetry. In the ground state, which is currently occupied in our Universe, the inflationary sector is both supersymmetric and R-symmetric and therefore gives a vanishing contribution to the vacuum energy today. However, during inflation, it is in an excited state that breaks supersymmetry and possibly R-symmetry, uplifting the theory to dS by an energy density $V_I$. 

\begin{figure}[t]
\begin{center}
\includegraphics[width=0.6\columnwidth]{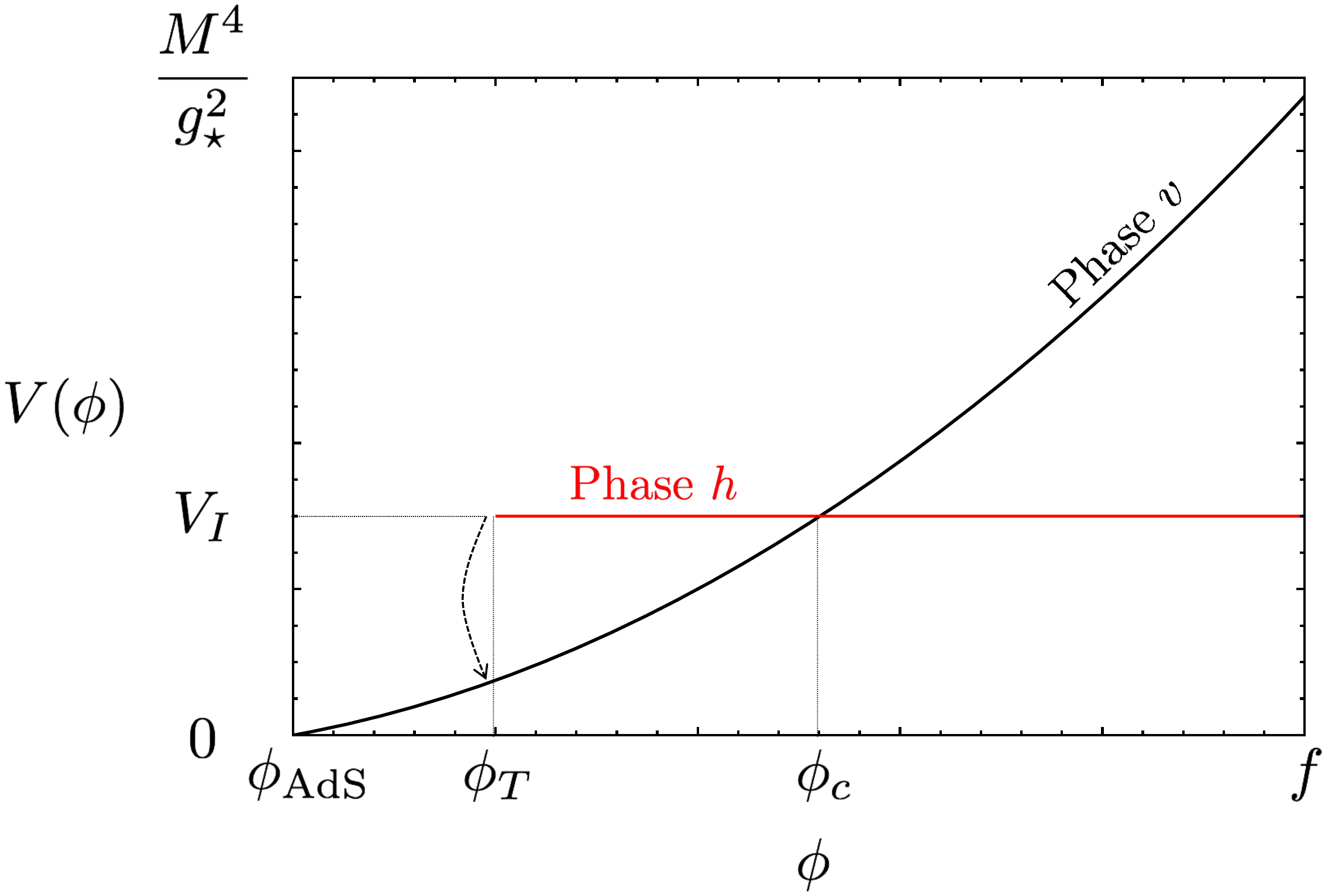} 
\end{center}
\caption{A sketch of the scalar potential $V(\phi )$ with the two branches corresponding to phase $h$ (with the microscopic theory residing in the `hidden' vacuum) and phase $v$ (the `visible' vacuum occupied by our Universe). The two branches are degenerate in energy at $\phi=\phi_c$ and transitions are only allowed at $\phi =\phi_T$, where vacuum $h$ decays into $v$.}
\label{fig:ccs}
\end{figure}

The apeiron potential in phases $v$ and $h$, as based on our assumptions, is sketched in \fig{fig:ccs}. With an appropriate choice of coordinates, we define $\phi_c =0$ to be the field point where the two phases are degenerate in energy. 

Phases $v$ and $h$ do not live in isolation, but phase transitions between them are possible. Transitions from vacuum $h$ to vacuum $v$ can occur for any $\phi <0$, although the tunnelling rate is likely to be slow since the height of the potential barrier separating the two vacua is characterised by some high-energy scale of the fundamental theory. Moreover, for $\phi < \phi_{\rm AdS}$, the tunnelling from vacuum $h$ to $v$ corresponds to a dS$\to$AdS transition. Therefore, any bubble created in this transition will rapidly crunch and any random-walk trajectory that ventures into the region $\phi < \phi_{\rm AdS}$ will not contribute to the final census. Note that it does not matter if the tunnelling is slow, since we are interested in the asymptotic behaviour of the system after extraordinary long times: any field configuration that belongs to phase $h$ and random walks in the region $\phi < \phi_{\rm AdS}$ will eventually tunnel to vacuum $v$ and collapse. Analogously, any field configuration in phase $v$ that explores the region $\phi < \phi_{\rm AdS}$ will rapidly crunch and cannot contribute to the volume-weighted field distribution in steady state. This is conveniently captured by an absorbing boundary condition $\pv_v(\phi_{\rm AdS})=0$. 
The situation in phase $h$ is different because of tunnelling $h\to v$. For modelling in a simple way the phase transition, we assume that tunnelling becomes relevant whenever $\phi <\phi_T$, where $\phi_T $ is somewhere in the range $\phi_{\rm AdS}<\phi_T< 0$ and we impose the condition $\pv_h(\phi_T)=0$.

Transitions $v\to h$ are also possible in principle, but
we assume that across the entire field range the lifetime to tunnel to vacuum $h$ is so slow as to be irrelevant, since the field in phase $v$ is more likely to slide into AdS during its exploration of the potential than to tunnel to vacuum $h$.  Note that this cannot happen in phase $h$, which has no AdS region, justifying our modelling assumption of having $h\to v$ transitions but not $v \to h$.

Finally we must define the conditions at the upper end of the field range, $\phi =f$, where the theory meets its UV cutoff. In phase $h$ we impose an absorbing boundary condition $\pv_h (f )=0$. For phase $v$, we impose a sourcing boundary condition $\pv_v^\prime (f)=-k_v/M_P$, where $k_v$ is a positive constant that parametrises a field flux injected from the UV theory. This could be the result of tunnelling from an additional UV vacuum. The role of this boundary condition is to preserve the trajectory of the distribution close to the diffusionless solution $\pvc$ (see \sec{sec:determ}) as the theory approaches the UV.  See \sec{sec:water} for a brief discussion about the absence of fine-tuning for this class of boundary conditions.

\subsection{SOL Dynamics}

Having defined the apeiron potential in the two phases and the corresponding boundary conditions, we can now solve the FPV. The analogy with the waterfall potential studied in \sec{sec:water} is manifest and indeed the solution of the FPV can be derived following the same procedure. We define the spectral decomposition as in \eq{declah}. 

In phase $h$ we find that the unique mode that satisfies positivity and the boundary conditions is, up to an overall normalisation,
\beq
\tpv_h(\phi ) =  \sin \bigg[ \sqrt{\frac{6(1-\lah )}{\hbar}} \, \frac{2\pi (\phi -\phi_T)}{H_I}  \bigg] ~,~~~~\lah = 1-\frac{\hbar\, H_I^2}{24(f-\phi_T)^2}~.
\label{lambb}
\eeq

In phase $v$ we consider a monotonically increasing potential which is degenerate in energy with the theory in phase $h$ at the field value $\phi =0$ and which scans the full range of cosmological constant values. Since naturalness requires the cosmological constant to be as large as $M^4$, where $M$ is the energy cutoff, we impose
\beq
V_v (\phi_{\rm AdS})  =0~,~~~V_v (0)  =V_I ~,~~~V_v (f ) = \frac{M^4}{g_*^2} ~.
\label{condbraa}
\eeq
We require that the potential satisfies slow roll in the full field range and positivity at least for $\phi \gsim 0$. This implies
\beq
\frac{V_v^{3/2}}{M_P^3}  <V_v^\prime <\frac{V_v}{M_P} ~~~~({\rm for}~\phi \gsim 0)~.
\label{slowro}
\eeq
If boundary conditions are such that the diffusionless solution $\pvc$ dominates over $\pvg$ in a neighbourhood of $\phi =0$, then the distribution will present a peak such that
\beq
V_v ({\bar \phi}) = V_I \, \lah^{2/\xi} ~,~~~
\sigma = \sqrt{\frac{2}{3\xi}} \, M_P~,
\label{autotutto}
\eeq
up to subleading corrections in the slow-roll parameters. As expected from \eq{relazzz}, in the classical slow-roll regime, the potential height at the location of the peak is determined by the eigenvalue.

Once the systems in the two phases are kept in `thermal equilibrium'  by the phase transition occurring at $\phi_T$, their steady-state distributions must have the same global expansion rate, which is measured by the eigenvalue $\lah$. Replacing in \eq{autotutto} the maximum eigenvalue determined in phase $h$, see \eq{lambb}, we find that the location of the peak is such that
\beq
V_v ({\bar \phi}) = V_I \Big( 1- \frac{\hbar \, H_I^2}{12\xi f^2}\Big)~,~~~
\frac{\bar \phi}{H_I} =-\frac{\hbar}{4\xi}  \frac{M_P^2}{f^2}  \frac{H_I^3}{V_v^\prime (0)}~,~~~
\sigma = \sqrt{\frac{2}{3\xi}} \, M_P~,
\eeq
taking $f\gg \phi_T,H_I$.
The first equation shows that, for $f\gg \xi^{-1/2}H_I$, the peak in phase $v$ occurs at a field point such that the potential energy matches the vacuum energy in phase $h$. The second and third equations show that, in the classical regime ($V_v^\prime (0)>H_I^3$) and for $f>\xi^{-1/4}(H_IM_P)^{1/2}$, the displacement of the peak from the critical point at $\phi =0$ is smaller than the width. Both conditions are satisfied in our theory. This result is remarkable, as $\phi =0$ is not a special point when phase $v$ is considered in isolation and there is no communication between the two phases around the critical point. 

\subsection{SOL Prediction}

The SOL mechanism ensures that, during inflation, the vacuum occupied by our Universe is almost exactly degenerate in energy with the supersymmetric and R-symmetric vacuum which has vanishing cosmological constant. Once inflation is over, the energy of both vacua will be reduced by the same amount $V_I$ and therefore the degeneracy will be preserved. After inflation, our Universe will undergo a reheating process, with various subsequent thermal phase transitions (e.g.~Higgs or quark-hadron transitions) which will alter its vacuum energy. However, once it reaches a temperature equal to about $H_I$, our Universe will return to a state with a vacuum energy almost  identical to the one it occupied during inflation, aside from the uplifting contribution from $V_I$. Since this contribution is identical in the two phases, we conclude that the cosmological constant in our Universe today remembers the SOL condition of being degenerate with the supersymmetric and R-symmetric vacuum, as long as $H_I$ is smaller than $10^{-3}$~eV. This condition is necessary for any residual difference between the vacuum energies of the dS state and the present state of the Universe to be smaller than the observed value of the cosmological constant. As we will see in the following, this condition is satisfied in our theory.

The prediction for the cosmological constant in phase $v$ today is obtained by subtracting the inflationary contribution $V_I$ from the vacuum energy $V_v(\phi)$ for a field value in the range $|\phi -{\bar \phi} |< n_\sigma \sigma$, where $n_\sigma$ is the number of standard deviations that we are willing to tolerate in the statistical distribution. Since, as shown before, $\sigma > |{\bar \phi}|$, the SOL prediction is dominated by the statistical uncertainty and is given by
\beq
|\Lambda_{\rm CC}^4| < \frac{V_v^\prime (0) M_P}{c_\xi } ~,~~~c_\xi =\sqrt{\frac{3\xi}{2n_\sigma^2}}~.
\label{predcc}
\eeq
The parameter $c_\xi$ is a measure of the intrinsic statistical nature of the SOL prediction and combines the probabilistic uncertainty ($n_\sigma$) with the uncertainty associated with the way we define the statistical ensemble over which we average the patches in the multiverse ($\xi$).
In the following, we will assume that the bound in \eq{predcc} is saturated and we will use it as an equality.

Combining \eq{predcc} with the CbQ condition in \eq{slowro}, we find upper bounds on the Hubble rate during inflation and on the reheating temperature after inflation
\beq
H_I < c_\xi^{1/3}(\Lambda_{\rm CC}^4/ M_P)^{1/3} \approx c_\xi^{1/3} \, 2\times 10^{-13}~{\rm eV}~,~~~
T_{\rm RH} < c_\xi^{1/6} (\Lambda_{\rm CC}^2 M_P)^{1/3} \approx c_\xi^{1/6}\, 25~{\rm MeV}~,
\label{duebou}
\eeq
where we have taken $g_R = 10.75$ as the number of degrees of freedom in the thermal bath.
The first bound in (\ref{duebou}) shows that, as previously advertised, the Hubble rate amply satisfies the requirement $H_I < \Lambda_{\rm CC}^{1/4}$. It also shows the remarkable fact that the field exchange between phase $v$ and $h$ occurs at a point $\phi_T$ in which the vacuum energy density is about 40 orders of magnitude different from the one corresponding to the actual cosmological constant of our Universe. For an observer who lives in phase $v$, like us, the field value where the cosmological constant vanish is truly unexceptional.

The second bound in (\ref{duebou}) shows that the reheating temperature is relatively low, but sufficiently high to allow for ordinary nucleosynthesis to proceed successfully. Indeed, nucleosynthesis only requires that $T_{\rm RH}$ be larger than about 4~MeV~\cite{Kawasaki:1999na,Kawasaki:2000en,Hannestad:2004px,Ichikawa:2006vm,deSalas:2015glj}. This leads only to the very weak constraint $c_\xi > 10^{-5}$.

The timescale for the system to reach the steady-state configuration is determined by the evolution in phase $h$, where the potential is in a quantum regime. The timescale can be estimated by recalling that, due to the field random walk, the spread of the distribution evolves as $\sigma^2 =H^3 t/(2\pi)^2$. Since the asymptotic distribution in \eq{lambb} has a spread $\langle \phi^2\rangle \approx f^2/\pi^2$, the number of $e$-folds necessary to reach the steady state is about $N\approx f^2/H_I^2$. Since $f$ is super-Planckian, $N$ exceeds the bound in \eq{entb} and inflation must be eternal. The same result can be obtained by estimating the characteristic evolution timescale as $\Delta t \approx 1/(3H_I \Delta \lah)$, where $\Delta \lah$ is the typical splitting between eigenvalues. In phase $h$, we find that two contiguous eigenvalues differ by ${\lah}_n -{\lah}_{n+1} =(2n+1)\hbar H_I^2/(24 f^2)$, leading to the same estimate on the number of required $e$-folds, $N\approx f^2/H_I^2$.

After inflation is over in our patch, the random-walk process terminates and the apeiron will simply roll down its potential by an amount
\beq
\delta \phi = - \frac{V_v^\prime (0 )}{3 H_{\rm now}^2} = -c_\xi M_P~.
\eeq
As long as $c_\xi <1/\sqrt{n_\sigma}$, the shift is within the statistical uncertainty ($|\delta \phi | <\sigma$) and therefore does not affect the SOL prediction. 

Although post-inflationary dynamics does not pose any significant constraint on the theory, it offers the possibility of an observational test of the mechanism since it modifies the dark energy EoS, see \eq{wdark},
\beq
w = -1+\left( \frac{V_v^\prime (0 )}{3 H_{\rm now} \Lambda_{\rm CC}^2}\right)^2 = -1 +\frac{c_\xi^2}{3} ~.
\eeq
Modifications from $w\!=\!-1$ (as predicted by an unchanging cosmological constant) are expected from SOL to be positive and possibly sizeable. However, the prediction is blurred by the statistical nature of SOL, as encoded in the parameter $c_\xi$, and the effect could  be unobservably small without conflict with nucleosynthesis. Nevertheless, the prediction is robust in the sense that it does not depend on model-dependent features such as the functional form of the apeiron potential. Present fits on cosmological parameters give~\cite{Aghanim:2018eyx} $w=-1.028 \pm 0.031$ and therefore $w+ 1<0.02$ at 95\% CL, although these results depend on prior modelling assumptions. Measurements of dark energy parameters are expected to become much more precise with future weak lensing data, galaxy surveys, CMB and large-scale structure observations. 

The results presented so far do not depend on the explicit form of $V_v(\phi)$ but only on its local properties around the critical point $\phi =0$. However, global information about the potential is necessary to determine the maximum cutoff energy $M$ up to which the classical slow-rolling conditions in \eq{slowro} are satisfied.
 A simple linear potential can reproduce the observed value of the cosmological constant only assuming an energy cutoff around some tens of MeV at most. This is already a remarkable result since SOL could justify a hierarchy $M^4/ \Lambda_{\rm CC}^4$ of 40 orders of magnitude, although it falls short of explaining the remaining 80 orders of magnitudes that are needed to have $M$ near the Planck scale. However, it is not difficult to think of potentials beyond linear that can extend the validity of classical slow roll up to very large values of $M$.
 
A simple example is a potential in phase $v$ which is exponential in $\phi$. When parameters are chosen to satisfy \eq{condbraa} and to reproduce the observed value of the cosmological constant as in \eq{predcc}, the exponential potential takes the form
\beq
V_v (\phi) = \frac{c_\xi \Lambda_{\rm CC}^4}{\kappa} \left( e^{\frac{\kappa \phi}{M_P}} -1\right) +V_I
~,~~~ \frac{f}{M_P} = \ln \left( \frac{\kappa M^4}{g_*^2 c_\xi \Lambda_{\rm CC}^4}+1\right)^{1/\kappa} ~,
\eeq
where $\kappa$ is a free constant.
With this potential, the apeiron scans a range of vacuum energies encompassing any positive value of the cosmological constant in the present Universe, up to the cutoff $M$.
The conditions for classical slow-roll in \eq{slowro} are satisfied for positive $\phi$, if we require
\beq
\frac{M^2}{g_* M_P^2} < \kappa <1 ~.
\eeq
For $M\approx M_P$ the scale $f$ is moderately super-Planckian, with $f/M_P\approx 280 $. If we reduce the cutoff, the range of $f$ expands: for instance, for $M=10^{16}$~GeV, we find $10^2\lsim f/M_P\lsim 10^7$.

The SOL mechanism, leading to field localisation at a point in phase $v$ nearly degenerate with phase $h$, works for the exponential potential in full analogy with the case of the linear potential that was studied in \sec{sec:water}. A difference is that the exponential potential is in the classical regime above the peak at $\phi =0$, but not below. However, this does not change the nature of the solution since the existence of the peak only requires classical slow roll for $\phi >0$.

In principle, with knowledge of the potential in the full field range, we could calculate the relative probability for a Universe to live in phase $v$ or $h$. However, the result depends on arbitrary boundary conditions at the upper endpoint and therefore a definite prediction requires knowledge of the UV completion. 

\section{Summary and Conclusions}
\label{sec:concl}
Seemingly we understand perturbations about our vacuum state, at least at currently accessible wavelengths, very well.  For example, we can predict Higgs boson production rates at hadron colliders with N$^3$LO accuracy~\cite{Anastasiou:2016cez} or compute the muon anomalous magnetic moment with precision at the level of a few $10^{-7}$~\cite{Aoyama:2020ynm,Borsanyi:2020mff}.  Yet, the structure of the vacuum itself still hides many mysteries.  We have, however, had some hints about the answers that may lurk within the vacuum.  For instance, the structure of the physical vacuum may be intimately related to the structure of physics in the deep UV, whatever the true nature of quantum gravity may be \cite{Vafa:2005ui}.  At more modest wavelengths, the measurement of IR parameters, such as the Higgs and top quark masses, have revealed that our current working hypothesis for describing perturbations about the vacuum (i.e.~the Standard Model) itself has a nontrivial vacuum structure with a deeper energy-density vacuum lying far from our current state.  Clues such as these have led to a modern-day incarnation akin to the days of the 
\ae ther:  when asking UV-motivated questions we increasingly seek answers from the vacuum. In this context, we have proposed a new concept that links the properties of the vacuum to physical observables: self-organised localisation.
SOL is a phenomenon that can drive physical parameters, measured at low energy, towards values determined by the global vacuum structure of the theory, even when this structure is inaccessible to direct low-energy observations.

The first basic ingredient of SOL is the hypothesis that one, or more, parameters of the microscopic theory (which can be identified with the SM or any of its field-theoretical extensions) are promoted to dynamical variables dependent on one, or more, scalar fields $\phi$. Generically referring to these parameters as $\mu$, the hypothesis is $\mu \to \mu(\phi)$, where the function $\mu(\phi)$ spans the full natural range of the parameter as $\phi$ varies in its domain. During inflation, the light scalar fields $\phi$ (where by `light' we mean relative to the expansion rate) are subject to large stochastic fluctuations and effectively scan the physical parameters $\mu$,
exploring the global vacuum structure of the theory, even reaching locales far from any local minima.  

The second ingredient is the hypothesis that the microscopic theory has a non-trivial vacuum structure as a function of $\mu$ with two, or more, quantum phases possibly coexisting for the same value of $\mu$ and terminating at some critical value $\mu_c$. Note that we refer to coexistence of phases as to the situation in which the theory has different local minima for the same $\mu$, but we do not necessarily require energy degeneracy among minima. Therefore, some of the vacua will be metastable, but this does not preclude them from being populated during inflation, even for timescales much longer than the tunnelling time, because the volume expansion prevents nucleated bubbles of true vacuum to percolate and take over the entire space. 

With these two ingredients and some appropriate hypotheses on the $\phi$ potential, we have observed the emergence of a rather universal phenomenon during inflation: points of criticality, where two phases are on the verge of ceasing to coexist, can act as global attractors for the scalar fields that govern the phase transition, with the volume-weighted field distribution typically becoming highly localised at these points.  We have referred to this phenomenon as SOL, since the scalar field localises itself around a value $\bar \phi$, intimately related to the critical point. As a result, the physical parameter $\mu$ is dynamically selected to take the value $\mu(\bar \phi)$ across the entire Universe, even if that value does not correspond to any enhanced symmetry and can appear to violate ordinary EFT reasoning. Due to SOL's probabilistic nature, the prediction $\mu =\mu(\bar \phi)$ is affected by an intrinsic uncertainty related to the width of the $\phi$ distribution around the localisation point, although in practice this does not significantly limit the predictive power.

The SOL condition for criticality can be realised in several different ways. The most straightforward example is the one in which varying $\mu(\phi)$ across a critical value $\mu_c$ induces a first-order phase transition in the microscopic theory. The variation of the background field $\phi$ is analogous to changing the magnetic field around a ferromagnet. As $\phi$ is varied, it may cross a point at which the two vacua of the microscopic theory are degenerate.  In the magnet analogy, this would be at $|B|=0$.  Continuing further,  the vacuum degeneracy will be broken such that one now resides in a metastable higher-energy vacuum until it reaches the point where decay to the lower vacuum state is inevitable.  This would be the critical magnetic field strength for a ferromagnet, beyond which all spins are guaranteed to flip to the lower-energy state. If the region of coexistence of the two vacua is small compared to the total field range and if the phase transition is sufficiently strong, the critical point  corresponds to a local maximum of the vacuum energy density and
then, in Landau's terminology, one will observe a first-order discontinuity of the gradient of the potential. However, resolving this cusp through the underlying microscopic physics exposes overlapping branches of metastable vacua, and the highest energy-density point will correspond to the field value at which the excited vacuum terminates. Ultimately, this field value determines the localisation point $\bar \phi$ and therefore links the prediction $\mu =\mu(\bar \phi)$ to the critical behaviour of the microscopic theory.

The case we have just illustrated is not, however, the only way for SOL to make use of criticality. We have studied other cases in which criticality could be a feature of a phase different from the one occupied by our Universe, thus remaining invisible to parameter variations in the local minimum, or it could emerge as a condition for degeneracy of the vacuum energy of different phases. Future studies may reveal other contexts in which SOL could relate field localisation to critical phenomena.

A crucial aspect for any of the quantitative results presented in this paper is the existence of steady-state 
solutions emerging from the stochastic equation that governs the time evolution of the volume-weighted field distribution. Much as thermodynamic systems reach thermodynamic equilibrium in a laboratory, these steady-state distributions give a fairly robust description of the properties of the system, independently of any initial condition of the Universe.

The question of how these steady-state distributions are reached dynamically is important to determine the timescale for the system to attain equilibrium. We have identified three relevant regimes of the $\phi$ potential, which are qualitatively distinct in their properties. When the potential is sufficiently steep, and the system is in what we call the C (Classical) regime, the field
distribution preferentially evolves towards configurations with lower energy density, although the ultimate behaviour depends on the form of the boundary conditions. In a second regime, termed QV (Quantum+Volume), quantum effects combined with volume expansion effects conspire to give large fluctuations to classical evolution and the volume-weighted distribution of the scalar field preferentially evolves towards configurations with high energy density. In the case of the multi-valued potentials characteristic of critical phenomena, this leads to a distribution with a peak of a specified width and position near the top of the metastable branch just before it terminates.  In a third parameter regime, termed Q$^2$V, the second-order quantum effects combined with volume effects give an even stronger preference for high values of the potential energy, leading to a final volume-weighted distribution sharply peaked at the critical point that marks the end of coexistence between the two phases.

The QV and Q$^2$V regimes show a natural attraction towards the critical points, whenever the phase transition is sufficiently strong to create a local maximum of the $\phi$ potential. However, in QV and Q$^2$V regimes, the characteristic timescale for reaching the steady state is so long that the dynamics needs to be supported by eternal inflation. Although it is possible to devise examples of SOL with shorter relaxation timescales, eternal inflation was an ingredient of all applications considered in this paper.  By choosing parameters in the C regime, there could be
physical applications where SOL occurs during non-eternal inflation, although additional field-theory ingredients such as post-inflationary mechanisms would be necessary to prevent the field from rolling too fast since reheating.  On the contrary, eternal inflation appears relatively generic in inflationary models~\cite{Guth:2007ng} and, despite the onset of the measure problem and the associated loss of calculability, seems well motivated.  Thus, to a large extent, eternal inflation appears to be a natural setting for SOL.

\bigskip

Our exploration shows that SOL is a rich and versatile phenomenon, which can take different forms and adapt to a variety of applications in cosmology. Indeed, SOL may have profound physical implications for our understanding of the vacuum.  For example, if the background of a light scalar field, here called {\it apeiron}, controls the values of Standard Model parameters, then we could expect those parameters to take values close to critical points at which a metastable vacuum is about to collapse. SOL implies a prevalence, in the multiverse, of vacua on the cusp between stability and metastability with respect to parameter variations.  
In this paper we have explored the relevance of SOL to three key questions surrounding the Standard Model vacuum.

\subsubsection*{1.  Near-criticality of the Higgs self-coupling}  
SOL is ideally suited to explain the observation that the measured values of the Higgs and top masses place the SM in a peculiarly balanced metastable state. As the apeiron scans the SM parameters, the theory develops two Higgs vacua. One is the ordinary low-field vacuum, which persists as long the Higgs self-coupling, renormalised at low energies, is positive. The other is a high-field vacuum, which starts to exist as soon as the Higgs self-coupling, renormalised at high energies, becomes negative. In an intermediate region of parameters, the two vacua coexist and SOL predicts that this is the most likely outcome after a long inflationary period, in perfect agreement with observations. The mechanism is detailed in \sec{sec:nearcrit}.

\subsubsection*{2.  Higgs naturalness}  
Since the electroweak quantum phase transition occurs smoothly as the Higgs mass is varied, the problem of Higgs naturalness seems less suited for SOL. However, the SM vacuum structure offers an intriguing 
possibility for exploiting vacuum coexistence. As the Higgs mass-squared parameter is scanned by the apeiron and its value becomes more negative, the vacuum moves to ever greater field values.  However, once it crosses a critical point, determined from the renormalisation group evolution to be the scale at which the effective quartic coupling passes through zero, the Higgs vacuum becomes metastable with respect to a deeper high-field vacuum and eventually vanishes entirely. SOL predicts that the apeiron localises during inflation in correspondence with the critical value for the Higgs mass. The mechanism dynamically creates an exponential hierarchy between the UV cutoff energy and the Higgs mass. This is due to dimensional transmutation which generates a natural separation between the cutoff and the instability scale where the quartic coupling vanishes. 

Although, on the back of dimensional transmutation, SOL creates a natural mass hierarchy, the result cannot be directly applied to the SM. New matter must be added to the SM in order to bring the instability scale closer to the electroweak scale. Moreover, the mechanism suffers from a little hierarchy problem which can be accommodated only at the price of a coincidence between the Hubble rate and the instability scale, or by adding further dynamical ingredients to the theory. This mechanism is discussed in \sec{sec:natur}.

\subsubsection*{3.  The cosmological constant}  
The vanishing of the cosmological constant does not appear to correspond to a phase transition and cannot be immediately linked to a question of vacuum coexistence.  However, if the Standard Model is embedded within an overarching framework, the situation may be propitious for SOL.  Consider a supergravity theory containing multiple moduli fields, of which the present vacuum, with the Standard Model low-energy field content, is just one of multiple possibilities.  If this theory also contains a supersymmetric and R-symmetric vacuum, then that vacuum will have vanishing energy density by dint of the symmetries. In \sec{sec:cc} we explored how SOL could transfer information from the hidden supersymmetric phase into our Universe.  

Due to the interplay between the two phases, the volume-weighted distribution is peaked at precisely the degeneracy point between the vacua, despite that this is far from where the tunnelling occurs.   Physically, the steady states in the two phases must inflate at the same rate, hence the system is dynamically attracted towards the point where the vacuum energies are degenerate.  The result is that the inflating multiverse tunes physical parameters towards vacuum degeneracy between two different phases which, locally, are entirely unaware of one another. 
 Thanks to the interplay with supersymmetry and R-symmetry, SOL has self-tuned the cosmological constant to be hierarchically smaller than what one expects generically in our present vacuum, where supersymmetry and R-symmetry are badly broken.

\bigskip

All these examples show that SOL is a sufficiently generic phenomenon that one can imagine to find other physical applications, beyond those considered here. Critical points are common in particle-physics theories and SOL can provide a dynamical explanation for a cosmological selection of near-critical behaviour.  Moreover, the example of the cosmological constant studied in  \sec{sec:cc} exhibits a feature qualitatively different from those of the Higgs examples in sects.~\ref{sec:nearcrit} and \ref{sec:natur}.
The key ingredient is the presence of two distinct phases of the theory. SOL singles out a specific value of a fundamental parameter in one phase, although that parameter is not special nor corresponds to any enhanced symmetry within that phase. The secret of the mechanism is that a value of a parameter which looks completely generic in one phase, may be special in the other phase. By relating parameters in the two phases, SOL makes a physical prediction which may appear inexplicable to observers who are confined to live in one vacuum and remain unaware of the existence of another phase of the theory, which is hidden to them.

It is interesting to draw an analogy between the steady-state configurations of the FPV solutions and thermal equilibrium in a gas. Microscopic properties of atoms in a gas (such as individual energies) change incessantly but, in thermal equilibrium, some macroscopic thermodynamical properties (such as temperature) remain constant and characterise the behaviour of the global system. Similarly, the Langevin trajectories describing the field value in a single patch of the Universe keep on evolving incessantly but, when steady-state is reached, the FPV solution identifies the field distribution that characterise the global behaviour in the multiverse. For atoms in a gas, it would be impossible to derive their microscopic properties from fundamental principles  by studying a single atom in isolation: any individual energy is as good as any other. Only when we consider the atom as part of a statistical ensemble, we can make probabilistic predictions on microscopic quantities based on macroscopic equilibrium properties. Similarly, attempts to derive SM parameters from fundamental principles may be futile: any parameter value is as good as any other. Only when we consider our Universe as part of a statistical ensemble, we may be able to make probabilistic predictions on SM parameters across the multiverse.

The analogy with a gas in thermal equilibrium takes a more concrete form when we think of the background Hubble rate as the Gibbons-Hawking temperature in de Sitter space. Particularly revealing is the SOL application to the cosmological constant, which shows how the physical result is linked to the macroscopic behaviour of the system. A single Langevin trajectory (or, in other words, a single universe) does not see anything special in the critical field value where the two phases are degenerate, just like a single atom in a gas does not see anything special in any particular energy value. And yet, just like, when two boxes of gas are put in thermal contact, atoms arrange themselves statistically to produce a single  temperature of the system, so field trajectories arrange themselves to reach a steady-state configuration in which the expansion rate is roughly the same in both phases. Since the expansion rate is proportional to the energy density, the energy density in the visible phase becomes nearly equal to the one in the hidden supersymmetric phase, even if the visible sector does not know anything about supersymmetry. According to SOL, the secret of the smallness of the cosmological constant in our Universe lies in the macroscopic statistical properties of the multiverse.

In our approach, the multiverse is not merely a setup suitable to anthropic selection. Instead, we propose to interpret the multiverse as a quantum statistical system, in which critical phenomena may play an essential role in the selection process that determines physical parameters.

\subsubsection*{Open questions}

A generic feature of SOL is that its probabilistic predictions refer to the end of inflation and not to the present epoch. Therefore, they can be modified, or even erased, by the subsequent thermal evolution of the Universe. This can be a virtue or a curse. In some cases, it can be beneficial because it allows for a certain flexibility of the SOL predictions. For instance, particle masses can be subjected to `AdS thermal' corrections proportional to the Hubble rate that evaporate at the end of inflation, thus affecting the original SOL prediction. In most cases, however, the requirement that the scalar field has not evolved significantly between reheating and the present day provides a strong constraint on the theory parameters, forcing the scalar field potential to be extraordinarily shallow and the dynamics to enter a regime of eternal inflation (see \sec{sec:post}).

Besides these aspects related to model building, a more important concern is that the viability of SOL still faces enormous challenges in the context of quantum cosmology. Some of the open issues have been commented upon in \sec{sec:time} and we summarise them here.

One issue is linked to our use of the volume-weighted Fokker-Planck equation, whose formulation requires a choice of space-time foliation, and therefore of gauge. The formation of peaks in the vicinity of critical points (which is a central element for SOL) is a gauge-independent result. However, the widths of the distributions (i.e.~the degree of localisation) are not. Their gauge dependence is not lethal for the success of SOL, unless one considers the somehow contrived case of $e$-folding gauge ($\xi =0$), which corresponds to a time slicing in which every patch undergoes identical expansion. 

Another open issue is related to our hypothesis that inflation is eternal. This brings in
 the `measure problem', which is endemic to any probabilistic interpretation  based on eternal inflation. Since eternal inflation is characterised by having an infinite reheating surface, defining probabilities for observables at the end of inflation is a task plagued with infinities. Although various proposals for dealing with this problem have been put forward in the literature (see e.g.~refs.~\cite{Winitzki:2008zz, Freivogel:2011eg} for reviews), the `measure problem' remains a serious concern.  This raises questions about quantitative results concerning, in particular, the predicted width of the field distribution, which  is not only a gauge-dependent quantity, but also measure-dependent.  On the other hand, we see that the qualitative behaviour of SOL, which is the central result of this work, is gauge-independent and seems to follow automatically in a volume-based measure.   Thus we expect that if the measure problem were to be understood better, the qualitative physics of SOL should remain, although quantitative aspects may change.

When in the QV or Q$^2$V regimes, the completion of the dynamical process that leads to SOL requires a number of inflationary $e$-folds much larger than $M_P^2/H_0^2$. This is not in conflict with the bound based on the maximum entropy available in de Sitter space~\cite{ArkaniHamed:2007ky,Creminelli:2008es,Dubovsky:2008rf,Dubovsky:2011uy}, which applies only to non-eternal inflation. However, it raises concerns~\cite{Dvali:2013eja,Dvali:2017eba} about the validity of the semi-classical approach that we have followed here.  As discussed in \sec{sec:time}, it is not clear how problematic this issue is for the generic predictions of SOL.

Finally, SOL requires super-Planckian field displacements and likely violates both the de Sitter Conjecture and the Distance Conjecture \cite{Ooguri:2006in,Klaewer:2016kiy}, placing most of its applications firmly in the swampland. If one feels the need, the theory could possibly be rescued from the swampland by employing mechanisms that make these super-Planckian displacements only a low-energy delusion~\cite{Silverstein:2008sg,Choi:2015fiu,Kaplan:2015fuy,Giudice:2016yja}.

\subsubsection*{Experimental tests}
In spite of these profound conceptual difficulties in the interpretation of predictions for physical observables, we believe that SOL offers a new route towards answering fundamental questions concerning the physical vacuum. A particularly appealing aspect of SOL is that, despite being a phenomenon apparently confined to an abstract reality, as it takes place in a multiverse populated by eternal inflation and whose expanse is largely beyond our causal contact, it has instead a clear smoking gun amenable to experimental scrutiny. 

The smoking gun of SOL is the prediction of coexistence of different vacua, with the Universe we inhabit being in a metastable state at the edge of collapse with respect to variations of some of the fundamental physical parameter. The experimental test does not rely on waiting for the Universe to collapse. Instead, it relies on making precise experimental measurements of fundamental parameters and performing theoretical extrapolations to study the vacuum structure as those parameters are altered. The case of near-criticality of the Higgs self-coupling is a perfect example. The precise experimental determination of the Higgs mass, top mass, and gauge coupling constants allowed a theoretical extrapolation that has revealed a striking peculiarity of the SM vacuum structure. That peculiarity could be interpreted as a hint for SOL. Similarly, the approach followed in \sec{sec:natur} to address Higgs naturalness requires the existence of new particles. If these particles were discovered and their couplings with the Higgs precisely measured, one could make a theoretical extrapolation to higher energies. Finding evidence for near-criticality of the vacuum with respect to variations of the Higgs mass-squared parameter would provide a clue in favour of SOL.

In some cases, it may be practically challenging to infer the existence of a different phase from experimental measurements and theoretical extrapolation. This is particularly true in cases where the visible phase (i.e.~the one occupied by our Universe) does not exhibit any sign of criticality, which is a feature only of a hidden phase, unaccessible to us. A good example is the SOL explanation of the cosmological constant proposed in \sec{sec:cc}, where it is hard to imagine how one could infer from measurements the existence of a hidden supersymmetric phase with nearly the same vacuum energy as our Universe. In these cases, one has to devise alternative observational tests, which depend on the particular SOL implementation. For the theory presented in \sec{sec:cc}, an experimental signal is provided by the time-dependence of the cosmological constant due to the residual motion of the apeiron. This effect leads to a modification of the dark-energy equation of state which could be within reach of the next generation of experiments in observational cosmology.

\bigskip

Decades of studying symmetries in quantum field theories in Minkowski space have yielded enormous progress in fundamental physics, yet stubbornly refuse to offer empirically viable and theoretically appealing answers to many fundamental questions concerning the microscopic world beyond our present reach.  New vistas and fresh opportunities arise when we look beyond this paradigm towards the interplay between cosmological evolution and the structure of the vacuum. In this work we have shown that SOL may select low-energy parameters to be naturally close to critical points as a result of the multi-valued vacuum structure. We are accustomed to believing that points of enhanced symmetry are preferred because they are stable under quantum corrections. On the contrary, SOL's assertion is that inflation is likely to deliver a Universe which is at the edge of collapse with respect to some parameter variation.  

\section*{Acknowledgments}
We thank Tim Cohen, David Curtin, Gia Dvali, Michael Geller, Daniel Green, Yonit Hochberg, Eric Kuflik, Marko Simonovi\'c, and Daniele Teresi for useful conversations and particularly Andrei Linde for detailed comments on the first version of this paper.  T.Y. is supported by a Branco Weiss Society in Science Fellowship and partially by the UK STFC via the grant ST/P000681/1.

\section*{Appendix}
\label{sec:app}
\addcontentsline{toc}{section}{\nameref{sec:app}}

\appendix

\section{The Volume-Weighted Fokker-Planck Equation}

\subsection{Structure of the Equation}

The volume-weighted Fokker-Planck equation (FPV) (for a review and a list of references, see~\cite{Winitzki:2008zz}) describes the time evolution of the
distribution $\pv (\phi,t)$, defined as the probability for a scalar field to take the value $\phi$ within a spatial patch, weighted by the corresponding volume-expansion factor in an inflating universe with Hubble rate $H$. It is the result of the average over a large number of random-walk trajectories of the classical field $\phi$ governed by a Langevin equation with Gaussian random noise. Expressed in terms of the proper time $t$, the FPV is 
\beq
\frac{\partial \pv (\phi,t)}{\partial t} =\frac{\partial}{\partial \phi}\left[ \frac{\hbar H^{a}}{8\pi^2} \frac{\partial}{\partial \phi} \Big( H^{3 -a}\pv (\phi,t)\Big)+\frac{d V}{d \phi} \frac{\pv(\phi,t)}{3H} \right] 
+3H \pv (\phi,t) ~.
\label{FPVp}
\eeq
The first term in the right-hand side is called the {\it diffusion term} in statistical mechanics and {\it quantum term} in cosmology, as it describes the effect of the random quantum fluctuations of the scalar field in de Sitter space. The power of $\hbar$ is required according to the dimensional analysis of ref.~\cite{Giudice:2016yja}, since $H$ has units of mass and $\phi$ units of scale. The second term is named {\it drift term} in statistical mechanics and {\it classical term} in cosmology, as it describes the field's tendency to roll down the classical potential $V=3H^2M_P^2$. 
The exponent $a$ parametrises the ambiguity in translating the Langevin equation into the FPV, with $a=0$ known as Ito ordering and $a=3/2$ as Stratonovich ordering. In the following, we will choose $a=0$, but any choice is perfectly equivalent since the difference amounts to corrections ${\mathcal O} (H^2/\pi^2 M_P^2)$ to the drift term, which are negligible in the regime where quantum-gravity effects can be ignored ($V\ll M_P^4$).
The third term, which will be called {\it volume term}, is what distinguishes FPV from the ordinary Fokker-Planck equation (FP) (for a review, see~\cite{Risken}) and it describes the effect of the expansion of physical space in de Sitter. Because of this term, FPV does not have a conserved current and its solutions do not retain their normalisations when integrated over $\phi$, unlike the ordinary FP. Hence, $\pv (\phi,t)$ does not describe a probability density, but the volume-weighted distribution of $\phi$ configurations at time $t$.

As we are interested in the $\phi$ dynamics on a de Sitter background, we split the scalar potential into a constant background value $V_0$ and a field dependent part defining
\beq
V(\phi ) =V_0 \left[ 1+\frac{2v(\vfup )}{\SdS }\right] ~,~~~\vfup \equiv \frac{\phi}{M_P}  ~,~~~h(\vfup )\equiv \frac{H(\phi )}{H_0}=\sqrt{1+\frac{ 2v(\vfup )}{\SdS}}
 ~,
\eeq
\beq
H_0^2\equiv\frac{V_0}{3M_P^2} ~,~~~
t_S \equiv \frac{\SdS}{ H_0}~,~~~ \SdS \equiv  \frac{8\pi^2 M_P^2}{\hbar \,H_0^2}~,~~~\tau \equiv \frac{t}{t_S}~,
\label{defpara}
\eeq
where $M_P$ is the reduced Planck mass and $\SdS$ is the maximum entropy of de Sitter space. We note that $\SdS^{-1} = \hbar V_0/(24 \pi^2M_P^4)$, showing that this quantity measures the size of quantum-gravity loops and thus our analysis can be valid only in the regime $\SdS \gg1$. The parameter $t_S$ is the time scale associated with de Sitter entropy.
With the above definitions, the FPV becomes 
\beq
\frac{\partial \pv}{\partial \tau} =
\left[ (h^{3} \pv )^\prime+2\SdS h^\prime  \pv \right]^\prime +3\SdS  h \pv ~,
\label{FPElesssenza}
\eeq
where primes denote derivatives with respect to $\vfup$. As already mentioned, it is always possible to pull $h^3$ outside the derivatives in the diffusion term since the difference only gives small corrections to the drift and volume terms suppressed by ${\mathcal O} (H^2/\pi^2 M_P^2)$.

\subsection{Time Reparametrisation}

To study the effect of different time parametrisations, let us consider the family of time coordinates $t_\GAUG$ such that
\beq
\frac{dt_\GAUG}{dt} = \Big( \frac{H}{H_0} \Big)^{1-\GAUG}~,
\label{talph}
\eeq 
where $\GAUG$, taken to vary between 0 and 1, will be called the gauge parameter. After the time-coordinate transformation in \eq{talph} and the redefinition $\pv \to (H/H_0)^{\GAUG -1} \pv$, the FPV in \eq{FPVp} becomes
\beq
\frac{\partial \pv (\phi,t_\GAUG)}{H_0^{1-\GAUG}\partial t_\GAUG} =\frac{\partial}{\partial \phi}\left[ \frac{\hbar H^{a}}{8\pi^2} \frac{\partial}{\partial \phi} \Big( H^{2+\GAUG -a}\pv (\phi,t_\GAUG)\Big)+\frac{d V}{d \phi} \frac{\pv(\phi,t_\GAUG)}{3H^{2-\GAUG}} \right] 
+3H^\GAUG \pv (\phi,t_\GAUG) ~,
\label{FPValpha}
\eeq
where we have used the property that $H$ depends on time only through $\phi$. Expressed in a general $\GAUG$-gauge, \eq{FPElesssenza} becomes
\beq
\frac{ \partial \pv}{ \partial \tau_\GAUG } =
\left[ (h^{\GAUG +2} \pv )^\prime +2\SdS h^{\GAUG -1}h^\prime  \pv \right]^\prime + 3\SdS h^\GAUG \pv ~,
\label{FPEless}
\eeq
where $\tau_\GAUG = t_\GAUG /t_S$.

The definition in \eq{talph} allows us to interpolate continuously between  {\it proper-time gauge} ($\GAUG =1$) and the case $\GAUG =0$, which is often referred to as {\it $e$-folding gauge}. The time coordinate $t_{\GAUG =0}$ measures the number of $e$-foldings accrued by each patch for any given value of $\phi$. In $e$-folding gauge, the coefficient of the volume term becomes field-independent, and the FPV solution is given simply by the solution of the corresponding FP equation times the field-independent factor $\exp (3H_0t_{\GAUG =0})$. This is because $e$-folding gauge corresponds to a time slicing in which every patch undergoes an identical expansion.

For simplicity, in the following we will drop the index $\GAUG$ in the time coordinate and simply call it $t$. It is implicit that the time coordinate refers to the corresponding choice of $\GAUG$.

\subsection{Spectral Representation}
\label{eigensec}

We express the FPV solution as a spectral representation by defining
\beq
\pv (\vfup , t )= \sum_{\LAMB} \, e^{3H_0t+{\LAMB}\tau } \, \tpv (\vfup , \LAMB ) ~,
\label{specrep}
\eeq
where $\LAMB$ and $\tpv$ are the eigenvalues and eigenfunctions determined by the FPV differential equation, which can be expressed as
\beq
\left( {\mathcal L}-\LAMB \right) \, \tpv (\vfup  ) = 0~,~~~
{\mathcal L} = \frac{\partial}{ \partial \vfup } h^{\GAUG +2} U^2 \frac{\partial}{\partial \vfup } U^{-2} +3\SdS (h^\GAUG -1) ~, ~~~ U= e^{\frac{\SdS}{2  h^2}} \, h^{-\frac{\GAUG +2}{2}}~.
\eeq

It is convenient to introduce the differential operator
\beq
{\mathcal L}_H = U^{-1}\, {\mathcal L} \, U ~.
\eeq
Assuming that $\pv$ vanishes at the boundaries of the physical field range at all times, the operator ${\mathcal L}_H$ is self-adjoint as can be easily shown by integration by parts. 
By construction, ${\mathcal L}_H$ has the same eigenvalue spectrum of ${\mathcal L}$ 
\beq
({\mathcal L}_H -\LAMB ) \Phi(\vfup ,\LAMB   ) =0 ~,~~~ \tpv (\vfup  ) =U \, \Phi(\vfup ,\LAMB   ) ~.
\eeq
Since ${\mathcal L}_H$ is self-adjoint, we conclude that the FPV eigenvalues $\LAMB$ must be real.

For eigenfunctions $\Phi$ normalised to one, we obtain
\beq
\LAMB= \int d\vfup  \, \Phi(\vfup ,\LAMB   ) \, {\mathcal L}_H \, \Phi(\vfup ,\LAMB   ) =
\int d\vfup  \, \left[ 3\SdS(h^\GAUG -1) \left| \Phi \right|^2-  \Big| h^{\frac{\GAUG +2}{2}} U \frac{\partial}{\partial \vfup }( U^{-1} \Phi )\Big|^2 \right]  
\label{eigenlimit}
\eeq
where we have assumed vanishing boundary terms. Therefore, we find that the maximum eigenvalue satisfies
\beq
\LAMB_{\rm max} \le 3\SdS \int d\vfup  \, (h^\GAUG -1) \left| \Phi(\vfup ,\LAMB_{\rm max}   ) \right|^2
\label{lamax}
 ~.
\eeq

In $e$-folding gauge ($\GAUG =0$), \eq{lamax} shows that the eigenvalues must be non-positive and $\LAMB_{\rm max} \le 0$. For a general gauge with $0 \le \GAUG \le 1$, we obtain  
\beq
\LAMB_{\rm max} \le 3\GAUG\, v_{\rm max} ~,
 \label{lmax}
\eeq
where $v_{\rm max}$ is the maximum of the dimensionless potential $v$ in the field range under consideration. We have assumed $v_{\rm max}\ge 0$, which is always possible after an appropriate redefinition of $H_0$. 

Due to the linearity of the FPV, each spectral mode evolves in time independently.
The value of $\LAMB_{\rm max}$ is especially important because it corresponds to the fastest inflating mode, which describes the asymptotic behaviour of $\pv$ at large times
\beq
\pv (\vfup , t  )
=e^{3H_0t+\LAMB_{\rm max}\tau } \, \tpv(\vfup , \LAMB_{\rm max}) ~~~~~({\rm for}~t\! \to\! \infty )~.
\label{asympdis}
\eeq

\subsection{Boundary Conditions}
\label{sec:boundary}

Using the results obtained in \sec{eigensec}, we can write the most general FPV solution as
\beq
\pv (\vfup , t )= \sum_{\LAMB} \, e^{3H_0t+{\LAMB}\tau+ \frac{\SdS}{2 h^2}} \, h^{-\frac{\GAUG +2}{2}}\left[ g_1(\LAMB ) \Phi_1(\vfup ,\LAMB ) +
g_2(\LAMB ) \Phi_2(\vfup ,\LAMB ) \right]
~,
\label{specrep2}
\eeq
where $g_{1,2}(\LAMB )$ are two arbitrary functions and $\Phi_{1,2}(\vfup ,\LAMB )$ are two linearly-independent solutions of the differential equation in Sturm-Liouville form 
\beq
 \left( h^{\GAUG +2} \Phi^\prime \right)^\prime +( q-\LAMB )  \Phi =0 ~,
~~~~
q =  \SdS h^{\GAUG -1} h''
-\SdS^2h^{\GAUG -4} {h'}^2 + 3\SdS(h^\GAUG-1) ~,
\eeq 
where we have dropped negligible corrections ${\mathcal O} (H^2/\pi^2 M_P^2)$.
The FPV solution is uniquely determined once $g_{1,2}(\LAMB )$ are computed from given boundary and initial conditions. 

The boundary conditions are two time-dependent equations that determine $\pv (\vfupud , t)$, where $\vfupu$ and $\vfupd$ are the upper and lower endpoints of the field range, respectively. Particularly convenient are the {\it absorbing} or {\it reflecting} boundary conditions
\beq
\pv (\vfupud , t)=0 ~~~~{\rm (absorbing~boundary~conditions)},
\eeq
\beq
\pv^\prime (\vfupud , t)=0 ~~~~{\rm (reflecting~boundary~conditions)}.
\eeq
In either case, these boundary conditions determine {\it (i)} the ratio $g_2/g_1$ and {\it (ii)} the eigenvalue spectrum, which is discrete and bounded from above for bounded potentials. For absorbing boundary conditions, we find
\beq
\frac{g_2(\LAMB )}{g_1(\LAMB )}=- \frac{ \Phi_1(\vfupd ,\LAMB ) }{ \Phi_2(\vfupd ,\LAMB ) } ~,
\eeq
while the eigenvalue spectrum is given by the discrete set of solutions of the equation
\beq
\Phi_1(\vfupu ,\LAMB )\, \Phi_2(\vfupd ,\LAMB ) = 
\Phi_2(\vfupu ,\LAMB )\, \Phi_1(\vfupd ,\LAMB )~.
\eeq
Analogous equations can be written for reflecting boundary conditions.

In particular, absorbing or reflecting boundary conditions determine the maximum eigenvalue $\LAMB_{\rm max}$, which fully specifies the behaviour of the asymptotic solution at large times. 
For any linear combination of absorbing and reflecting boundary conditions, initial conditions affect only an overall constant of the distribution $\pv$ and therefore the asymptotic state is effectively independent of initial conditions. This may no longer be true for boundary conditions which are non-homogeneous in $\pv$ or its derivative.

\subsection{Initial Conditions}
\label{sec:initial}

The initial condition is a field-dependent equation that determines $\pv (\vfup , 0)$ at $t=0$. While boundary conditions determine $g_2/g_1$ and the eigenvalue spectrum, the initial condition fixes the remaining combination of $g_1$ and $g_2$, thus fully specifying a unique FPV solution.

If the potential is continuous and differentiable everywhere in the field range (a hypothesis that does not always hold in the examples we will consider), the Sturm-Liouville theorem ensures that, for absorbing or reflecting boundary conditions (or for any linear combination of the two), there is a single eigenfunction $\Phi (\vfup , \LAMB )$ for each eigenvalue, and together they form an orthonormal set
\beq
\int d\vfup \, \Phi (\vfup ,\LAMB_1 ) \Phi (\vfup ,\LAMB_2 ) = \delta_{\LAMB_1 \LAMB_2 } ~~~\forall \,\LAMB_{1,2}~.
\label{orthon}
\eeq
With the help of this relation, we can determine $g(\LAMB)$ for any initial condition of $\pv$ at $t=0$
\beq
g(\LAMB)=\int d\vfup \,  e^{-\frac{\SdS}{2h^2}}\, \, h^{\frac{\GAUG +2}{2}}\, \Phi(\vfup ,\LAMB  )\, \pv (\vfup ,0) ~.
\eeq

A special case is when the field range is unbounded, with absorbing boundary conditions set at infinite field values and with an initial condition such that the scalar field is infinitely localised at a value $\vfup_0$
\beq
\pv (\vfup, 0)= \delta (\vfup -\vfup_0 ) ~.
\label{deltaf}
\eeq
For this initial condition, the FPV solution is  
\beq
\pv (\vfup , t )= \sum_{\LAMB} \, e^{3H_0t+{\LAMB}\tau+ \frac{\SdS}{2}(\frac{1} {h^2}-\frac{1}{ h_0^2})} \,\left( \frac{h_0}{h}\right)^{\frac{\GAUG +2}{2}} \Phi(\vfup_0 , \LAMB ) \Phi(\vfup , \LAMB)~,
\eeq
where $h_0=h(\vfup_0)$. 

\subsection{Schr\"odinger Form}

An alternative form of the FPV is obtained by making a field redefinition in \eq{FPEless} such that the coefficient of the second derivative with respect to the field becomes constant and the one of the first derivative vanishes
\beq
\pv (\vfup , t)= e^{\frac{\SdS}{2h^2}}\, \, h^{-\frac{3(\GAUG +2)}{4}}\, \Psi (x,t)  ~,~~~~\frac{dx}{d\vfup} = h^{-\frac{\GAUG +2}{2}}~,
\eeq
\beq
\frac{\partial\Psi}{H_0\, \partial t}= \frac{\partial^2\Psi}{\SdS \, \partial x^2}+  {\hat V}(x)\Psi
~,~~~~
{\hat V}(x) = \big(  h^{\GAUG -1}h'' -  \SdS h^{\GAUG -4}{h'}^2+ 3 h^\GAUG \big)_{\vfup =\vfup(x)} ~,
\label{schr}
\eeq
where primes denote derivatives with respect to $\vfup$ and corrections ${\mathcal O} (H^2/\pi^2M_P^2)$ have been neglected. Equation (\ref{schr}) exhibits an obvious analogy with the Schr\"odinger equation in Euclidean time. Its general solution can be written as a spectral expansion in stationary modes, following the same procedure described in~\sec{eigensec}.

\subsection{Perturbative Expansion}

In many physical applications, we are interested in the behaviour of the FPV solutions in a neighbourhood of a generic field point which, with an appropriate coordinate shift, can be chosen to be $\vfup =0$ with $V(0)=V_0$. The perturbative domain is defined as the neighbourhood around $\vfup=0$ in which the variation of the energy density due to its field dependence is only a small perturbation of $V_0$ (i.e.~$|V(\vfup )-V_0|\ll V_0$). Therefore, the conditions for perturbative expansion, together with slow roll, are
\beq
v(\vfup),v'(\vfup),v''(\vfup) \ll \SdS ~.
\eeq
In the perturbative domain, we can expand \eq{FPEless} in powers of $v/\SdS$ and, at leading order, we obtain
\beq
t_S \Big( \frac{\partial }{\partial t}-3H_0\Big) \pv = \pv'' +(2v' \pv )' + 3\GAUG v \pv ~,
\label{FPVancora}
\eeq
where $v(0)=0$. Here we have dropped the ${\mathcal O} (v/\SdS )$ correction to the diffusion term, which is a subleading effect in our perturbative expansion, but we have kept the ${\mathcal O} (v)$ contribution to the volume term, which is a truly leading effect since the field-independent constant can be simply reabsorbed in the definition of $\pv$.

It is instructive to rewrite for a moment \eq{FPVancora} in terms of the original dimensionful variables 
\beq
\Big( \frac{\partial }{H_0 \partial t}-3\Big) \pv =( R_Q + R_C + R_V )\, \pv   ~,
\eeq
\beq
R_Q= \frac{\hbar\, H_0^2}{8\pi^2} \frac{\partial^2}{\partial \phi^2} ~,~~~
R_C= \frac{\partial}{\partial \phi} \frac{V'}{3H_0^2} ~,~~~
R_V = \frac{ \GAUG  (V-V_0)}{2 H_0^2M_P^2}
 ~,
\eeq
where $R_{Q,C,V}$ represent, at leading order in the perturbative expansion, the differential operators of the quantum, classical and volume terms, respectively. This manifestly shows the physical origin of the three terms. Working at fixed de Sitter radius, we note that $R_Q$ is a purely quantum effect (as it vanishes for $\hbar \to 0$), $R_C$ derives from classical mechanics (as it is independent of $\hbar$ and $M_P$) and $R_V$ is a purely gravitational effect related to space expansion (as it vanishes for $M_P \to \infty$). In other words, the FPV dynamics originates from an intimate interplay of different physical effects, when quantum mechanics, classical physics and general relativity come together in the description of phenomena occurring during the inflationary epoch.

In terms of the spectral modes $\tpv$ defined in \eq{specrep}, the perturbative FPV equation in \eq{FPVancora} becomes
\beq
({\tpv}^{\prime}+ 2v' {\tpv})^{\prime}
+ (  3\GAUG v -\LAMB ) \tpv   =0 ~.
\label{FPVmod}
\eeq
This equation describes the stationary modes of the FPV distribution for a scalar field 
whose potential gives only a small field-dependent modulation of the background vacuum energy.
All local information related to the microscopic dynamics of the scalar field is encoded in the function $v$, while the eigenvalues $\LAMB$ contain information related to physics beyond the range of the effective theory, which is described by the boundary conditions at the field endpoints. 

If a mode $\tpv$ has a local Gaussian peak at $\vfup = {\bar \vfup}$ (such that ${\tpv}^\prime ( {\bar \vfup} ) =0$) with width $\sigma^2 \equiv -{\tpv}( {\bar \vfup} ) /{\tpv}^{\prime\prime}( {\bar \vfup} )$, then \eq{FPVmod} gives a relation between the peak parameters and the corresponding eigenvalue $\LAMB$
\beq
3\GAUG v({\bar \vfup} ) + 2v''({\bar \vfup} )=  \LAMB +\frac{1}{\sigma^2} ~.
\label{relazpeak}
\eeq

\subsection{Junction Conditions}
 
\subsubsection*{Potentials with kinks}
 
In applications to critical phenomena, one is confronted with scalar potentials that are not differentiable at a critical point $\vfup_c$
\beq
v (\vfup) =v_-(\vfup)\, \Theta (\vfup_c-\vfup ) + v_+(\vfup )\, \Theta (\vfup -\vfup_c) ~.
\label{potpm}
\eeq
Here $v_\pm (\vfup)$ describe the potential on either side of the critical point and $\Theta$ is the Heaviside step function 
\beq
\Theta (x) = \left\{ \begin{array}{l}1~~{\rm for}~x>0 \\ 0~~{\rm for}~x<0  \end{array}\right.  ~.
\label{heaviside}
\eeq
We will first consider the case in which the potential is continuous at the critical point, $v_+(\vfup_c ) = v_-(\vfup_c )$,  while its first derivative is not, $v^\prime_+(\vfup_c ) \ne v^\prime_-(\vfup_c )$. 

Since the potential is continuous, the FPV distribution and its time derivative must be continuous as well, but its field derivative may not. The corresponding junction condition at the critical point, for each spectral mode $\tpv (\vfup, \LAMB )$, is obtained by integrating \eq{FPVmod} in a neighbourhood of $\vfup_c$
\beq
\lim_{\epsilon \to 0} \int_{\vfup_c -\epsilon}^{\vfup_c +\epsilon} d\vfup \, ({\tpv}^{\prime}+ 2v' {\tpv} )^\prime =0 ~~~\Rightarrow ~~~
\frac{\Delta \tpv^\prime}{\tpv (\vfup_c)} =-2\Delta v' ~,
\label{junk}
\eeq
where $\Delta v' \equiv v_+^\prime (\vfup_c) -v_-^\prime (\vfup_c)$ and $\Delta \tpv^\prime
\equiv \tpv^{\mathsmaller{\mathsmaller{\mathsmaller{\mathsmaller{\mathsmaller{\mathsmaller (+)}}}}}\, \prime} (\vfup_c)-
\tpv^{\mathsmaller{\mathsmaller{\mathsmaller{\mathsmaller{\mathsmaller{\mathsmaller (-)}}}}}\, \prime} (\vfup_c)
$, with $\tpv^{\mathsmaller{\mathsmaller{\mathsmaller{\mathsmaller{\mathsmaller{\mathsmaller (\pm)}}}}}}$ being the spectral modes on either side of the critical point.

On both sides of the critical point, the spectral modes can be written as a linear combination of two independent solutions of \eq{FPVmod} for the corresponding potentials $v_\pm$
\beq
\tpv^{\mathsmaller{\mathsmaller{\mathsmaller{\mathsmaller{\mathsmaller{\mathsmaller (\pm)}}}}}}
 (\vfup )=  e^{-v_\pm(\vfup)}\left[ \gupm (\LAMB )\, \Phiupm (\vfup ,\LAMB ) +\gdpm (\LAMB )\, \Phidpm (\vfup ,\LAMB )\right] ~.
\eeq
While one pair of functions $\gudpm$ is fixed by the boundary conditions at the endpoints of the field range and by initial conditions, as explained in sects.~(\ref{sec:boundary}) and (\ref{sec:initial}), the other pair is determined by the junction condition in \eq{junk}  together with the continuity condition $\Delta \tpv (\vfup_c) =0$, which give
\beq
\! \! \! 
\gudp (\LAMB )=\left. \frac{\gudm W[\Phiudm ,\Phidup ]
+\gdum W[\Phidum ,\Phidup ]
+(\gudm \Phiudm +\gdum \Phidum )\Phidup \, \Delta v'}
{W[\Phiudp ,\Phidup ]}  \right|_{(\vfup_c,\LAMB)} 
\label{match}
\eeq
where the Wronskian is defined as
$W[\Phi_1,\Phi_2]= \Phi_1 \Phi_2^{\prime} - \Phi_1^{ \prime} \Phi_2$. Note that $W[\Phiupm ,\Phidpm ]$ is independent of $\vfup$ because $\Phiudpm$ satisfy linear second-order differential equations without first derivatives.

\subsubsection*{Multivalued potentials}

Another example of junction conditions, which is relevant to physical applications, is the case of a multivalued potential with two branches corresponding to two different phases. The potentials $v_{a,b}(\vfup)$ on the two branches are both continuous and differentiable, but branch $b$ has a termination point and exists only for $\vfup \ge \vfup_c $. Assume that the field cannot make any transition between the two branches, except at the critical point $\vfup_c$ where the field undergoes an instantaneous phase transition from $b$ to $a$, and take
  $v_b (\vfup_c ) > v_a (\vfup_c )$.

This physical setup imposes the junction condition 
\beq
\tpv^{(b)} (\vfup_c ) =0
~.
\label{appeqquno}
\eeq 
The junction conditions on the $a$ branch are obtained from considering the sum of the two FPV in \eq{FPVmod} on the two branches
\beq
(\tpv^{(a)\, \prime}+ \tpv^{(b)\, \prime}+2 v'_a \tpv^{(a)}
+2 v'_b \tpv^{(b)})^{\prime} +3 \GAUG (v_a \tpv^{(a)}
+ v_b \tpv^{(b)}) -\LAMB (\tpv^{(a)}+ \tpv^{(b)})=0 ~.
\eeq
By integrating this equation in a neighbourhood of the critical point, as before, and exploiting the continuity of $\tpv^{(a)}$ and $\tpv^{(a)}+\tpv^{(b)}$, we find the junction conditions
\beq
\Delta \tpv^{(a)\, \prime} +\tpv^{(b)\, \prime} (\vfup_c ) =0 ~,~~~~
\Delta \tpv^{(a)} =0~,
\label{appeqqdue}
\eeq
where $\Delta \tpv^{(a)\, \prime} $ is the discontinuity across the critical point, defined as before. 

Equation~(\ref{appeqqdue}) describes flux conservation. Although the FPV does not have a conserved current, there is an effective conservation law because the volume term does not enter the discontinuity.

\section{Solutions for Linear and Quadratic Potentials}

Linear and quadratic potentials offer interesting tools for FPV studies because {\it (i)} they allow for exact analytical solutions and {\it (ii)} they serve as prototypes or reference points for broader classes of theories. In particular, linear potentials are valid local approximations of any monotonic potential and give the leading term of the general class of EFT potentials considered in this paper. Quadratic potentials provide simple models to describe phenomena around minima or maxima of more general potentials. 

\subsection{Linear Potential}
\label{sec:lin}

Consider a potential linear in the field with coupling constant $\Gamma$ 
\beq
V(\phi ) = V_0 + \Gamma \, \phi ~,~~~v(\vfup ) = \gamma \, \vfup  ~,~~~ \gamma \equiv  \frac{\Gamma  \SdS}{6M_P H_0^2}~,
\label{potlin}
\eeq
where, with no loss of generality, we take $\Gamma >0$. 
The perturbative regime, where $\vfup$ gives only a small modulation of the energy density, is valid for field excursions around $\vfup =0$ of size $\Delta \vfup$ with
\beq
\Delta \vfup \ll \frac{\SdS}{\gamma} ~,
\label{linearrange}
\eeq
while the slow-roll condition requires $\gamma < \SdS$.
In this range, we can use the perturbative FPV in \eq{FPVmod}, which becomes
\beq
{\tpv}^{\prime\prime}+ 2\gamma \, {\tpv}^\prime
+ ( 3\GAUG \, \gamma \vfup -\LAMB ) \tpv   =0 ~.
\label{supeqlin}
\eeq

Following the EFT analysis (see \sec{sec:EFT}), we define the following regimes for the strength of the coupling constant $\gamma$.
\beq
\begin{array}{lc}
{\mbox {\bf Classical (C) :}} & ~~~ \gamma \gg \GAUG  \Delta \vfup \\[5pt]
{\mbox{\bf Quantum+Volume (QV) :}}  & ~~~ \sqrt{\GAUG} \ll \gamma \ll \GAUG \Delta \vfup \\[5pt]
{\mbox{\bf Quantum{\boldmath$^2$}+Volume (Q{\boldmath$^2$}V) :}} & ~~~  \frac{1}{\GAUG  \Delta \vfup^3} \ll \gamma \ll \sqrt{\GAUG} \\[5pt]
{\mbox {\bf Quantum (Q) :}} & ~~~\gamma \ll \frac{1}{\GAUG  \Delta \vfup^3}
\end{array}
\nonumber
\eeq

The most general solution of \eq{supeqlin} is (see \sec{sec:airy} for definitions)
\beq
\pv (\vfup ,t )=\sum_{\LAMB} \, e^{3H_0t+{\LAMB}\tau
 -\gamma \vfup}\left[ g_1( \LAMB)\, \Ai(x) + g_2( \LAMB) \,  \Bi(x) \right] ~,
\eeq
\beq 
x=\frac{\gamma^2+\LAMB}{c^2} -c\, \vfup~,~~~c=(3\GAUG \gamma )^{1/3}~.
\label{deftanti}
\eeq

The first solution has a peak at $\vfup = {\bar \vfup}$ with width $\sigma$ such that 
\beq
{\bar \vfup} = \frac{\LAMB}{3\GAUG \gamma} ~,~~~~\sigma =\sqrt{\frac{2}{3\GAUG}} ~~~({\rm C~or~QV~regime})~,
\eeq
\beq
{\bar \vfup} = \frac{\LAMB}{3\GAUG \gamma} -\frac{a_1'}{c}~,~~~~\sigma =\frac{1}{\sqrt{-a_1'}\, c} ~~~({\rm Q}^2{\rm V~regime})~,
\eeq
and turns negative, entering a subsequent oscillatory regime, for \beq
\vfup >  \frac{\LAMB}{3\GAUG \gamma} +\frac{\gamma}{3\GAUG}
- \frac{a_1}{c}~,
\eeq
where $a_1 = -2.34$ and $a_1'=-1.02$ are the first zeros of the function $\Ai(x)$ and its derivative. Because of the bound on the maximum eigenvalue in \eq{lmax}, the peak location ${\bar \vfup}$ is always below the upper endpoint of the field range under consideration. In the C and QV regimes, the peak is well approximated by a Gaussian with a universal width $\sigma$, independent of the eigenvalue $\LAMB$ and the coupling constant $\gamma$. In the Q$^2$V regime, the peak corresponds to the first oscillation of $\Ai(x)$ before the function turns negative.

The second solution is monotonically decreasing with $\vfup$ until it turns negative entering a subsequent oscillatory regime for  
\beq
\vfup > \frac{\LAMB}{3\GAUG \gamma} +\frac{\gamma}{3\GAUG}- \frac{b_1}{c}~,
\eeq
where $b_1 = -1.17$ is the first zero of the function $\Bi(x)$.

In the case of $e$-folding gauge ($\GAUG =0$), the parameter $c$ vanishes and the solution is
\beq
\pv (\vfup ,t )=\sum_{\LAMB} \, e^{3H_0t+{\LAMB}\tau
 -\gamma \vfup}\left[ g_1( \LAMB)\, e^{-\sqrt{\gamma^2+\LAMB}\,\vfup} + g_2( \LAMB) \,  e^{\sqrt{\gamma^2+\LAMB}\,\vfup}  \right] ~.
\eeq

\subsubsection{${\boldmath \delta}$-function Initial Condition}
\label{sec:lineardelta}

Consider absorbing boundary conditions set at infinite field values 
\beq
\pv (\vfup \! =\! \pm \infty, t) =0 ~~~~~{\rm (boundary~conditions)}
\label{boundcl}
\eeq
and choose an initial condition with a $\delta$-function as in \eq{deltaf}. The FPV solution is (see identities in \sec{sec:airy})
\beq
\pv (\vfup ,t )= \int \frac{d \LAMB}{c} \,  e^{3H_0t+{\LAMB}\tau -\gamma (\vfup-\vfup_0)}  \, \Ai(x_0) \Ai(x)~,
\label{linboh}
\eeq
where $x_0=x(\vfup_0)$. The boundary conditions in \eq{boundcl} leave the eigenvalue spectrum unconstrained and therefore the summation over $\LAMB$ is replaced here by an integral.

Setting boundary conditions at infinity is inconsistent with our perturbative expansion, which is valid only in the finite field range defined by \eq{linearrange}. 
The choice in \eq{boundcl} is an approximation that allows for a simpler analytical treatment of the realistic case in which the absorbing boundary conditions are set at finite field values $\vfup =\pm \vfupE$. This approximation gives sensible results as long as the FPV solution is localised in a field region $|\vfup |\ll \vfupE$ and 
vanishes exponentially well before reaching the endpoints, thus remaining insensitive to the field value where the absorbing boundary condition is set.  Whenever this is not the case, the approximation of using \eq{boundcl} breaks down and its results can be trusted only up to a time cutoff. To obtain the distribution at times larger than this cutoff, boundary conditions at finite field values must be reinstated. This caveat about setting boundary conditions at infinity applies also to other examples of potentials that we study in following sections.

The spectral integral in \eq{linboh}, which can be performed using \eq{airyid}, gives a Gaussian function peaked at ${\bar \vfup}(t)$, with variance $\sigma (t)$ and normalisation factor $\chi (t)$, defined such that $e^{3H_0 t +\chi}$ measures how much the inflationary expansion makes $\pv$ differ from a true probability distribution,
\beq
\pv (\vfup ,t )=\frac{1}{\sqrt{2\pi \sigma^2}} \exp \left[ 3H_0 t+\chi -\frac{(\vfup -{\bar \vfup})^2}{2\sigma^2}\right] ~,
\label{gauss}
\eeq
\beq
\sigma^2 =  2\tau~,~~~~ {\bar \vfup}=\vfup_0 -\gamma \tau ( 2-3\GAUG \tau )~,~~~~
\chi = 3\GAUG \gamma \tau [ \vfup_0+\gamma \tau (\GAUG \tau -1) ] ~.
\label{parlinear}
\eeq 
where the dimensionless time coordinate $\tau$ is defined in \eq{defpara}.

The variance $\sigma$ grows as the square root of time, with the familiar random-walk behaviour $\langle \phi^2\rangle = H_0^3 t/ (2\pi )^2$. In $e$-folding gauge ($\GAUG =0$), the peak of the distribution ${\bar \vfup}$ slides down the potential, tracking the classical equation of motion. 
The same happens in the C regime for a general gauge ($\GAUG \ne 0$), with the peak rolling down until the time $\tau \sim \vfupE /\gamma \ll 1/\GAUG$, when the approximation of using \eq{boundcl} breaks down. The field value at which the peak eventually stops depends on the exact location of boundary conditions.

In the QV and Q$^2$V regimes, ${\bar \vfup}$ starts by rolling down but, when $\sigma$ reaches the value $\sqrt{2/3\GAUG}$ (and for $|\vfup_0 -\gamma /3 \GAUG |\ll \vfupE$), the peak of the distribution reverses its motion and climbs back the slope. This is the effect of the volume term, which favours patches with large $\phi$, since they inflate faster. The time for ${\bar \vfup}$ to get back where it started is
\beq
t_{\rm idle} =\frac{2 {t_S}}{3\GAUG} ~.  
\label{idle}
\eeq
Note that \eq{idle} fully depends on initial conditions, and this result corresponds to the choice of a $\delta$-function. In practice, $t_{\rm idle}$ can be made arbitrarily short by starting with a broader field distribution. For $t > t_{\rm idle}$, the peak of the distribution continues to climb the potential, eventually growing with constant acceleration, until the approximation breaks down at $\tau \sim \sqrt{ \vfupE / \GAUG \gamma }$. The final location of the peak depends on the boundary conditions and therefore lies beyond the validity of the approximation.

The time $\Delta t$ for ${\bar \vfup}$ to travel a distance $\Delta \vfup$, measured after the end of the idling phase for the quantum regime, is 
\beq
\frac{\Delta t}{t_S} 
\approx
\left\{ 
\begin{array}{cc}
\frac{\Delta \vfup }{2\gamma} &({\rm C~regime}) \vspace{0.2cm}  \\
\sqrt{ \frac{\Delta \vfup}{3\GAUG \gamma}}  &({\rm QV/Q}^2{\rm V~regimes} )
\end{array}   \right. ~.
\eeq
When the motion is in the C regime ($\gamma\gg \GAUG \Delta \vfup$), the time $\Delta t$ for the peak to explore the field range $\Delta \vfup$ is smaller than $t_S$. When the motion is in the QV or Q$^2$V regime ($\gamma\ll \GAUG \Delta \vfup$), the time $\Delta t$ is larger than $t_S$. 

The approximation of \eq{boundcl} allows for a simple analytical treatment of the early stages of the time evolution, starting from a sharply localised distribution. It makes manifest the difference between the dynamics in the classical and quantum regimes. In the classical regime, the dynamics follows the corresponding FP behaviour, with the peak of the distribution tracking the classical equation of motion and its width growing as the square root of time. The typical timescale of the evolution is smaller than $t_S$. In the quantum regime, there is a sharp change of behaviour at $t\sim t_S$, when the volume term in the FPV starts playing a crucial role and the peak steers in a direction opposite to classical motion, towards maximising the potential energy. The hallmark in the change of behaviour is the width, with $\sigma \ge \sqrt{2/3\GAUG}$ (i.e.~having super-Planckian size, when measured in physical units) signalling the new dynamical FPV regime. Therefore, the time scale to reach the asymptotic configuration is larger than $t_S$.

While the approximation of \eq{boundcl} gives us information about early dynamics, it is unable to describe the approach to the asymptotic state in the case of a linear potential. The reason is that, for a linear potential with absorbing boundary conditions, the asymptotic state is always affected by the endpoints of the field range (by the lower endpoint for the classical regime and the upper endpoint for the quantum regime). To determine the asymptotic distribution, we need to reinstate the boundary conditions at a finite field point, as done in the following section.

\subsubsection{Bounded Field Range}

Consider a bounded field range with absorbing boundary conditions at the endpoints
\beq
\pv (\vfup \! =\! \pm \vfupE,t)=0 ~~~~~{\rm (boundary~conditions)},
\label{boundarclas}
\eeq
where $ \vfupE  \ll \SdS /\gamma$ so that the perturbative approach is valid for all field values. 

The most general FPV solution that satisfies the boundary conditions in \eq{boundarclas} is
\beq
\pv (\vfup ,t )=\sum_{\LAMB} \, e^{3H_0t+{\LAMB}\tau
 -\gamma \vfup}\, g(\LAMB) \left[ \frac{ \Ai(x)}{\Ai (x_{\! \mathsmaller E_-})} - \frac{ \Bi(x)}{\Bi (x_{\! \mathsmaller E_-})}  \right] ~,
\label{sumlinno}
\eeq
\beq
x=\frac{\gamma^2 +\LAMB}{c^2} -c\vfup~,~~~x_{\! \mathsmaller E_\pm} =x(\vfup \!=\! \pm  \vfupE ) ~,
\eeq
where $g(\LAMB )$ are arbitrary constants and the eigenvalues are determined by the equation $\Ai (x_{\! \mathsmaller E_+})/\Ai (x_{\! \mathsmaller E_-}) = \Bi (x_{\! \mathsmaller E_+} )/\Bi (x_{\! \mathsmaller E_-})$.

As long as we are not in the Q regime, we can consider the limit $c\, \vfupE\! \gg \! 1$, in which \eq{sumlinno} becomes
\beq
\pv (\vfup , t) =\sum_{n=1}^\infty  e^{3H_0t+\LAMB_n\tau-\gamma \vfup} \, g_n \left[ 2\, \Ai (x_n) - e^{-\frac43 x_{\! \mathsmaller E_-}^{3/2}} \, \Bi (x_n) \right]~,
\label{sumlin}
\eeq
\beq
x_n=a_n+c\, (\vfupE - \vfup )~,~~~x_{\! \mathsmaller E_-} =a_n+2c\, \vfupE ~,
\eeq
where, for the eigenvalues that satisfy $|a_n|\! \ll \! c\, \vfupE$ and up to corrections exponentially suppressed in $x_{\! \mathsmaller E_-}$,
\beq
\LAMB_n =  a_n (3\GAUG \gamma)^{\frac23}+ 3\GAUG \gamma \vfupE -\gamma^2  ~,
\label{eigenlibou}
\eeq
with $a_n$ being the zeroes of the Airy function $\Ai (x)$. 

Whenever the distribution is localised away from the lower endpoint, the contribution from $\Bi (x)$ in \eq{sumlin} is negligible and can be safely dropped. Effectively, this corresponds to setting the lower endpoint of the field range to minus infinity, without affecting the result. Only when the distribution is sizeable in the neighbourhood of the lower endpoint, the location of the boundary condition matters.

The asymptotic state at large times is obtained by retaining in \eq{sumlin} only the contribution from the maximum eigenvalue, which corresponds to $n\! =\! 1$ ($a_1\!=\! -2.34$). The asymptotic distribution is peaked at the field value ${\bar \vfup}$, which is given by the solution of the equation
\beq
\frac{\Ai^{\, \prime} ({\bar x} )}{\Ai({\bar x} )} =-\frac{\gamma}{c} +\Big( \frac{\Bi^{\, \prime} ({\bar x} )}{\Bi({\bar x} )} +\frac{\gamma}{c} \Big) \frac{\Bi({\bar x} )\, e^{-\frac43 \xE^{3/2}}}{2\, \Ai({\bar x} )} ~,
~~~~
{\bar x}= a_1+c\, (\vfupE-{\bar \vfup}) ~.
\eeq
The width of the distribution, defined as $\sigma^2 =- \pv ({\bar \vfup})/\pv^{\prime \prime} ({\bar \vfup})$, is given by
\beq
\sigma^2 = \frac{1}{\gamma^2 -c^2 \, {\bar x} } ~.
\eeq
Whenever ${\bar \vfup}$ is far from the lower endpoint, one can simply take the limit $\exp ( {-\frac43 \xE^{3/2}}) \to 0$. The solution behaves differently in the three following regimes.

\subsubsection*{Q\boldmath$^2$V regime}
For $\gamma \ll \sqrt{\GAUG}$, the peak of the asymptotic distribution is such that
\beq
{\bar \vfup} = \vfupE- \frac{a_1^\prime -a_1 }{(3\GAUG \gamma)^{\frac13}}
~,~~~
\sigma = \frac{ 1}{\sqrt{-a_1^\prime}\, (3\GAUG \gamma)^{\frac13}} 
~.
\eeq
As expected from \eq{relazpeak}, this result satisfies $3\GAUG v({\bar \vfup} )= \LAMB_{\rm max}+\sigma^{-2}$. 
As the potential gets shallower, the peak moves away from the endpoint, although its width grows proportionally. As a result, in the Q$^2$V regime, the peak is always located within a fixed-width distance from the endpoint, since $(\vfupE-{\bar \vfup})/\sigma \approx 1.3$.

\subsubsection*{QV regime}
For $\sqrt{\GAUG}\ll \gamma \ll \GAUG \vfupE$, the peak of the asymptotic distribution is such that
\beq
{\bar \vfup} = \vfupE- \frac{\gamma}{3\GAUG}
~,~~~
\sigma = \sqrt{\frac{ 2}{3\GAUG}}  ~,
\label{peakinQV}
\eeq
which also satisfies \eq{relazpeak}. The width $\sigma$ is independent of the coupling $\gamma$ and has Planckian size in physical units. Note that the asymptotic value of $\sigma$ is smaller than the maximum value of the width $\sigma \sim (\vfupE /\GAUG \gamma )^{1/4}$ that was found in \sec{sec:lineardelta} from the time evolution following a $\delta$-function initial condition. A numerical analysis of \eq{sumlinno} with a localised initial condition indeed shows that $\sigma$ grows as the square root of time, reaching super-Planckian values, but eventually decreases and converges to its asymptotic value given in \eq{peakinQV}.

The peak location moves away from the endpoint as the potential gets steeper. However, its relative distance from the upper endpoint of the field range can be made arbitrarily small in the QV regime, as long as $\gamma$ is much smaller than $\vfupE$. Nonetheless, the peak is always separated from the endpoint by a distance that is parametrically larger then the width, since $(\vfupE-{\bar \vfup})/\sigma ={\mathcal O} (\gamma /\! \sqrt{\GAUG}) \gg 1$. 

\subsubsection*{C regime}
As we increase the value of $\gamma$ and enter the C regime ($\gamma \gg \GAUG \vfupE$), the peak slides down the potential and the result becomes sensitive on how we set boundary conditions at the lower endpoint of the field range. In this case, the contribution from $\Bi (x)$ in \eq{sumlin} is important and peak of the asymptotic distribution is such that
\beq
{\bar \vfup} = -\vfupE + \frac{1}{\gamma} ~,~~~~ \sigma = \frac{1}{\gamma} ~.
\eeq
The peak gets closer to the lower endpoint as the potential becomes steeper and it is always only 1-$\sigma$ away from it. However, our definition of $\sigma$ describes only the local property of the peak at its maximum, while in reality the distribution is asymmetric and more spread than $1/\gamma$ on the side opposite to the lower endpoint.

\subsection{Quadratic Potential (Positive Curvature)}
\label{sec:quadp}

Consider a potential quadratic in the field with positive curvature ($M^2>0$)
\beq
V(\phi )=V_0+\frac{M^2\phi^2}{2} ~,~~~v(\vfup) = \frac{m^2\vfup^2}{2}  ~,~~~
m^2 \equiv  \frac{\SdS  M^2}{6H_0^2}~.
\label{potquad}
\eeq
The perturbative regime, where $\phi$ gives only a small modulation of the energy density, is valid for field excursions around $\vfup =0$ of size $\Delta \vfup$ with
\beq
\Delta \vfup \ll \sqrt{ \frac{\SdS}{m^2}} ~,
\label{quadrrange}
\eeq
while the slow-roll condition requires $m^2 < \SdS$, which corresponds to $M<H_0$.
In this range, we can use the perturbative FPV in \eq{FPVmod}, which becomes
\beq
 \tpv^{\prime \prime}+ 2 m^2 \vfup \, \tpv^{\prime}+\Big(  
\frac32 \,{\GAUG \, m^2\vfup^2} +2m^2 -\LAMB \Big) \tpv =0 ~.
\eeq

Following the EFT analysis (see \sec{sec:EFT}), we define four different regimes for $m^2$.
\beq
\begin{array}{lcc}
{\mbox {\bf Classical (C) :}} & ~~~ \GAUG \ll m^2 \ll \frac{\SdS}{\Delta \vfup^2}  & \\[5pt]
{\mbox{\bf Quantum+Volume (QV) :}}  & ~~~ \frac{\sqrt{\GAUG}}{\Delta \vfup}\ll m^2 \ll \GAUG &
({\rm with~}  \frac{\GAUG}{\SdS} \ll m^2  ) \\[5pt]
{\mbox{\bf Quantum{\boldmath$^2$}+Volume (Q{\boldmath$^2$}V) :}} & ~~~ \frac{1}{\GAUG {\Delta \vfup^4}} \ll   m^2 \ll \frac{\sqrt{\GAUG}}{\Delta \vfup} & \\[5pt]
{\mbox {\bf Quantum (Q) :}} & ~~~  m^2 \ll \frac{1}{\GAUG {\Delta \vfup^4}}  &
\end{array}
\nonumber
\eeq
Combining the definition of the QV regime with \eq{quadrrange}, we find that $m^2$ must necessarily satisfy $\GAUG/\SdS \ll m^2 \ll \GAUG$, whenever the system is in the QV regime. Therefore, when $m^2< \GAUG/\SdS$ or equivalently $M < \sqrt{3 \GAUG /2}\, \hbar H_0^3/( 4 \pi^2 M_P^2)$, the system must necessarily be in the Q$^2$V regime, regardless of the field range $\Delta \vfup$.

The most general FPV solution, for $m^2 > 3\GAUG /2$, is (see \sec{sec:weber} for definitions)
\beq
\pv (\vfup , t )=\sum_{\LAMB} \, e^{3H_0t+{\LAMB}\tau-m^2(1+R)\frac{ \vfup^2}{2}} \Big[
g_1(\LAMB) \,  H_\nu ( \WAV \vfup ) + g_2(\LAMB)\,  {H}_{\nu } (-\WAV \vfup )\Big] ~,
\eeq
\beq
\WAV = \sqrt{{m^2R}}~,~~~R=\sqrt{1-\frac{3\GAUG}{2m^2}}~,~~~
\nu = \frac{m^2(1-R)-\LAMB}{2m^2R}
~.
\eeq
Sufficiently away from the minimum ($|\vfup | \gg 1/\WAV$), the solution can be approximated as
\beq
\pv (\vfup ,t )\approx \sum_{\LAMB} \, e^{3H_0t+{\LAMB}\tau} \Big[
{\tilde g}_1(\LAMB) \, e^{-m^2(1+R)\frac{ \vfup^2}{2}}\, |\vfup|^\nu + {\tilde g}_2(\LAMB)\,  e^{m^2(R-1)\frac{ \vfup^2}{2}}\, |\vfup|^{-\nu -1} \Big] ~,
\eeq
with 
\beq
{\tilde g}_1 =(2 \WAV )^\nu g_\Theta ~,~~~
{\tilde g}_2 =\frac{\sqrt{\pi} \,  g_1 g_2}{\Gamma (-\nu) \WAV^{\nu +1}g_\Theta }~,~~~
g_\Theta =\left\{ \begin{array}{cc}
g_1 & {\rm for~} \vfup >0 \\ g_2 & {\rm for~} \vfup <0 \end{array} \right. ~.
\label{condtilde}
\eeq

For $m^2 = 3\GAUG /2 $, the solution is
\beq
\pv (\vfup ,t )= \sum_{\LAMB} \, e^{3H_0t+{\LAMB}\tau-\frac{m^2\vfup^2}{2}} \Big[
g_1(\LAMB) \, e^{-\sqrt{\LAMB -m^2}\,  \vfup} + g_2(\LAMB)\, e^{\sqrt{\LAMB -m^2}\,  \vfup} \Big] ~.
\eeq

For $m^2 < 3\GAUG /2 $, the parameter $R$ becomes imaginary and the solutions, which are obtained by analytic continuation, exhibit an oscillatory behaviour. 

\subsubsection{${\boldmath \delta}$-function Initial Condition}
\label{sec:dqua}

Consider absorbing boundary conditions set at infinite field values (with the proviso explained at the beginning of \sec{sec:lineardelta})
\beq
\pv (\vfup \! =\! \pm \infty, t) =0 ~~~~~{\rm (boundary~conditions)}
\label{initquad}
\eeq
and choose an initial condition with a $\delta$-function as in \eq{deltaf}. The FPV solution for $m^2 > 3\GAUG /2 $ is (see identities in \sec{sec:weber})
\beq
\pv (\vfup ,t)= \frac{\WAV}{\sqrt{\pi}}\, \sum_{n=0}^\infty  \frac{H_n(\WAV \vfup_0)H_n(\WAV \vfup)}{2^{n}\, n!} \, e^{3H_0t+\LAMB_n\tau-m^2(1+R)\frac{ \vfup^2}{2} +m^2(1-R)\frac{ \vfup_0^2}{2}} ~,
\label{pvperqp}
\eeq
where the eigenvalue spectrum is 
\beq
\LAMB_n= m^2 (1-R-2nR )
~,~~ n \in \mathbb{N}_0 ~.
\label{eigendqua}
\eeq
The maximum eigenvalue is obtained for $n=0$ and it is such that
\beq
\LAMB_{\rm max}=m^2(1-R) ~~\Rightarrow~~ \frac{3\GAUG}{4}\le \LAMB_{\rm max}\le \frac{3\GAUG}{2} ~.
\label{lamaxqua}
\eeq

The summation over spectral modes, which can be performed using \eq{herid}, gives the same Gaussian function as in \eq{gauss} with
\beq
\sigma^2 = \frac{1-u^2}{ m^2\, S}
~,~~~
{\bar \vfup} =\frac{2R\, u\, \vfup_0}{S}
~,~~~
\chi=  \ln \sqrt{\frac{2R}{S}} + \frac{R-1}{R}\ln u
+\frac{3\GAUG (1-u^2) \vfup_0^2}{4 S}~,
\label{uffa1}
\eeq
\beq
S=R+1+(R-1)u^2~,~~~ u=\exp (\! -2m^2 R \tau )~.
\eeq
As expected, the asymptotic behaviour at large times satisfies \eq{asympdis}.

In all $\GAUG$-gauges, the peak of the distribution ${\bar \vfup}$ falls to the bottom of the parabolic potential exponentially fast. The width starts by growing as $\sigma \propto t^{1/2}$, but  eventually freezes at the value
\beq
\sigma^2_\infty =\frac{1}{m^2(R+1)} ~.
\label{sigqp}
\eeq

For $m^2<3\GAUG/2$, we are entering the quantum regime and the distribution eventually climbs the potential. Therefore the approximation of using \eq{initquad} is valid only up to a finite time cutoff and eventually breaks down. The asymptotic behaviour can be computed by evaluating $\LAMB_{\rm max}$ from the boundary condition $\pv (\vfup =\pm \vfupE ) =0$, following the same procedure as in the case of the linear potential. Numerical calculations or an analysis based on the discriminant condition discussed in \sec{sec:PS} show that the asymptotic distribution has a symmetric double-peak structure located in the vicinity of the endpoints.

\subsection{Quadratic Potential (Negative Curvature)}
\label{sec:quadn}

Consider a potential quadratic in the field with negative curvature 
\beq
V(\phi )=V_0-\frac{M^2\phi^2}{2} ~,~~~v(\vfup) = -\frac{m^2\vfup^2}{2} ~,
\label{potquadn}
\eeq
with $M^2>0$ and $m^2$ is defined as in \eq{potquad}. The most general solution of the perturbative FPV is 
\beq
\pv (\vfup , t )=\sum_{\LAMB} \, e^{3H_0t+{\LAMB}\tau+m^2(1-{\hat R})\frac{ \vfup^2}{2}} \Big[
g_1(\LAMB) \,  H_{\hat \nu} ( {\hat \WAV} \vfup ) + g_2(\LAMB)\,  {H}_{{\hat \nu} } (-{\hat \WAV} \vfup )\Big] ~,
\eeq
\beq
{\hat \WAV} = \sqrt{{m^2{\hat R}}}~,~~~{\hat R}=\sqrt{1+\frac{3\GAUG}{2m^2}}~,~~~
{\hat \nu} = \frac{-m^2(1+{\hat R})-\LAMB}{2m^2{\hat R}}
~.
\eeq
Sufficiently away from the minimum ($|\vfup | \gg 1/{\hat \WAV}$), the solution can be approximated as
\beq
\pv (\vfup ,t )\approx \sum_{\LAMB} \, e^{3H_0t+{\LAMB}\tau} \Big[
{\tilde g}_1(\LAMB) \, e^{m^2(1- {\hat R})\frac{ \vfup^2}{2}}\, |\vfup|^{\hat \nu} + {\tilde g}_2(\LAMB)\,  e^{m^2(1+ {\hat R})\frac{ \vfup^2}{2}}\, |\vfup|^{-{\hat \nu} -1} \Big] ~,
\eeq
where ${\tilde g}_{1,2}(\LAMB)$ are given by \eq{condtilde} after the replacement $\nu ,\WAV \to {\hat \nu},{\hat \WAV}$.

\subsubsection{${\boldmath \delta}$-function Initial Condition}

Consider absorbing boundary conditions set at infinite field values (with the proviso explained at the beginning of \sec{sec:lineardelta})
\beq
\pv (\vfup \! =\! \pm \infty, t) =0 ~~~~~{\rm (boundary~conditions)}
\eeq
and choose an initial condition with a $\delta$-function as in \eq{deltaf}. The FPV solution is 
\beq
\pv (\vfup ,t)= \frac{{\hat \WAV}}{\sqrt{\pi}}\, \sum_{n=0}^\infty  \frac{H_n({\hat \WAV} \vfup_0)H_n({\hat \WAV} \vfup)}{2^{n}\, n!} \, e^{3H_0t+\LAMB_n\tau+m^2(1-{\hat R})\frac{ \vfup^2}{2} -m^2(1+ {\hat R})\frac{ \vfup_0^2}{2}} ~.
\label{pvperqn}
\eeq
The eigenvalue spectrum is 
\beq
\LAMB_n= - m^2 ( {\hat R}+1+2 n  {\hat R} )
~,~~ n \in \mathbb{N}_0 
\label{eigendquan}
\eeq
with a maximum eigenvalue
\beq
\LAMB_{\rm max}=- m^2 ({\hat R}+1) \approx
\left\{ 
\begin{array}{lc}
 -2m^2 &({\rm C~regime~or}~\GAUG =0 )  \vspace{0.1cm}  \\
 -\sqrt{\frac{3\GAUG m^2}{2}} &({\rm QV~regime}) 
\end{array}   \right. ~.
\label{lamaxquan}
\eeq

The summation over spectral modes gives the same Gaussian function as in \eq{gauss} with
\beq
\sigma^2 = \frac{{\hat u}^2-1}{ m^2\, {\hat S}}
~,~~~
{\bar \vfup} =\frac{2{\hat R}\, {\hat u}\, \vfup_0}{{\hat S}}
~,~~~
\chi=  \ln \sqrt{\frac{2{\hat R}}{{\hat S}}} - \frac{{\hat R}+1}{{\hat R}}\ln {\hat u}
+\frac{3\GAUG ({\hat u^2}-1) \vfup_0^2}{4 {\hat S}}~,
\label{uffauffa1}
\eeq
\beq
{\hat S}={\hat R}+1+({\hat R}-1){\hat u}^2~,~~~ {\hat u}=\exp ( 2m^2 {\hat R}\tau )~.
\eeq
As expected, the asymptotic behaviour at large times satisfies \eq{asympdis}.

In $e$-folding gauge ($\GAUG =0$), the peak of the distribution (${\bar \vfup}$) falls down the hillside slope exponentially fast, unless it starts exactly at the top of the hump ($\vfup_0 =0$). The width starts increasing as $\sigma \propto t^{1/2}$, but then grows exponentially with time. 

In all other gauges ($\GAUG \ne 0$),  $\bar \vfup$ starts by rolling down the hill but then, once $\sigma$ reaches the value $(2/3\GAUG)^{1/2}$, which is of Planckian size in physical units, it reverses its motion climbing back the potential. The `idling time' necessary for $\bar \vfup$ to get back to the initial point is
\beq
\frac{t_{\rm idle}}{t_S}  =\frac{1}{2 m^2{\hat R}}\ln \Big( \frac{{\hat R}+1}{{\hat R}-1}\Big)
\approx
 \left\{ \begin{array}{c} \frac{1}{2m^2} \ln \frac{8{m}^2}{3\GAUG}
  \vspace{0.1cm} \\ 
 \frac{2}{3\GAUG }
 \end{array}\right. 
\begin{array}{c}   ({\rm C~regime}) 
 \vspace{0.2cm} \\
   ({\rm QV~regime}) \end{array} ~.
\eeq
As mentioned before, $t_{\rm idle}$ is only a feature of the chosen initial condition.

At later times, $\bar \vfup$ continues to climb the hill exponentially fast, asymptotically reaching its top. 
The time $\Delta t$ for ${\bar \vfup}$ to travel from $\vfup_0$ to a field value $\vfup_0/\delta$ (with $\delta\gg1$ and $\vfup_0<0$), measured from the end of the idling phase, is 
\beq
\frac{\Delta t}{t_S}  =\frac{\ln \delta}{ 2m^2{\hat R}}
\approx
 \left\{ \begin{array}{c} 
 \frac{\ln \delta}{2m^2}
  \vspace{0.1cm} \\
 \frac{\ln \delta}{\sqrt{6\GAUG m^2}}
 \end{array}\right. 
\begin{array}{c}   ({\rm C~regime}) 
 \vspace{0.2cm} \\
   ({\rm QV~regime}) \end{array} ~.
\eeq
This shows that the typical time scale for the peak to explore the field range is smaller than $t_S$ in the C regime and larger than $t_S$ in the QV regime. 

A cautionary note is warranted.  It would appear from this analysis that even in the C regime the field wishes to ultimately climb the potential.  However, this outcome is the result of the symmetry of the potential and boundary conditions.  The final stationary solution must be symmetric and there is no special point on either slope, since they are infinitely long, hence it can only rest atop the potential.  This is not, however, due to a volume contribution overcoming the classical rolling, as it is in the Q regimes.  Thus, in the C regime, were the potential or the boundary conditions not exactly symmetric, the solution would not come to rest at the top.  As a result, it seems that the `climbing' behaviour with a sub-$t_S$ time scale is possible only as a result of symmetry, rather than dynamics.

The width of the distribution starts growing as $\sigma \propto t^{1/2}$, but then freezes at the asymptotic value
\beq
\sigma^2_\infty = \frac{1}{{m}^2({\hat R}-1)}
\approx
 \left\{ \begin{array}{c}
 \frac{4 }{3\GAUG}
    \vspace{0.2cm} \\
  \sqrt{ \frac{2}{3\GAUG m^2}}
 \end{array}\right. 
\begin{array}{c}   ({\rm C~regime}) 
 \vspace{0.2cm} \\
   ({\rm QV~regime}) \end{array} ~.
   \label{sigqn}
\eeq
The asymptotic width in the QV regime is always larger than in the C regime, where it is independent of $m^2$. 

\subsection{Linear-Linear Potential}

To illustrate the role of junction conditions, in this section we consider FPV solutions for systems with different branches of linear potentials.

\subsubsection{Pyramid Potential}

Consider two linear potentials with different slopes matched together at a critical point, where they form a cusp. With an appropriate choice of coordinates we choose the critical point at the origin and, for simplicity, we take equal and opposite slopes on the two sides of the critical point. Therefore, the potential is
\beq
v (\vfup ) = - \gamma |\vfup | ~.
\eeq

The junction condition in \eq{junk} requires that, at the critical point,
\beq
\frac{\Delta \tpv^\prime}{\tpv (0)}= 4 \gamma ~~~~~{\rm (junction~condition)}.
\label{junk2}
\eeq
 We choose absorbing boundary conditions for a bounded field range
\beq
\pv (\vfup \! =\! \pm \vfupE,t)=0 ~~~~~{\rm (boundary~conditions)},
\label{boundpyrbounded}
\eeq
where $\vfupE  \ll  \SdS /\gamma$ so that the perturbative approach is valid for all field values.

We start by considering the case $\vfupE \to \infty$, which is a valid approximation as long as the distribution is localised away from the endpoints. Using the results obtained in \sec{sec:lin} for the linear potential, we find that the general FPV solution satisfying continuity and boundary conditions at infinity is
\beq
\pv (\vfup ,t )=\sum_{n=1}^\infty \, e^{3H_0t+\LAMB_n\tau
 +\gamma |\vfup |}\, g_n\,  \Ai (Q_n+ c| \vfup |) ~,~~~
Q_{n} \equiv  \frac{\gamma^2 +\LAMB_n }{c^2}~,~~~ c\equiv (3\GAUG \gamma )^{\frac13} ~,
\label{superuf}
\eeq
where $g_n$ are arbitrary constants. The eigenvalues $\LAMB_n$ are determined by the junction condition in \eq{junk2} and are given by the solutions of the equation
\beq
 \frac{\Ai^{\, \prime} (Q_n)}{\Ai (Q_n)} =\frac{\gamma}{c} ~,
 \label{eigenap}
\eeq
with the maximum eigenvalue $\LAMB_{\rm max}$ corresponding to $n=1$. 

The distribution $\pv$ at asymptotically large times has two peaks, one at each side of the critical point, located at the field values ${\bar \vfup}_\pm$ given by the solutions of the equation
\beq
 \frac{\Ai^{\, \prime} (Q_1+ c|{\bar \vfup}_\pm |)}{\Ai (Q_1+ c|{\bar \vfup}_\pm )|} =-\frac{\gamma}{c} ~.
\label{eigenap2}
\eeq
The width of each peak, defined as $\sigma^2 = -\pv ({\bar \vfup}_\pm)/\pv^{\prime \prime} ({\bar \vfup}_\pm)$, is
\beq
\sigma^2 = \frac{1}{\gamma^2 -c^2 (Q_1\pm c{\bar \vfup}_\pm  )} ~.
\label{eigenap3}
\eeq
Equations (\ref{eigenap})--(\ref{eigenap3}) can be solved analytically with power expansions, in the appropriate regimes depending on the size of $\gamma$. 

When ${\bar \vfup}_\pm$ are in the vicinity of the endpoints, the approximation of taking $\vfupE \to \infty$ is no longer valid and we need to extend \eq{superuf}. Taking for simplicity $c\, \vfupE \! \gg \! 1$, the stationary solution corresponding to the maximum eigenvalue is
\beq
\tpv(\vfup , \LAMB_{\rm max})= 
e^{\gamma | \vfup |} \left[ 2\, \Ai (x) -e^{-\frac43 \xE^{3/2}}\, \Bi (x) \right] ~,~~~\LAMB_{\rm max}= a c^2 -\gamma^2~,
\label{distrconB}
\eeq
\beq
x = a + c | \vfup |  ~,~~~ \xE =a + c\vfupE ~,
\eeq
where the parameter $a$, up to negligible corrections proportional to $\exp (-\frac43 \xE^{3/2})$, is the largest solution to the equation
\beq
\frac{\Ai^{\, \prime}(a)}{\Ai (a)}=\frac{\gamma}{c} ~~~\Rightarrow ~~~
a=\left\{ \begin{array}{cc}
a_1 & ({\rm for~}\gamma \gg \sqrt{\GAUG}) \\
a_1^\prime & ({\rm for~}\gamma \ll \sqrt{\GAUG}) 
\end{array} \right. ~.
\eeq
Therefore, $a$ always lies in the range $-2.34\! < \! a\! < \!  -1.02$. 

\subsubsection*{Q\boldmath$^2$V regime}

For $\gamma \ll \sqrt{\GAUG}$, the eigenvalues are
\beq
\LAMB_n = a_n^\prime (3\GAUG\gamma)^{\frac23}+ \frac{(3\GAUG)^{\frac13}\gamma^{\frac43}}{a_n^\prime} + {\mathcal O}(\gamma^2 ) ~,
\eeq
with $\LAMB_{\rm max}$ corresponding to $n=1$. Working at leading order in the $\gamma/\!\sqrt{\GAUG}$ expansion, the location and width of the two peaks of the asymptotic distribution are
\beq
{\bar \vfup}_\pm = \mp \frac{2  \gamma^{\frac13}}{ a_1^\prime (3\GAUG)^{\frac23}}   ~,~~~~
\sigma =\frac{1}{ \sqrt{ -a_1^\prime}\, (3\GAUG \gamma)^{\frac13}} ~.
\eeq
In the Q$^2$V regime, the two peaks are much broader than their separation, since $ ({\bar \vfup}_+-{\bar \vfup}_-)/ \sigma ={\mathcal O}(\gamma^{2/3}/\GAUG^{1/3}) \ll 1 $, and therefore in practice they form a single peak centred at the critical point. The shallower the potential, the broader the distribution becomes.

\subsubsection*{QV regime}

For $\gamma \gg \sqrt{\GAUG}$, the eigenvalues are
\beq
\LAMB_n = -\gamma^2+ a_n (3\GAUG\gamma )^{\frac23} +3\GAUG+ {\mathcal O}(\GAUG^{\frac43} \gamma^{-\frac23} ) 
~,
\label{autov}
\eeq
with $\LAMB_{\rm max}$ corresponding to $n=1$. 
Working at leading order in the $\sqrt{\xi}/\gamma$ expansion, the location and width of the two peaks of the asymptotic distribution are
\beq
{\bar \vfup}_\pm = \pm \frac{ \gamma}{3\GAUG}
~,~~~~
\sigma = \sqrt{\frac{2}{3\GAUG}}~.
\label{sigmauf}
\eeq
For the peaks to lie within the perturbative region defined in \eq{linearrange} we must require $\gamma \ll \sqrt{\xi \SdS}$ (which corresponds to $\Gamma \ll \sqrt{\GAUG} H_0^3$). As $\gamma$ is decreased and the potential gets shallower, the location of the peaks becomes closer to the critical point. However, the two peaks are always well separated in units of width since $ ({\bar \vfup}_+-{\bar \vfup}_-)/ \sigma ={\mathcal O}(\gamma /\! \sqrt{\GAUG}) \gg 1 $.

\subsubsection*{C regime}

As $\gamma$ grows larger, the peaks move further away from the critical point and eventually the approximation $\vfupE \to \infty$ is inconsistent with perturbativity, but the properties of the asymptotic distribution can be derived from \eq{distrconB}. For $\gamma/ \!\sqrt{\GAUG} \! \gg \!  \sqrt{\GAUG} \vfupE \! \gg \! 1$, the asymptotic locations and width of the two peaks are
\beq
{\bar \vfup}_\pm = \pm \Big(\vfupE - \frac{1}{\gamma} \Big) ~,~~~~ \sigma = \frac{1}{\gamma} ~,
\eeq
and the maximum eigenvalue is $\LAMB_{\rm max} = -\gamma^2 +a_1 (3\GAUG \gamma)^{2/3}$.
The peaks settle as low as possible in the potential, compatibly with boundary conditions.

\bigskip

Notice the difference between the outcomes in the C regime for the pyramid potential and the negative-curvature quadratic potential studied in \sec{sec:quadn}. For the pyramid, the peak of the distribution asymptotically reaches the maximum of the potential only in the Q$^2$V regime, while it settles somewhere along the slope but close to the top in the QV regime and slides down in the C regime. Instead, for the quadratic potential with negative curvature, the peak of the distribution asymptotically reaches the maximum in all regimes. The difference is related to the fact that in the quadratic potential with negative curvature the peak is forced to the top even in the C regime by the symmetry of the boundary conditions, which are placed at infinity, and the form of the potential whose vanishing gradient at the top requires, through symmetry, a vanishing gradient for the solution also.  This does not occur for the pyramid, despite the symmetry, since the junction condition allows for a non-vanishing gradient at the top of the slope, consistent with the symmetry of the system.

\subsubsection{Cliff Potential}

Consider two potentials juxtaposed on the two sides of a critical point $\phi_c$, chosen to be at the origin
\beq
V(\phi ) =   \Theta (-\phi)\, V_-(\phi) + \Theta (\phi)\, V_+(\phi)  ~.
\label{potlinlinlin}
\eeq
Assume that $V_+$ is in the classical regime, while $V_-$ is in a quantum regime, and impose absorbing or reflecting boundary conditions at some values of the field well separated from the critical point. In a neighbourhood of the critical point, we can always approximate $V_\pm$ as two linear potentials with corresponding couplings
\beq
\gamma_\pm = \frac{V'_\pm (0)  \SdS}{6M_P H_0^2} ~.
\eeq
Under these assumptions, the general study of the FPV in the classical regime presented in \sec{sec:determ} shows that the solution proportional to $\Bi(x)$ is singled out just above the critical point, while the solution proportional to $\Ai(x)$ dominates below. Therefore, in the vicinity of the critical point, the FPV solution that satisfies boundary conditions and continuity takes the form
\beq
\pv (\vfup ,t )=\sum_{\LAMB} \, e^{3H_0t+\LAMB \tau}\, g(\LAMB )\, \bigg[  \Theta (-\vfup)\, e^{-\gamma_- \vfup } \, \frac{\Ai(x_-)} {\Ai(Q_{-})} 
+  \Theta (\vfup)\, e^{-\gamma_+ \vfup } \, \frac{ \Bi(x_+)}{ \Bi(Q_{+})}
 \bigg] ~,
\eeq
\beq
x_\pm =Q_{\pm} - c_\pm\vfup 
~,~~~
Q_{\pm} \equiv  \frac{\gamma_\pm^2 +\LAMB }{c_\pm^2}~,~~~ c_\pm \equiv (3\GAUG \gamma_\pm )^{\frac13} 
~,
\eeq
where $g(\LAMB )$ are arbitrary constants. The eigenvalues $\LAMB$ are determined by the junction condition in \eq{junk}, $\Delta \tpv^\prime / \tpv (0)= 2( \gamma_- -\gamma_+)$,
which corresponds to
\beq
c_+ \, \frac{\Bi^\prime (Q_{+})}{\Bi (Q_{+})} - c_- \, \frac{\Ai^\prime (Q_{-})}{\Ai (Q_{-})} =\gamma_+ - \gamma_- ~.
\label{cpiueg}
\eeq

Since the system is in the C regime for positive $\vfup$, the volume-weighted distribution will preferentially populate low-potential field values and therefore will be pushed towards the critical point. If the maximum eigenvalue is positive, the distribution for negative $\vfup$ will favour high-potential field values. As a result, the distribution will accumulate at the transition between the two regimes to form a peak. 

The condition to have a positive eigenvalue and keep the C regime above the critical point up to a value $\vfup_E$ is
\beq
\sqrt{\GAUG} \ll \GAUG \vfup_E \ll \gamma_+ \ll (\GAUG^2 / \gamma_- )^{\frac13} ~.
\eeq
Under this condition, we can expand \eq{cpiueg}. Since $Q_+$ and $Q_-$ are both large and positive when evaluated at the maximum eigenvalue, we can use the asymptotic expansions for the Airy functions finding that, at leading order, the maximum eigenvalue is
\beq
\LAMB_{\rm max} =\left( \frac{3\GAUG}{4\gamma_+}\right)^2 -2\gamma_+ \gamma_-~.
\eeq

As expected, the asymptotic distribution is monotonically increasing for negative $\vfup$ and decreasing for positive $\vfup$ with a cusp at the critical point. The peak at ${\bar \vfup} =0$ is asymmetric and the spread on either side of the critical point is characterised by the quantities
\beq
\sigma_- \equiv \frac{\pv(\vfup\! =\! 0_-,t_\infty )}{\pv^\prime (\vfup\! =\! 0_-,t_\infty )}= \frac{4\gamma_+}{3\GAUG}~,~~~
\sigma_+ \equiv -\frac{\pv(\vfup\! =\! 0_+,t_\infty )}{\pv^\prime (\vfup\! =\! 0_+,t_\infty )}= \frac{1}{2\gamma_+}~.
 \eeq

The distribution is well localised in the negative $\vfup$ region because the spread is much smaller than the full perturbative field range ($\sigma_-/\Delta \vfup \approx \gamma_+ \gamma_- /(\GAUG \SdS) \ll 1$). It is also well localised in the positive $\vfup$ region because $\sigma_+/\vfupE \approx 1/(\gamma_+ \vfup_E)\ll 1$.  The spread of the distribution in the negative $\vfup$ region largely exceeds the field interval in the positive $\vfup$ region where we can trust the perturbative classical regime of the potential. However, this is just an artefact of our linear approximation of the potential. By taking a potential $V_+$ that satisfies the classical behaviour beyond the perturbative domain, one can extend the validity of the model up to arbitrarily large field values. 

Because of the strong asymmetry in the distribution, the average field value is not at the origin but at $\langle \vfup \rangle \approx \sigma_+\! -\! \sigma_-$. Since $\sigma_- \gg \sigma_+$, the average field value occurs for large and negative $\vfup$, although the most probable value is $\vfup =0$. The probability of finding the field in the negative region is also much larger than in the positive region, since 
\beq
\frac{\int_{\vfup <0} d \vfup \, \pv}{  \int_{\vfup >0}d \vfup \, \pv} \approx \frac{8\gamma_+^2}{3\GAUG} \gg 1 ~.
\eeq 

In the cliff potential, the origin of the localisation is directly linked to the junction condition at the critical point $\phi_c$. Indeed, if the distribution to the left of $\phi_c$, called $\pv_-$,  has a gradient at the critical point satisfying
\beq
0 < \frac{\pv_-^\prime (\phi_c,t)}{\pv_-(\phi_c,t)} <  \frac{24 \pi^2 M_P^4\, \Delta V'}{\hbar \, V^2(\phi_c)} ~,
\label{eq:SOLmatch}
\eeq
then the junction condition implies that the solution $\pv_+$ on the right of $\phi_c$ must  take a value
\beq
\frac{{\pv_+}'(\phi_c,t)}{\pv_+ (\phi_c,t)} < 0 ~.
\eeq
Since the gradient of $\pv$ changes sign at $\phi_c$, a local peak must exist at the critical point, at any time during which \eq{eq:SOLmatch} is satisfied. In particular, if \eq{eq:SOLmatch} is satisfied at asymptotically large times, the peak will persist indefinitely.  

A remarkable feature of the cliff potential is that the localisation of the field occurs at a point which is neither a minimum nor a maximum of the potential. The localisation is entirely driven by the junction conditions imposed by a discontinuity of the potential gradient. However, the cliff potential has limited interest for critical phenomena, since it requires a positive discontinuity $\Delta V' >0$, while a first-order phase transition usually leads to a negative gradient discontinuity.

\subsection{EFT Notation}
\label{sec:EFT}

In the body of this paper we have used a compact parametrisation of the potential which is particularly suitable for an EFT treatment. The potential is expressed in terms of an overall coupling $\ett$ and the field range $f$, according to
\beq
V(\phi ) = V_0 +\ett^2 f^4 \fun (\vf ) ~,~~~~ \vf \equiv \frac{\phi}{f} ~,
\eeq
where $\fun (\vf )$ is a generic function such that $\fun(0)=0$ and $|\fun (\vf)|\le 1$. The perturbative domain corresponds to $\ett^2 f^4 \ll V_0$. 

The parameters that characterise the behaviour of the FPV solutions are
\beq
\alpha \equiv \frac{3 \hbar H_0^4}{4 \pi^2 \ett^2 f^4} ~,~~~~ \beta \equiv \frac{3\GAUG f^2}{2 M_P^2}~,~~~~   t_R \equiv \frac{3 H_0}{\ett^2 f^2} = \frac{\alpha \beta \SdS}{3\GAUG H_0}
~.
\eeq
 A further advantage of this parametrisation is that the gauge variable $\xi$ is completely reabsorbed into the definition of $\beta$.

The parameters $\alpha$ (which measures the strength of the diffusion term in units of the drift term) and $ \beta $ (which measures the strength of the volume term in units of the drift term) characterise the following regimes, valid for $\beta >1$.
\beq
\begin{array}{lc}
{\mbox {\bf Classical (C) :}} & ~~~ \frac{1}{\SdS}\ll \alpha    \ll \frac{1}{\beta} \\[8pt]
{\mbox{\bf Quantum+Volume (QV) :}}  & ~~~ \frac{1}{\beta} \ll \alpha   \ll \frac{1}{\sqrt{\beta}}  
\\[8pt]
{\mbox{\bf Quantum{\boldmath$^2$}+Volume (Q{\boldmath$^2$}V) :}} & ~~~  \frac{1}{\sqrt{\beta}}  \ll \alpha   \ll \beta  \\[8pt]
{\mbox {\bf Quantum (Q) :}} & ~~~  \alpha    \gg \beta
 \end{array}
\nonumber
\eeq
For convenience, we summarise in table~1 how to convert the notations used in this appendix to those of the EFT approach. 

\medskip

\begin{table}[h]
\begin{center}
\begin{tabular}{c|c|c|c|}
& & Appendix & EFT \\[5pt]
 \hline 
 & &  & \\[-12pt]
Scalar field  & (dimensionful)& $\phi = {\vfup} \, M_P$ & $\phi =  \vf \, f$ \\
& (dimensionless) & $\vfup$ & $ \sqrt{\frac{2\beta}{3\GAUG}} ~\vf $ \\[4pt]
\hline
& &  & \\[-12pt]
Potential & (dimensionful) & $V(\phi ) = V_0 \left[ 1+ \frac{2v(\vfup )}{\SdS}\right] $ & $ V(\phi ) = V_0 +\ett^2 f^4 \fun(\vf )$ \\
& (dimensionless)  & $v(\vfup ) $ & $ \frac{\fun(\vf )}{\alpha}$ \\[4pt]
\hline
& &  & \\[-12pt]
Eigenvalues & (dimensionful) & $ \LAMB /t_S$ & $\lambb /t_R $ \\
& (dimensionless) & $ \LAMB $ & $ \frac{3\GAUG}{\alpha \beta} \, \lambb$ \\[4pt]
\hline
& &  & \\[-12pt]
Linear  & 
Appendix: $v(\vfup )=  \gamma \vfup$  & $ \gamma$ & $  \sqrt{\frac{3\GAUG}{2\alpha^2 \beta}}$ \\
potential &  EFT: $\fun(\vf )= \vf$  & & \\[4pt]
\hline
& &  & \\[-12pt]
Quadratic & 
 Appendix: $v(\vfup )= \frac{m^2 \vfup^2}{2}$ & $ m^2$ & $  \frac{3\GAUG}{2\alpha \beta}$ \\
potential  &  EFT: $\fun(\vf )= \frac{\vf^2}{2}$  & & \\[4pt]
\hline
\end{tabular}
\end{center}
\caption{The conversion between the notations used in this appendix and those of the EFT approach. Quantities in the same line of the table are equal, but expressed in the corresponding convention.} 
\end{table}

\section{Useful Identities of Some Special Functions}

\subsection{Airy's Equation}
\label{sec:airy}

Consider the equation
\beq
{p}^{\prime \prime} +2A {p}^{\prime} +(B+C\vfup) {p} =0 ~,
\label{inita}
\eeq
where $A,B,C$ are generic constants. First, take the case $C\ne 0$. With the definitions
\beq
{p} = e^{-A\vfup}\, \Phi ~,~~~ x=\frac{k}{c^2}-c\, \vfup~,~~~k = A^2-B~,~~~c =C^{1/3} ~,
\label{defaba}
\eeq
\eq{inita} turns into
\beq
\frac{d^2\Phi}{dx^2} -x\, \Phi =0 ~,
\label{airyeq}
\eeq
which is known as Airy's differential equation~\cite{Vallee}.

Two linearly-independent solutions of \eq{airyeq} are given by the Airy functions $\Ai (x)$ and $\Bi (x)$. 
Their power expansions for small argument are
\beq
\Ai (x) =  f(x)- \frac{ g(x)}{\sqrt{3}}~, ~~~\Bi (x) = \sqrt{3}\,  f(x)+g(x)~,
\eeq
\beq
f(x)=a_0\left[ 1+\frac{x^3}{6}+{\mathcal O}(x^6)\right] ~,~~~g(x)=\frac{1}{2\pi a_0}\left[ x+\frac{x^4}{12}+{\mathcal O}(x^7)\right] ~,
\eeq
with $a_0 \equiv \Ai (0) =3^{-2/3}/\Gamma(2/3)=0.355$.

Their asymptotic behaviours are
\beq
\Ai (x) ~\stackrel{x\to +\infty}{\longrightarrow} ~\frac{\exp \big( -\frac23\, x^{\frac32} \big)}{2\sqrt{\pi}\, x^{\frac14}}
\Big[ 1-\frac{5x^{-\frac32}}{48} +{\mathcal O}(x^{-3})\Big] ~,
\eeq
\beq
\Bi (x)~ \stackrel{x\to +\infty}{\longrightarrow} ~\frac{\exp \big( \frac23\, x^{\frac32} \big)}{\sqrt{\pi}\, x^{\frac14}} 
\Big[ 1+\frac{5x^{-\frac32}}{48} +{\mathcal O}(x^{-3})\Big]~,
\eeq
\beq
\Ai (x) ~\stackrel{x\to -\infty}{\longrightarrow} ~\frac{1}{\sqrt{\pi}\, |x|^{\frac14}} \Big[ \cos \Big( \frac23\, |x|^{\frac32}-\frac{\pi}{4}\Big) +\frac{5}{48 |x|^{\frac32}} \sin \Big( \frac23\, |x|^{\frac32}-\frac{\pi}{4}\Big) \Big]
\eeq
\beq
\Bi (x) ~\stackrel{x\to -\infty}{\longrightarrow} ~\frac{1}{\sqrt{\pi}\, |x|^{\frac14}} \Big[ - \sin \Big( \frac23\, |x|^{\frac32}-\frac{\pi}{4}\Big) +\frac{5}{48 |x|^{\frac32}} \cos \Big( \frac23\, |x|^{\frac32}-\frac{\pi}{4}\Big) \Big]
\eeq
\beq
\frac{\Ai^{\, \prime} (x)}{\Ai (x)} ~\stackrel{x\to +\infty}{\longrightarrow} ~ -\sqrt{x} -\frac{1}{4x}+\frac{5x^{-\frac52}}{32} +{\mathcal O}(x^{-4})
~,
\eeq
\beq
\frac{\Bi^{\, \prime} (x)}{\Bi (x)} ~\stackrel{x\to +\infty}{\longrightarrow} ~ \sqrt{x} -\frac{1}{4x}-\frac{5x^{-\frac52}}{32} +{\mathcal O}(x^{-4})
\eeq
and their Wronskian is
\beq
W[\Ai , \Bi ]=\Ai (x) \, \Bi^{\, \prime} (x)- \Ai^{\, \prime} (x)\,  \Bi (x) =\frac{1}{\pi} ~.
\eeq

The Airy functions and their derivatives can be expressed in terms of the modified Bessel functions 
\beq
\Ai (x)=\frac{\sqrt{x}}{3} \left[ I_{-\frac13}(t)-I_{\frac13}(t) \right] ~,~~~
\Bi (x)=\sqrt{\frac{x}{3}} \left[ I_{-\frac13}(t)+I_{\frac13}(t) \right] ~,
\eeq
\beq
\Ai^{\, \prime} (x)=-\frac{x}{3} \left[ I_{-\frac23}(t)-I_{\frac23}(t) \right] ~,~~~
\Bi^{\, \prime}  (x)=\frac{x}{\sqrt{3}} \left[ I_{-\frac23}(t)+I_{\frac23}(t) \right] ~,
\eeq
where $t=(2/3)x^{3/2}$.

The zeroes of the Airy functions and their derivatives 
\beq
\Ai (a_n) =0 ~,~~~\Bi (b_n) =0 ~,~~~  \Ai^{\, \prime} (a_n^\prime) =0 ~,~~~\Bi^{\, \prime} (b_n^\prime) =0
\eeq
are negative numbers approximately given by 
\begin{center}
\begin{tabular}{c|cccc}
 $n$ & 1 & 2 & 3 & ... \\
 \hline
 $a_n$ & $-2.34$ & $-4.09$ & $-5.52$ & ...\\
 $b_n$  & $ -1.17$ & $ -3.27$ & $ -4.83 $ & ...\\
 $a_n^\prime$  & $ -1.02$  & $ -3.25$  & $ -4.82$ & ...\\
 $b_n^\prime $  & $ -2.29$  & $-4.07 $  & $ -5.51$& ...
 \end{tabular}
 \end{center}

The Airy functions of the first kind provide an orthonormal basis 
\beq
\delta(x-y) = \int_{-\infty}^{+\infty} dk \, \Ai(k+x)\, \Ai(k+y)  ~.
\eeq
They also satisfy the identity
\beq
\int_{-\infty}^{+\infty} dk \, e^{kt}\, \Ai(k+x)\Ai(k+y)=\frac{1}{2\sqrt{\pi t}} \exp \left[\frac{t^3}{12}-\frac{(x-y)^2}{4t}-\frac{(x+y)t}{2}\right] ~.
\label{airyid}
\eeq

In summary, the most general solution of \eq{inita} for $C\ne 0$
is
\beq
p (\vfup )= g_1\, e^{-A\vfup}\, \Ai(x) + g_2 \, e^{-A\vfup}\,  \Bi(x) ~,
\eeq
where $x$ is defined in \eq{defaba} and $g_{1,2}$ are two arbitrary constants. 

For $C=0$, the solution is
\beq
p (\vfup )= g_1\, e^{-(A+\sqrt{A^2-B})\vfup} + g_2\, e^{-(A-\sqrt{A^2-B})\vfup} ~.
\eeq

\subsection{Hermite's Equation}
\label{sec:weber}

Consider Hermite's differential equation
\beq
{p}^{\prime \prime} +2A\vfup {p}^{\prime} +(B+C\vfup^2) {p} =0 ~,
\label{initw}
\eeq
where $A,B,C$ are generic constants. First, consider the case $A^2>C$. With the definitions
\beq
{p} = e^{-\frac{A\vfup^2}{2}}\, \Phi ~,~~~ x=\WAV \vfup~,~~~\WAV = (A^2-C)^{1/4}~,~~~\nu = \frac{B-A}{2\, \WAV^2}-\frac12 ~,
\label{defabw}
\eeq
\eq{initw} turns into
\beq
\frac{d^2\Phi}{dx^2} +(2\nu +1 -x^2)\Phi = 0 ~,
\label{weber}
\eeq
which is known as Weber's differential equation.

Two linearly-independent (for non-integer $\nu$) solutions of \eq{weber} are\footnote{The function $\Phi_1(\nu ,x)$ is related to other special functions
$$
\frac{1}{\sqrt{2^\nu}\, c_{\nu}}\Phi_1(\nu ,x)=  D_\nu (\sqrt{2}\, x) =U(-\nu -\frac12 , \sqrt{2}\, x)=\sqrt{\frac{2^\nu}{x}}\, W_{\frac{\nu}{2}+\frac14 , -\frac14}(x^2) ~,
$$
where $D_\nu (z)$ is the Whittaker function defined as {\tt ParabolicCylinderD[{\it nu},z]} in Mathematica,  $U(a,z)$ is the parabolic cylinder function defined in ref.~\cite{Abramowitz}, and $W_{k,m}(z)$ is Whittaker's confluent hypergeometric function defined as {\tt WhittakerW[k,m,z]} in Mathematica.
}
\beq
\Phi_1(\nu ,x) = c_\nu\, e^{-\frac{x^2}{2}} \, H_\nu (x) ~,~~~ \Phi_2(\nu ,x) = c_\nu \,e^{-\frac{x^2}{2}} \, H_{\nu} (-x) ~,
\label{solwebnon}
\eeq
\beq
c_\nu = \left[ 2^{\nu}\sqrt{\pi} \, \Gamma(\nu+1) \right]^{-\frac12} ~,
\eeq
\beq
H_\nu (x)=2^\nu \sqrt{\pi} \left[ \frac{_1F_1\big(-\frac{\nu}{2},\frac12,x^2 \big)}{\Gamma\big( \frac{1-\nu}{2}\big)}-
2x\,\frac{ _1F_1\big(\frac{1-\nu}{2},\frac32,x^2 \big)}{\Gamma\big( -\frac{\nu}{2}\big)}\right] ~.
\label{defher}
\eeq
Here ${_1F_1}(a,b,z)$ is Kummer's confluent hypergeometric function of the first kind, $\Gamma(z)$ is Euler's gamma function and $H_\nu (x)$ is the Hermite function defined in Mathematica as {\tt HermiteH[nu,x]}. 

The function $H_\nu (x)$ satisfies the relations
\beq
H_{\nu +1} (x)=2x H_\nu (x)-2\nu H_{\nu -1} (x)~,
\eeq
\beq
H_\nu (ix)=\frac{2^{\nu}\, \Gamma (\nu +1)}{\sqrt{\pi}}\,  e^{-x^2}\left[ e^{-\frac{i\pi \nu}{2}} H_{-\nu-1}(x) + e^{\frac{i\pi \nu}{2}} H_{-\nu-1}(-x) \right]
~,
\eeq
\beq
H_\nu (x)  { H}_{\nu +1} (-x) +H_{\nu +1} (x)  { H}_\nu  (-x) = 2 \, e^{x^2}b_\nu {\bar b}_\nu~,~~~~~~
b_\nu = 2^\nu ~,~~~{\bar b}_\nu = \frac{\sqrt{\pi}}{\Gamma (-\nu )} ~.
\eeq

The asymptotic behaviours are
\beq
H_\nu (x) ~~\stackrel{x\to +\infty}{\longrightarrow}~~ b_\nu \, x^\nu ~,~~~~
H_\nu (x) ~\stackrel{x\to -\infty}{\longrightarrow}~{\bar b}_\nu \, e^{x^2} |x|^{-\nu -1}~,
\eeq
\beq
H_\nu (x) ~~\stackrel{\nu\to -\infty}{\longrightarrow}~~ b_\nu {\bar b}_{\frac{\nu -1}{2}}\,  \, e^{\frac{x^2}{2}-|x|\sqrt{-2\nu -1}} ~,
\eeq
\beq
\left. \begin{array}{l}
\Phi_1(\nu ,x) ~~\stackrel{x\to +\infty}{\longrightarrow} \\
\Phi_2(\nu ,x) ~~\stackrel{x\to -\infty}{\longrightarrow} 
\end{array} \right\} =c_\nu b_\nu \, e^{-\frac{x^2}{2}}|x|^\nu
~~~~
\left. \begin{array}{l}
\Phi_1(\nu ,x) ~~\stackrel{x\to -\infty}{\longrightarrow} \\
\Phi_2(\nu ,x) ~~\stackrel{x\to +\infty}{\longrightarrow} 
\end{array} \right\}
=c_\nu {\bar b}_\nu \, e^{\frac{x^2}{2}}|x|^{-\nu -1} ~.
\eeq
The power expansion for small $x$ is
\beq
H_\nu (x) =b_\nu {\bar b}_{\frac{\nu -1}{2}} \left[ 1-\nu x^2 -{\scriptstyle \frac{\nu (2-\nu)}{6}}x^4\right]
-b_{\nu +1}  {\bar b}_{\frac{\nu}{2}} \left[ x+{\scriptstyle \frac{1-\nu}{3}}x^3 +{\scriptstyle \frac{(1-\nu)(3-\nu)}{30}}x^5 \right] +{\mathcal O}(x^6)~.
\eeq
The power expansion for small $\nu$ is
\beq
H_\nu (x) ~~\stackrel{\nu\to 0}{\longrightarrow}~~
\left\{ \begin{array}{lc}
1+\nu (\psi +\sqrt{\pi}\, x ) & ({\rm for~} x \to 0) \\
1+\nu \log 2x & ({\rm for~} x \to +\infty )\\
\frac{\nu \sqrt{\pi} \, e^{x^2}}{|x|} [ -1 +\nu (\gamma +\log |x| ) ] & ({\rm for~} x \to -\infty ) 
\end{array}
\right. ~,
\label{exppicnu}
\eeq
where $\psi = \log 2 +\psi^{(0)}(1/2)/2 = -0.29$, $\psi^{(n)}(x)$ is the polygamma function, and $\gamma =0.58$ is the Euler-Mascheroni constant.

The derivative of the function $H_\nu (x)$ is
\beq
\frac{d H_\nu (x)}{dx} = 2\nu H_{\nu -1} (x) ~.
\eeq
The Wronskian of the two solutions is
\beq
W[\Phi_2,\Phi_1]=\Phi_2(\nu ,x)\frac{d \Phi_1(\nu ,x)}{dx}-\frac{d \Phi_2(\nu ,x)}{dx}\Phi_1(\nu ,x)=
\frac{2\,\sin \pi \nu}{\pi} ~,
\eeq
which shows that $\Phi_{1,2}(\nu , x)$ are no longer linearly independent for integer $\nu$. In this case, an independent solution is found by choosing a linear combination of the hypergeometric functions different from the one defined by \eq{defher}.

For non-negative integer $\nu =n$, $H_n (x)$ are the Hermite polynomials and the functions $\Phi_1(n,x)$ form an orthonormal basis
\beq
\int_{-\infty}^{+\infty}dx\, \Phi_1 (n,x)\, \Phi_1 (m,x) = \delta_{nm} ~.
\eeq
They also satisfy the properties
\beq
\sum_{n=0}^{\infty}  \Phi_1 (n,x)\Phi_1 (n,y)=\delta(x-y) ~,
\eeq
\beq
\sum_{n=0}^{\infty} t^n\, \Phi_1 (n,x)\Phi_1 (n,y)=\frac{1}{\sqrt{\pi (1-t^2)}}\exp \left[ \frac{4txy-(1+t^2)(x^2+y^2)}{2(1-t^2)}\right] ~~(0<t<1).
\label{herid}
\eeq

In summary, the most general solution of \eq{initw} for $A^2>C$
is
\beq
p (\vfup )=  e^{-\frac{\vfup^2}{2}(A+\WAV^2)}\left[ g_1\, H_\nu (\WAV \vfup ) + g_2\,  {H}_{\nu } (-\WAV \vfup ) \right]~,
\eeq
where $\WAV$ and $\nu$ are defined in \eq{defabw} and $g_{1,2}$ are two arbitrary constants.  

For $A^2=C$, the solution is
\beq
p (\vfup )= g_1\, e^{-\frac{A\vfup^2}{2}- \sqrt{A-B}\, \vfup }+g_2\, e^{-\frac{A\vfup^2}{2}+ \sqrt{A-B}\, \vfup } ~.
\eeq

For $A^2<C$, we define
\beq
{p} = e^{-\frac{A\vfup^2}{2}}\, {\bar \Phi} ~,~~~ {\bar x}={\bar \WAV} \vfup~,~~~{\bar \WAV} = (C-A^2)^{1/4}~,~~~{\bar \nu} = \frac{B-A}{2\, {\bar \WAV}^2}-\frac12 ~
\label{defabwdark}
\eeq
and \eq{initw} turns into
\beq
\frac{d^2{\bar \Phi}}{d{\bar x}^2} +(2{\bar \nu} +1 +{\bar x}^2){\bar \Phi} = 0 ~,
\label{weberdark}
\eeq
which is Weber's differential equation of the second kind.
The solutions of this equation are obtained from \eq{solwebnon} with the replacements $\Phi \to {\bar \Phi}$, $\WAV \to \sqrt{i}\, {\bar \WAV}$ and $2\nu +1 \to -i (2{\bar \nu} +1)$.

The solutions to \eq{weberdark} have an oscillatory behaviour. For instance, in the limit of fixed ${\bar \nu}$ and positive and large ${\bar x}$, two linearly-independent solutions have the asymptotic behaviour (for ${\bar x}\to +\infty$)
\beq
{\bar \Phi}_1({\bar \nu} ,{\bar x}) =\frac{1}{\sqrt{{\bar x}}} \left( s_1 \cos {\bar X} -s_2 \sin {\bar X} \right)~,~~~
{\bar \Phi}_2({\bar \nu} ,{\bar x}) =\frac{1}{\sqrt{{\bar x}}} \left( s_1 \sin {\bar X} +s_2 \cos {\bar X} \right)~,
\eeq
\beq
{\bar X}=\frac{{\bar x}^2 +(2{\bar \nu}+1)\ln  {\bar x}+\omega}{2} ~,
~~~ \omega = \frac{\pi}{2}+ (2{\bar \nu}+1)\ln \sqrt{2} + \arg  \Gamma \left[ \frac{1-i(2{\bar \nu}+1)}{2}\right] ~,
\eeq
\beq
s_1 =1-\frac{2{\bar \nu} +1}{4{\bar x}^2} + {\mathcal O}( {\bar x}^{-4}) ~,~~~
s_2 = \frac{2 {\bar \nu}({\bar \nu} +1)-1}{8{\bar x}^2} + {\mathcal O}( {\bar x}^{-4}) ~.
\eeq
In the limit of large and negative ${\bar \nu}$, the solutions ${\bar \Phi}_{1,2}$ can be expressed in terms of Airy functions.

\addcontentsline{toc}{section}{References}
\bibliographystyle{mine}
\bibliography{biblio}

\end{document}